\documentclass[paper,toc,nohyper]{JHEP3}
\usepackage{epsfig}
\usepackage{amsbsy,ams} 
\usepackage{subfigure,epsfig}
\usepackage{graphicx}
\usepackage{pifont}
\usepackage{amsmath}
\usepackage{booktabs}
\usepackage{cite}
                 
\def\dsigma{{\rm d} \hat\sigma}
\def\ph#1{\phantom{.}}
\def\JET{J}

\def\Finite{{\cal F}inite}
\def\bom#1{{\mbox{\boldmath $#1$}}}
\def\d{\hbox{d}}

\def\doubletilde#1{\widetilde{\vphantom{\raise 1.5pt \hbox{#1}}\smash{\kern -2pt\widetilde{#1}}}}
\def\e{\epsilon}

\def\eps{\epsilon}

\def\h{\mathrm{H}}

\def\spa#1.#2{\left\langle\mskip-1mu#1\,#2\mskip-1mu\right\rangle}
\def\spb#1.#2{\left[\mskip-1mu#1\,#2\mskip-1mu\right]}

\def\Li{\mathop{\rm Li}\nolimits}

\def\NLO{\rm NLO}

\def\calF{{\cal F}_{3}^{0}}

\def\calS{{\cal S}}

\def\l{\left}
\def\r{\right}

\def\wt{\widetilde}
\def\t{\tilde}

\def\dPSxx#1#2{\int{\rm d}\Phi_{#1}(p_3,\ldots,p_{#2};x_1 p_1,x_2 p_2)\frac{1}{S_{#1}}\,\frac{{\rm d}x_1}{x_1}\frac{{\rm d}x_2}{x_2}}
\def\dPS#1#2{{\rm d}\Phi_{#1}(p_3,\ldots,p_{#2};p_1,p_2)\,\frac{1}{S_{#1}}}
\def\PSs{{\rm d}\Phi_3(p_3,\dots,p_5;p_1,p_2)}

\def\PSh{\int\frac{{\rm d}x_{1}}{x_{1}} \frac{{\rm d}x_{2}}{x_{2}}{\rm d}\Phi_3(p_3,\dots,p_5;\bar{p}_1,\bar{p}_2)}

\def\NLO{{\cal N}_{LO}}

\def\NNNLORV{{\cal N}_{NNLO}^{RV}}
\def\NNNLORR{{\cal N}_{NNLO}^{RR}}

\def\S#1#2#3#4{{\cal S}(s_{#1#2},s_{#3#4},x_{#1#2,#3#4})}

\newcommand{\beq}{\begin{equation}}
\newcommand{\eeq}{\end{equation}}
\newcommand{\bea}{\begin{eqnarray}}
\newcommand{\eea}{\end{eqnarray}}
\def\z#1{{\zeta_{#1}}}

\def\n2f{{n^{\,2}_{\! f}}}
\def\pgg(#1){p_{\rm{gg}}(#1)}
\def\H(#1){{\rm{H}}_{#1}}
\def\Hh(#1,#2){{\rm{H}}_{#1,#2}}

\allowdisplaybreaks

\preprint{
  IPPP/11/78\\
  \\}

\title{Real-Virtual corrections for gluon scattering at NNLO}

\author{Aude Gehrmann-De Ridder$^a$, E.W.N. Glover$^b$, Joao Pires$^{a}$
	\\
$^a$Institute for Theoretical Physics, ETH, CH-8093 Z\"urich, Switzerland\\
	$^b$Institute for Particle Physics Phenomenology, University of Durham,
South Road,\\ Durham DH1 3LE, England
}	

\abstract{We use the antenna subtraction method to isolate the mixed real-virtual infrared singularities present in gluonic scattering amplitudes at next-to-next-to-leading order. In a previous paper, we derived the subtraction term that rendered the double real radiation tree-level process finite in the single and double unresolved regions of phase space.  Here, we show how to construct the real-virtual subtraction term using antenna functions with both initial- and final-state partons which removes the explicit infrared poles present in the one-loop amplitude, as well as the implicit singularities that occur in the soft and collinear limits.  As an explicit example, we write down the subtraction term that describes the single unresolved contributions from the five-gluon one-loop process. The infrared poles are explicitly and locally cancelled in all regions of phase space prior to integration, leaving a finite remainder that can be safely evaluated numerically in four-dimensions.
We show numerically that the subtraction term correctly approximates the matrix elements in the various single unresolved configurations.
\\
}

\begin{document}
\newpage

\section{Introduction}
\label{sec:intro}

In hadronic collisions, the most basic form of the strong interaction at short
distances is the scattering of a coloured parton off another coloured parton.
Experimentally, such scattering can be observed via the production of one or
more jets of hadrons with large transverse energy. In QCD, the (renormalised and mass factorised) inclusive
cross section has the form,
\begin{equation}
\label{eq:totsig}
{\rm d}\sigma =\sum_{i,j} \int   
\frac{d\xi_1}{\xi_1} \frac{d\xi_2}{\xi_2} f_i(\xi_1,\mu_F^2) f_j(\xi_2,\mu_F^2) \dsigma_{ij}(\alpha_s(\mu_R),\mu_R,\mu_F) \nonumber
\end{equation}
where the probability of finding a parton of type $i$ in the proton carrying a momentum fraction $\xi$ is described by the parton distribution function $f_i(\xi,\mu_F^2)d\xi$ and the parton-level scattering cross section ${\rm d}\hat\sigma_{ij}$  for parton $i$ to scatter off parton $j$ normalised to the hadron-hadron flux\footnote{The partonic cross section normalised to the parton-parton flux is obtained by absorbing the inverse factors of $\xi_1$ and $\xi_2$ into $\dsigma_{ij}$.} is summed over the possible parton types $i$ and $j$. As usual $\mu_R$ and $\mu_F$ are the renormalisation and factorisation scales.

The infrared-finite partonic cross section for a parton of type $i$ scattering off parton of type $j$, ${\rm d}\hat\sigma_{ij}$, has the perturbative expansion 
\begin{equation}
\label{eq:sigpert}
\dsigma_{ij} = {\rm d}\hat\sigma_{ij}^{LO}
+\left(\frac{\alpha_s(\mu_R)}{2\pi}\right)\dsigma_{ij}^{NLO}
+\left(\frac{\alpha_s(\mu_R)}{2\pi}\right)^2\dsigma_{ij}^{NNLO}
+{\cal O}(\alpha_s^3)
\end{equation}
where the next-to-leading order (NLO) and next-to-next-to-leading order (NNLO) strong corrections are identified. Note that the leading order cross section is ${\cal O}(\alpha_s^2)$.

The single-jet inclusive and dijet cross sections have been studied at
NLO \cite{Ellis:1988hv,Ellis:1990ek,Ellis:1992en,Giele:1993dj,Giele:1994gf,Nagy:2001fj} and successfully compared with data from the high energy frontier at the TEVATRON \cite{CDF:2007ez,D0:2008hua,CDF:2008eq} and at the LHC \cite{ATLAS:2010wv,CMS:2011me}. A particular success is the determination of the strong coupling constant as a function of the jet transverse energy~\cite{Giele:1995kb,CDF:2001hn,D0:2009nc}.

The theoretical prediction may be improved by including the
NNLO perturbative predictions. This has the
effect of (a) reducing the renormalisation scale dependence and (b) improving
the matching of the parton level theoretical jet algorithm with the hadron
level experimental jet algorithm because the jet structure can be modelled by
the presence of a third parton. The resulting theoretical uncertainty at NNLO
is estimated to be at the few per-cent level \cite{Glover:2002gz}.

Any calculation of these higher-order corrections requires a systematic
procedure for extracting the infrared singularities that arise when one or more
final state particles become soft and/or collinear. These singularities are
present in the real radiation contribution at NLO, and
in double real radiation and mixed real-virtual contributions at NNLO.

Subtraction schemes are a well established solution to this problem and several methods for systematically constructing general subtraction terms have been proposed in the literature at NLO \cite{Catani:1996vz,Frixione:1995ms,Nagy:1996bz,Frixione:1997np,Somogyi:2006cz} and at NNLO \cite{GehrmannDeRidder:2005cm, Weinzierl:2003fx,Kilgore:2004ty,Frixione:2004is,Somogyi:2005xz,Somogyi:2006da,Somogyi:2006db,Somogyi:2008fc,Aglietti:2008fe,Somogyi:2009ri,Bolzoni:2009ye,Bolzoni:2010bt,Czakon:2010td,Anastasiou:2010pw,Czakon:2011ve,Anastasiou:2011qx,Boughezal:2011jf}. 
Another NNLO subtraction type scheme has been proposed in~\cite{Catani:2007vq}. 
It is not a general subtraction scheme, but can nevertheless deal with an entire class of processes, those without coloured final states,  in hadron-hadron collisions.  It has been explicitly applied to several observables~\cite{Catani:2007vq,Grazzini:2008tf,Catani:2009sm,Catani:2010en,Ferrera:2011bk,Catani:2011qz}.
In addition, there is the completely independent sector decomposition
approach which avoids the need for analytical integration and 
which has been developed for virtual~\cite{Binoth:2000ps,Binoth:2003ak,Heinrich:2008si,Carter:2010hi} 
and real radiation~\cite{Heinrich:2002rc,Anastasiou:2003gr, Binoth:2004jv, Heinrich:2006ku,Carter:2010hi} corrections to
NNLO, and applied to several observables already~\cite{Anastasiou:2004qd,Anastasiou:2004xq,Anastasiou:2005qj,Melnikov:2006di}.

We will follow the NNLO antenna subtraction method which has been fully developed for the production of
massless partons in electron-positron annihilation
~\cite{GehrmannDeRidder:2005cm,GehrmannDeRidder:2005hi,GehrmannDeRidder:2005aw},
leading to the successful description of the infrared structure of three-jet
events at NNLO \cite{GehrmannDeRidder:2007jk,Weinzierl:2009nz} and the
subsequent numerical calculation of the NNLO corrections to event shape
distributions~\cite{GehrmannDeRidder:2007bj,GehrmannDeRidder:2007hr,Weinzierl:2008iv,Weinzierl:2009ms}, the
moments of event shapes~\cite{GehrmannDeRidder:2009dp,Weinzierl:2009yz} and jet
rates~\cite{GehrmannDeRidder:2008ug,Weinzierl:2010cw}.  

The formalism has been extended to include massive fermions at NLO~\cite{GehrmannDeRidder:2009fz,Abelof:2011jv} and recently Bernreuther et al applied the antenna subtraction scheme to electron-positron annihilation with massive final state quarks at NNLO~\cite{Bernreuther:2011jt}.  

Here, we are more interested in the application of the antenna subtraction
scheme to hadron-hadron collisions, and specifically to processes such as $pp \to {\rm jet}+X$, $pp \to V+{\rm jet}+X$ (with $V=W^\pm$ or $Z$) and $pp \to H+{\rm jet}+X$.  The formalism has been fully worked out
for initial state hadrons at NLO~\cite{Daleo:2006xa} and further developed at
NNLO~\cite{Daleo:2009yj,Boughezal:2010mc,Gehrmann:2011wi}. Within this method, 
the subtraction terms are constructed from so-called antenna functions which
describe all unresolved partonic radiation (soft and collinear) between a hard pair of radiator partons. The hard radiators may be in the initial or in the final state, and
in the most general case, final-final (FF), initial-final  (IF) and initial-initial  (II) antennae need to be considered.
Within the antenna subtraction method, the subtraction terms and therefore the antennae
also need to be integrated over the unresolved phase space, which is different in the three configurations.
All of the integrals for the final-final and initial-final configurations are fully known.   Work is presently underway to determine the few remaining integrals of tree-level four parton antenna for the initial-initial configuration. A first step in that direction has been realised in  \cite{Boughezal:2010mc}.  Once completed, the computation of the NNLO corrections for jet production at the LHC is in range. 

Suppressing the labels of the partons in the initial state of the hard scattering, the general form for the subtraction terms for an $m$-particle final state at NNLO is given by~\cite{GehrmannDeRidder:2005cm}:
\begin{eqnarray}
\dsigma_{NNLO}&=&\int_{{\rm{d}}\Phi_{m+2}}\left(\dsigma_{NNLO}^{RR}-\dsigma_{NNLO}^S\right)
+\int_{{\rm{d}}\Phi_{m+2}}\dsigma_{NNLO}^S\nonumber\\
&+&\int_{{\rm{d}}\Phi_{m+1}}\left(\dsigma_{NNLO}^{RV}-\dsigma_{NNLO}^{VS}\right)
+\int_{{\rm{d}}\Phi_{m+1}}\dsigma_{NNLO}^{VS}
+\int_{{\rm{d}}\Phi_{m+1}}\dsigma_{NNLO}^{MF,1}\nonumber\\
&+&\int_{{\rm{d}}\Phi_m}\dsigma_{NNLO}^{VV}
+\int_{{\rm{d}}\Phi_m}\dsigma_{NNLO}^{MF,2}.
\end{eqnarray}
Here, $\dsigma^{S}_{NNLO}$ denotes the subtraction term for the $(m+2)$-parton final state which behaves like the double real radiation contribution
$\dsigma^{RR}_{NNLO}$ in all singular limits. 
Likewise, $\dsigma^{VS}_{NNLO}$ is the one-loop virtual subtraction term 
coinciding with the one-loop $(m+1)$-final state $\dsigma^{RV}_{NNLO}$ in all singular limits. 
The two-loop correction 
to the $m$-parton final state is denoted by $\dsigma^{VV}_{NNLO}$.  In addition, when there are partons in the initial state, 
there are two mass factorisation contributions, 
$\dsigma^{MF,1}_{NNLO}$ and $\dsigma^{MF,2}_{NNLO}$, for the $(m+1)$- and $m$-particle final states respectively.

In order to construct a numerical implementation of the NNLO cross section, 
the various contributions must be reorganised according to the number of final state particles,
\begin{eqnarray}
\dsigma_{NNLO}&=&\int_{{\rm{d}}\Phi_{m+2}}\left[\dsigma_{NNLO}^{RR}-\dsigma_{NNLO}^S\right]
\nonumber \\
&+& \int_{{\rm{d}}\Phi_{m+1}}
\left[
\dsigma_{NNLO}^{RV}-\dsigma_{NNLO}^{T}
\right] \nonumber \\
&+&\int_{{\rm{d}}\Phi_{m\phantom{+1}}}\left[
\dsigma_{NNLO}^{VV}-\dsigma_{NNLO}^{U}\right],
\end{eqnarray}
where the terms in each of the square brackets is finite and well behaved in the infrared singular regions.  More precisely,
\begin{eqnarray}
\label{eq:Tdef}
\dsigma_{NNLO}^{T} &=& \phantom{ -\int_1 }\dsigma_{NNLO}^{VS}
- \int_1 \dsigma_{NNLO}^{S,1} - \dsigma_{NNLO}^{MF,1},  \\
\label{eq:Udef}
\dsigma_{NNLO}^{U} &=& -\int_1 \dsigma_{NNLO}^{VS}
-\int_2 \dsigma_{NNLO}^{S,2}
-\dsigma_{NNLO}^{MF,2}.
\end{eqnarray}
Note that because integration of the subtraction term $\dsigma_{NNLO}^S$ gives contributions to both the $(m+1)$- and $m$-parton final states, we have explicitly decomposed the integrated double real subtraction term into a piece that is integrated over one unresolved particle phase space and a piece that is integrated over the phase space of two unresolved particles,
\begin{equation}
\int_{{\rm{d}}\Phi_{m+2}}\dsigma_{NNLO}^S
= \int_{{\rm{d}}\Phi_{m+1}} \int_1 \dsigma_{NNLO}^{S,1}
+\int_{{\rm{d}}\Phi_{m}} \int_2 \dsigma_{NNLO}^{S,2}.
\end{equation}

In a previous paper~\cite{Glover:2010im}, the subtraction term $\dsigma_{NNLO}^S$ corresponding to the leading colour pure gluon contribution to dijet production at hadron colliders was derived.   The subtraction term was shown to reproduce the singular behaviour present in $\dsigma_{NNLO}^{RR}$  in all of the single and double unresolved limits.   

It is the purpose of this paper to construct the appropriate subtraction term
$\dsigma_{NNLO}^{T}$ to render the leading colour five-gluon contribution
$\dsigma_{NNLO}^{RV}$ explicitly finite and numerically well behaved in all
single unresolved limits.  Our paper is organised in the following way.    In
Section~\ref{sec:RVsub}, the general structure of $\int_1\dsigma^{S,1}_{NNLO}$
and $\dsigma^{VS}_{NNLO}$ are discussed and analysed.  The coupling constant
renormalisation to remove the UV singularities and the mass factorisation of the
initial-state singularities into the parton distributions are described in
Section~\ref{sec:massfac}. In Section~\ref{sec:RVgluon} we turn our attention to
the specific process of gluon scattering at NNLO.   Our notation for gluonic
amplitudes is summarised in Section~\ref{subsec:gluon} and the one-loop
five-gluon amplitudes discussed in Section~\ref{subsec:five}.   

There are two separate configurations relevant for $gg \to ggg$ scattering
depending on whether the two initial state gluons are colour-connected or not.
We denote the configuration where the two initial state gluons are
colour-connected (i.e. adjacent) by IIFFF, while the configuration where the
colour ordering allows one final state gluon to be sandwiched between the
initial state gluons is denoted by IFIFF. Explicit forms for the real-virtual
subtraction term $\dsigma_{NNLO}^{T}$ for these two configurations are given in
Section~\ref{sec:DsigmaVSNNLO} where the cancellation of explicit poles is made
manifest.  The validity of the subtraction term is tested numerically in
Section~\ref{sec:numerical} by studying the subtracted one-loop matrix elements
in all of the single unresolved limits. In particular, we show that in all cases
the ratio of the finite part of the real-virtual and subtraction terms
approaches unity. Finally, our findings are summarised in
Section~\ref{sec:conc}.

Four appendices are also enclosed.  Appendix~A summarises the phase space
mappings for the final-final, initial-final and initial-initial configurations. 
Appendix~B gives a description of the tree-level antennae appearing in the
subtraction terms in both their unintegrated form
(Appendix~\ref{subsec:unintegratedthree}) and after integration over the
unresolved phase space (Appendix~\ref{subsec:integratedthree}).  The
unintegrated one-loop antenna is given in Appendix~\ref{subsec:oneloopantenna}.
 Appendix~C contains a modified form for the wide angle soft subtraction
terms present in $\dsigma^{S}_{NNLO}$, while formulae relating to the mass factorisation contribution are given in Appendix~D.

\section{Real-Virtual antenna subtraction at NNLO}
\label{sec:RVsub}
In this paper, we focus on the kinematical situation of the scattering 
of two massless coloured partons to produce massless coloured partons, 
and particularly the production of jets from gluon scattering in hadronic collisions.
To establish some notation, consider the leading-order parton-level contribution from the $(m+2)$-parton processes to the $m$-jet cross section at LO in $pp$ collisions,
\begin{equation}
\label{eq:process}
pp \to m~{\rm jets}
\end{equation}
which is given by  
\begin{eqnarray}
{\rm d}\hat\sigma_{LO}&=&\NLO \sum_{\textrm{perms}}{\rm d}\Phi_{m}(p_3,\hdots,p_{m+2};p_1,p_2)\frac{1}{S_{m}}\nonumber\\
&&\times|{\cal M}_{m+2}(\hat{1},\hat{2}\ldots,m+2)|^2 J_{m}^{(m)}(p_3,\hdots,p_{m+2})
\label{eq:LOcross}.
\end{eqnarray}
To make the subsequent discussion more general, we denote a
generic tree-level $(m+2)$-parton colour ordered amplitude by the symbol 
${\cal M}_{m+2}(\hat{1},\hat{2}\ldots,m+2)$,
where $\hat{1}$ and
$\hat{2}$ denote the initial state partons of momenta $p_1$ and  $p_2$
while the $m$-momenta in the final state are labeled 
$p_3,\hdots,p_{m+2}$. For convenience, and where the order of momenta does not matter, we will often denote the set of $(m+2)$-momenta $\{p_1,\hdots,p_{m+2}\}$ by $\{p\}_{m+2}$. The symmetry factor $S_{m}$ accounts for the production of identical particles in the final-state. 

At leading colour, the colour summed squared matrix elements 
are determined by the squares of the individual colour ordered amplitudes where the sum runs over the various colour ordered amplitudes.\footnote{The sub-leading colour contributions are implicitly contained in the squared matrix elements.}  For gluonic amplitudes this is the sum over the group of non-cyclic permutations of $n$ symbols.   
The normalisation factor ${\cal N}_{LO}$ includes the hadron-hadron
flux-factor, spin and colour summed and averaging factors as well as
the dependence on the renormalised QCD coupling
constant $\alpha_s$. 

${\rm d}\Phi_{m}$ is the $2\to m$ particle phase space:
\begin{eqnarray}
&&{\rm d}\Phi_{m}(p_3,\ldots,p_{m+2}; p_1, p_2)=\nonumber\\
&&\frac{{\rm d}^{d-1}p_3}{2E_3(2\pi)^{d-1}}\ldots\frac{{\rm d}^{d-1}p_{m+2}}{2E_{m+2}(2\pi)^{d-1}}
(2\pi)^d\delta^d(p_1+p_2-p_3-\ldots-p_{m+2}).
\end{eqnarray}
The jet function $J_{m}^{(n)}(\{p\}_{n+2})$ defines the
procedure for building $m$-jets from $n$ final state partons.  
Any initial state momenta in the set $\{p\}_{n+2}$ are simply ignored
by the jet algorithm.  The key property of
$J_{m}^{(n)}$ is that the jet observable is collinear 
and infrared safe.

In a previous paper~\cite{Glover:2010im}, we have discussed the NNLO
contribution coming from processes where two additional partons are
radiated - the double real contribution  $\dsigma^{RR}_{NNLO}$ and its
subtraction term $\dsigma^{S}_{NNLO}$. $\dsigma^{RR}_{NNLO}$ involves
the $(m+4)$-parton process at tree level and is given by, 
\begin{eqnarray}
{\rm d}\hat\sigma^{RR}_{NNLO}&=&\NNNLORR
\sum_{\textrm{perms}}\dPS{m+2}{m+4} \nonumber \\
&&\times|{\cal M}_{m+4}(\hat{1},\hat{2}\ldots,m+4)|^2 J_{m}^{(m+2)}(p_3,\{p\}_{m+4})
\label{eq:RRcross}.
\end{eqnarray}
In this paper, we are concerned with the NNLO contribution coming from single
radiation at one-loop, i.e. the  $(m+3)$-parton process, $\dsigma^{RV}_{NNLO}$.

In our notation, the one-loop $(m+3)$-parton contribution to $m$-jet
final states at NNLO in hadron-hadron collisions is given by
\begin{eqnarray}
\dsigma^{RV}_{NNLO}
&=& \NNNLORV \sum_{\textrm{perms}}\dPS{m+1}{m+3} \nonumber \\ &&\times
|{\cal M}^1_{m+3}(\hat{1},\ldots,m+3)|^2\;
J_{m}^{(m+1)}(\{p\}_{m+3})\;,
\label{eq:nnloonel}
\end{eqnarray}
where we introduced a shorthand notation for the interference of one-loop and tree-amplitudes, 
\begin{equation}
\label{eq:m1def}
|{\cal M}^1_{m+3}(\hat{1},\ldots,m+3)|^2 
= 2 \,{\rm Re}\, \left({\cal M}^{1}_{m+3}(\hat{1},\ldots,m+3)\,
{\cal M}^{{0},*}_{m+3}(\hat{1},\ldots,m+3)\right)\;,
\end{equation}
which explicitly 
captures the colour-ordering of the leading colour contributions.  The
subleading contributions in colour are implicitly included in
\eqref{eq:m1def} but will not be considered in 
detail in this paper. 
As usual, the sum is the appropriate combination of colour ordered
amplitudes. 
For gluonic amplitudes, this is the sum over the group of 
non-cyclic permutations of $n$ symbols.  

The normalisation factor ${\cal N}_{LO}$ depends 
on the specific process and parton channel under consideration.
Nevertheless, $\NNNLORR$ and $\NNNLORV$ are related both to each other and to ${\cal N}_{LO}$ for any number of jets and for any partonic process by
\begin{eqnarray}
\NNNLORV&=&{\cal N}_{LO}  \left(\frac{\alpha_s N}{2\pi}\right)^{2}
\frac{\bar{C}(\epsilon)^2}{C(\epsilon)},\\
\NNNLORR&=&{\cal N}_{LO}  \left(\frac{\alpha_s N}{2\pi}\right)^{2}
\frac{\bar{C}(\epsilon)^2}{C(\epsilon)^2},\\
\NNNLORV&=&\NNNLORR \times C(\epsilon),
\end{eqnarray}
where
\begin{eqnarray}
\label{eq:Cdef}
C(\epsilon)=(4\pi)^{\epsilon}\frac{e^{-\epsilon\gamma}}{8\pi^2},\\
\label{eq:Cbar}
\bar{C}(\e)=(4\pi)^{\e}e^{-\e\gamma}.
\end{eqnarray}
As expected, each power of the (bare) coupling is accompanied by a factor of $\bar{C}(\e)$.
In this paper, we are mainly concerned with the NNLO corrections to \eqref{eq:process} when
$m=2$ and for the pure gluon channel.
The normalisation factor ${\cal N}_{LO}$ will be given for this special case in Section 4.

Eq.~\eqref{eq:nnloonel}   contains two types of infrared singularities. The 
renormalised one-loop virtual correction  ${\cal M}^{{1}}_{m+3}$  to the
$(m+3)$-parton matrix contains explicit infrared  poles, which can be expressed
using the infrared singularity operators defined  in
\cite{Catani:1998bh,Sterman:2002qn}. On the other hand, the requirement of
building $m$-jets from $(m+1)$-partons allows one of the  final state partons
to become unresolved, leading to implicit local infrared singularities which become explicit
only after integration over the unresolved patch of the  final state
$(m+1)$-parton phase space.  The single unresolved infrared singularity
structure of one- and two-loop amplitudes has been studied in
\cite{Bern:1994zx,Bern:1998sc,Kosower:1999xi,Kosower:1999rx,Bern:1999ry,Catani:2000pi,Bierenbaum:2011gg,Catani:2003vu,Kosower:2002su,Kosower:2003cz,Weinzierl:2003ra,Bern:2004cz,Badger:2004uk}.

As discussed in Section~\ref{sec:intro}, in order
to carry out the numerical integration over the $(m+1)$-parton phase,
weighted by the appropriate jet function, we have to construct an 
infrared subtraction term\footnote{Strictly speaking, $\dsigma^{T}_{NNLO}$ is not a subtraction term since it adds back part of the the double radiation subtraction term $\dsigma^{S}_{NNLO}$ integrated over the phase space of a single unresolved particle.  Nevertheless, since it contains all the terms needed to render the $(m+1)$-particle final state finite, it is convenient to call it the real-virtual subtraction term.} $\dsigma^{T}_{NNLO}$ which 
\begin{itemize}
\item[(a)] removes the explicit infrared poles of the virtual one-loop
$(m+3)$-parton matrix element.
\item[(b)] correctly describes the single unresolved limits of the  virtual one-loop $(m+3)$-parton matrix element.
\end{itemize}

The subtraction term has three components;
\begin{eqnarray}
\label{eq:dsigmaTbreak}
\dsigma_{NNLO}^{T} &=& \phantom{ -\int_1 }\dsigma_{NNLO}^{VS}
- \int_1 \dsigma_{NNLO}^{S,1} - \dsigma_{NNLO}^{MF,1},
\end{eqnarray}
where $\int_1 \dsigma_{NNLO}^{S,1}$ is derived from the double real radiation subtraction term $\dsigma_{NNLO}^{S}$ integrated over the phase space of one unresolved particle. Part of this contribution cancels the explicit poles in the virtual matrix element, while the real-virtual subtraction term   
$\dsigma_{NNLO}^{VS}$ accounts for the single unresolved limits of the
virtual matrix element.
 
In the following subsections we shall present the general structure of $\int_1
\dsigma_{NNLO}^{S,1}$ and $\dsigma_{NNLO}^{VS}$.
The remaining poles are associated with the initial state collinear
singularities and are absorbed by the mass factorisation counterterm
$\dsigma_{NNLO}^{MF,1}$ which will be presented explicitly in Section~\ref{sec:massfac}.  

A key element of the antenna subtraction scheme is the factorisation of the matrix elements and phase space in the singular limits where one or more particles are unresolved. In determining the various contributions to $\dsigma_{NNLO}^{T}$, we shall therefore specify the unintegrated and/or
integrated antennae and the reduced colour ordered matrix-element squared involved.   The factorisation is guaranteed by the momentum mapping described in the Appendix~\ref{sec:appendixA}. For conciseness, only the redefined hard radiator momenta will be specified in the functional dependence
of the matrix element squared. The other momenta will simply be denoted by ellipsis. 

In order to combine the subtraction terms and real-virtual matrix elements, it is convenient to slightly modify the phase space, such that
\begin{eqnarray}
\dsigma^{RV}_{NNLO}
&=& \NNNLORV\,\sum_{\textrm{perms}}\dPSxx{m+1}{m+3} \nonumber \\ &&\times
|{\cal M}^1_{m+3}(\hat{1},\ldots,m+3)|^2\;\delta(1-x_1)\delta(1-x_2)\,
J_{m}^{(m+1)}(\{p\}_{m+3})\;.
\label{eq:nnloonelalt}
\end{eqnarray}
The integration over $x_1$ and $x_2$ reflects the fact that the subtraction terms contain contributions due to radiation from the initial state such that the parton momenta involved in the hard scattering carry only a fraction $x_i$ of the incoming momenta. 
In general, there are three regions:   
the soft ($x_1 = x_2 = 1$), collinear ($x_1 = 1$, $x_2 \neq 1$ and $x_1 \neq 1$, $x_2 = 1$) 
and hard ($x_1 \neq 1$, $x_2 \neq 1$).  The real-virtual matrix elements only contribute in the soft region, as indicated by the two delta functions.  
 
In Sections~\ref{subsec:RVS1} and \ref{subsec:RVVS1}, we discuss the first two terms that contribute to $\dsigma_{NNLO}^{T}$ given in Eq.~\eqref{eq:dsigmaTbreak}, namely $\int_1 \dsigma_{NNLO}^{S,1}$ and $  \dsigma_{NNLO}^{VS}$. The final contribution $\dsigma_{NNLO}^{MF,1}$ is discussed in Section~\ref{sec:massfac}.

\subsection{Contribution from integration of double real subtraction term:  $\int_1 \dsigma_{NNLO}^{S,1}$}
\label{subsec:RVS1}

There are five different types of contributions to ${\rm{d}}\hat\sigma_{NNLO}^S$
according to the colour connection of the unresolved partons:
\begin{itemize}
\item[(a)] One unresolved parton but the experimental observable selects only
$m$ jets.
\item[(b)] Two colour-connected unresolved partons (colour-connected).
\item[(c)] Two unresolved partons that are not colour-connected but share a common
radiator (almost colour-connected).
\item[(d)] Two unresolved partons that are well separated from each other 
in the colour 
chain (colour-unconnected).
\item[(e)] Compensation terms for the over subtraction of large angle soft emission.
\end{itemize}
Each type of contribution takes the form of antenna functions multiplied by colour ordered matrix elements.   The various types of contributions are summarised in Table~1.  
\begin{table}[t!]
\begin{center}
{\small
\begin{tabular}{|c|ccccc|}
\hline
 & $a$ & $b$ & $b,c$ & $d$ & $e$   \\\hline
$\dsigma_{NNLO}^{S}$ 
& $X_3^0 |{\cal M}^0_{m+3}|^2$ 
& $X_4^0 |{\cal M}^0_{m+2}|^2$ 
& $X_3^0 X_3^0 |{\cal M}^0_{m+2}|^2$
& $X_3^0 X_3^0 |{\cal M}^0_{m+2}|^2$ 
& $S X_3^0 |{\cal M}^0_{m+2}|^2$   \\
$\int_1 \dsigma_{NNLO}^{S,1}$ 
& ${\cal X}_3^0 |{\cal M}^0_{m+3}|^2$  
& -- 
&  ${\cal X}_3^0 X_3^0 |{\cal M}^0_{m+2}|^2$ 
& -- 
& ${\cal S} X_3^0 |{\cal M}^0_{m+2}|^2$ \\
$\int_2 \dsigma_{NNLO}^{S,2}$ 
& -- 
&  ${\cal X}_4^0 |{\cal M}^0_{m+2}|^2$ 
& -- 
&${\cal X}_3^0 {\cal X}_3^0 |{\cal M}^0_{m+2}|^2$ 
&  --   \\ \hline
\end{tabular}
}
\end{center}
\label{tab:S1breakdown}
\caption{Type of contribution to the double real subtraction term ${\rm{d}}\hat\sigma_{NNLO}^{S}$, together with the integrated form of each term.   The unintegrated antenna and soft functions are denoted as $X_3^0$, $X_4^0$ and $S$ while their integrated forms are ${\cal X}_3^0$, ${\cal X}_4^0$ and ${\cal S}$ respectively.  ${\cal M}^0_{n}$ denotes an $n$-particle tree-level colour ordered amplitude.  }
\end{table}
We see that the $a$, $c$ and $e$ types of subtraction term, as well as the $b$-type terms that are products of three-particle antenna, can be integrated over a single unresolved particle phase space and therefore contribute to the $(m+1)$-particle final state so that,
\begin{equation}
\label{eq:S1}
\int_1 \dsigma_{NNLO}^{S,1} = 
\int_1 \dsigma_{NNLO}^{S,a}
+\int_1 \dsigma_{NNLO}^{S,(b,c)} 
+\int_1 \dsigma_{NNLO}^{S,e}.
\end{equation}
with, 
\begin{equation}
\int_1 \dsigma_{NNLO}^{S,(b,c)}=
\int_1 \dsigma_{NNLO}^{S,b,3\times 3}
+\int_1 \dsigma_{NNLO}^{S,c}
\end{equation}

On the other hand, the double unresolved antenna functions $X_4^0$ in $\dsigma_{NNLO}^{S,b}$ and the colour-unconnected $X_3^0 X_3^0$ terms in $\dsigma_{NNLO}^{S,d}$ 
can immediately be integrated over the phase space of both unresolved particles and appear directly in the $m$-particle final state, 
\begin{equation}
\label{eq:S2}
\int_2 \dsigma_{NNLO}^{S,2} = 
\int_2 \dsigma_{NNLO}^{S,b,4}
+\int_2 \dsigma_{NNLO}^{S,d}.
\end{equation}
These integrated contributions will be discussed elsewhere.

We now turn to a detailed discussion of each of the terms in Eq.~\eqref{eq:S1}.

\subsubsection{Cancellation of explicit infrared poles in
  $\dsigma^{RV}_{NNLO}$:  $\int_1 \dsigma_{NNLO}^{S,a}$}
\label{sec:RVaff}
 
In the antenna subtraction approach, the single unresolved configuration 
coming from the tree-level process with two additional particles, i.e. the
double-real process  involving $(m+4)$-partons, is subtracted
using a three-particle antenna function - two hard radiator partons emitting
one unresolved parton. Once the unresolved phase space is integrated over,
one recovers an $(m+3)$-parton contribution that precisely cancels the
explicit pole structure in the virtual one-loop $(m+3)$-parton matrix
element.

The integrated subtraction term formally written as  $\int_1
\dsigma_{NNLO}^{S,a}$ is split into three different contributions,
depending on whether the hard radiators are in the initial or final state.

When both hard radiators $i$ and $k$  are in the final state,
then $X_{ijk}$ is a final-final (FF) antenna function that describes 
all singular configurations (for this colour-ordered amplitude) where
parton $j$ is unresolved.  
The subtraction term for this single unresolved configuration, summing
over all possible positions of the unresolved parton, reads,

\begin{eqnarray}
&&\dsigma_{NNLO}^{S,a(FF)}=\NNNLORR\sum_{\textrm{perms}}\dPS{m+2}{m+4}\nonumber\\
&&\times\sum_j X_{ijk}^0 |{\cal M}_{m+3}(\hdots,I,K,\hdots)|^2
J_{m}^{(m+1)}(\{p\}_{m+3}).
\label{eq:sub1}
\end{eqnarray}
Besides the three parton antenna function $X^0_{ijk}$ which depends only on
$p_i$, $p_j$ and $p_k$,
the subtraction term involves an $(m+3)$-parton amplitude depending on 
the redefined on-shell momenta $p_I$ and $p_K$, whose
definition in terms of the original momenta are given in Appendix~\ref{sec:appendixAFF}. 
The $(m+3)$-parton amplitude also depends
on the other final state momenta which, in the final-final map, are not redefined and
on the two initial state momenta $p_1$ and $p_2$. This dependence is
manifest as ellipsis in \eqref{eq:sub1}.  The jet function is applied to the $(m+1)$-momenta that remain after the mapping, i.e. $p_3, \ldots, p_I, p_K, \ldots, p_{m+4}$.

To perform the integration of the subtraction term in Eq.~\eqref{eq:sub1} and make its infrared poles explicit, we exploit the following factorisation of the phase space,
\begin{eqnarray}
\label{eq:psx1}
\lefteqn{{\rm d} \Phi_{m+2}(p_{3},\ldots,p_{m+4};p_1,p_2)  }
\nonumber \\ &=&
{\rm d} \Phi_{m+1}(p_{3},\ldots,p_I,p_K,\ldots,p_{m+4};p_1,p_2)
\cdot 
{\rm d} \Phi_{X_{ijk}} (p_i,p_j,p_k;p_I+p_K) \\
&\equiv & 
{\rm d} \Phi_{m+1}(p_{3},\ldots,p_{m+3};x_1p_1,x_2p_2) \,\frac{{\rm d}x_1}{x_1}\frac{{\rm d}x_2}{x_2} \nonumber \\
&&\times \delta(1-x_1)\delta(1-x_2)
{\rm d} \Phi_{X_{ijk}} (p_i,p_j,p_k;p_I+p_K), 
\label{eq:FFPSfact}
\end{eqnarray}
where we have simply relabeled the final-state momenta in passing 
from \eqref{eq:psx1} to \eqref{eq:FFPSfact}.
In \eqref{eq:FFPSfact} the antenna phase space ${\rm d} \Phi_{X_{ijk}}$ is proportional 
to the three-particle phase space relevant to a $1\to 3$ decay, and one
can define the integrated final-final antenna by
\begin{equation}
\label{eq:x3intff}
{\cal X}^0_{ijk}(s_{IK},x_1,x_2) = \frac{1}{C(\epsilon)} 
\int {\rm d} \Phi_{X_{ijk}}\;X^0_{ijk}\,\delta(1-x_1)\,\delta(1-x_2)\, ,
\end{equation}
where $C(\epsilon)$ is given in \eqref{eq:Cdef}.

The integrated single-unresolved contribution necessary to cancel the
explicit infrared poles from virtual contributions in this final-final
configuration then reads,  
\begin{eqnarray}
\int_1 \dsigma_{NNLO}^{S,a,(FF)}
&=&   \NNNLORV\,
\sum_{\textrm{perms}}\dPSxx{m+1}{m+3}
\nonumber \\
&\times& \,
\sum_{ik}\;  
{\cal X}^0_{ijk}(s_{ik},x_1,x_2)\,
|{\cal M}_{m+3}(\ldots,i,k,\ldots)|^2\,
\JET_{m}^{(m+1)}(\{p\}_{m+3})\;,
 \label{eq:subv1aff} 
\end{eqnarray}
where the sum runs over all colour-connected pairs of final state
momenta $(p_i,p_k)$ and the final state momenta $I,K$ have been
relabelled as $i,k$. 
Expressions for the integrated final-final three-parton antennae are
available in Ref.~\cite{GehrmannDeRidder:2005cm}.

When only one of the hard radiator partons is in the initial state,
$X_{i,jk}$ is a initial-final (IF) antenna function that describes 
all singular configurations (for this colour-ordered amplitude) 
where parton $j$ is unresolved between the initial state parton
denoted by $\hat{i}$ (where $\hat{i}=\hat{1}$ or $\hat{i}=\hat{2}$) 
and the final state parton $k$. The antenna only depends on these
three parton momenta $p_{i},p_{j}$ and $p_{k}$.  
The subtraction term for this single unresolved configuration, summing
over all possible positions of the unresolved parton, reads,
\begin{eqnarray}
\dsigma_{NNLO}^{S,a,(IF)}&=&\NNNLORR\sum_{\textrm{perms}}\dPS{m+2}{m+4}\nonumber\\
&&\times \sum_{ i=1,2} \sum_j X_{i,jk}^0 |{\cal M}_{m+3}(\ldots,\hat{I},K,\ldots)|^2
J_{m}^{(m+1)}(\{p\}_{m+3}).\nonumber\\
\label{eq:sub2}
\end{eqnarray}
As in the previous final-final case, the reduced $(m+3)$-parton matrix element
squared involves the mapped momenta $\hat{I}$ and $K$ which are defined in Appendix~\ref{sec:appendixAIF}. Likewise, the jet algorithm acts on the $(m+1)$-final state momenta that remain after the mapping has been applied.

In this case, the phase space in \eqref{eq:sub2} can be 
factorised into the convolution 
of an $(m+1)$-particle phase space, involving only the redefined momenta, 
with a 2-particle phase space \cite{Daleo:2006xa}. For
the special case $i=1$, it reads,
\begin{eqnarray}
\label{eq:psx2}
\lefteqn{
{\rm d}\Phi_{m+2}(p_3,\dots,p_{m+4};p_1,p_2)}\nonumber \\
&=&{\rm d}\Phi_{m+1}(p_3,\dots,p_K,\dots,p_{m+4};x_1p_1,p_2)\,\frac{{\rm d} x_1}{x_1}\,\delta(x_1-\hat{x}_1)\,\frac{Q^2}{2\pi}{\rm d}\Phi_{2}(p_j,p_k;p_1,q)\\
&\equiv
&{\rm d} \Phi_{m+1}(p_3,\ldots,p_{m+3};x_1p_1,x_2p_2) \,\frac{{\rm d} x_1}{x_1}\,\frac{{\rm d}x_2}{x_2}\, \delta(x_1-\hat{x}_1)\,\delta(1-x_2)\frac{Q^2}{2\pi}{\rm d}\Phi_{2}(p_j,p_k;p_1,q)\, ,\nonumber \\
\label{eq:IFPSfact}
\end{eqnarray}
with $Q^2=-q^2$ and $q=p_j+p_k-p_1$ and where we have also relabelled the final-state momenta in passing from \eqref{eq:psx2} to \eqref{eq:IFPSfact}.
The quantity $\hat{x}_1$ is defined in Eq.~\eqref{eq:xiIF}.

Using this factorisation property one can carry out the 
integration over the unresolved phase space of the antenna function in $\eqref{eq:sub2}$ analytically. We define 
the integrated initial-final antenna function by,
\begin{equation}
\label{eq:x3intif}
{\cal X}_{1,jk}^0(s_{\bar{1}K},x_1,x_2)=\frac{1}{C(\epsilon)}\int {\rm d}\Phi_2 \frac{Q^2}{2\pi}\,\delta(x_1-\hat{x}_1)\,\delta(1-x_2) X_{1,jk}^0\,,
\end{equation}
where $C(\epsilon)$ is given in \eqref{eq:Cdef}. Similar expressions are obtained when $i=2$ via exchange of $x_1$ and $x_2$,  
\begin{equation}
{\cal X}_{2,jk}^0(s_{\bar{2}K},x_1,x_2)=\frac{1}{C(\epsilon)}\int {\rm d}\Phi_2 \frac{Q^2}{2\pi}\,\delta(x_2-\hat{x}_2)\,\delta(1-x_1) X_{2,jk}^0\,.
\end{equation}

The explicit poles present in 
$\dsigma^{RV}_{NNLO}$ (defined in \eqref{eq:nnloonelalt}) associated
with the colour-connected initial-final radiators $\hat{i}$ and $k$
can therefore be removed with the
following form,
\begin{eqnarray}
\lefteqn{\int_1 \dsigma_{NNLO}^{S,a,(IF)}=
  \NNNLORV\,
\sum_{\textrm{perms}}\dPSxx{m+1}{m+3}}\nonumber \\
& & \times \, \sum_{i=1,2}\, 
\sum_{k}\;    {\cal X}^0_{i,jk}(s_{\bar{i}k},x_1,x_2)
|{\cal M}_{m+3}(\ldots,\hat{I},k,\ldots)|^2\,
\JET_{m}^{(m+1)}(\{p\}_{m+3})\;. 
\label{eq:subv1aif} 
\end{eqnarray}
In this expression, only the redefined momentum $K$  has been relabelled
$(K\to k)$. The rescaled initial state radiator $\hat{I} (=\bar{i})$ is
not relabelled and appears in the functional dependence of the
integrated antenna and in the matrix-element squared.
Explicit forms for the integrated initial-final three-parton antennae 
are available in Ref.~\cite{Daleo:2006xa}.

If we consider the case where the two hard radiator partons $i$ and
$k$ are in the initial state finally, 
then $X_{ik,j}$ is a initial-initial (II) antenna function that describes all singular configurations (for this colour-ordered amplitude) where parton $j$ is unresolved.  
The subtraction term for this single unresolved configuration, summing over all possible positions of the unresolved parton, reads,
\begin{eqnarray}
&&\dsigma_{NNLO}^{S,a,(II)}=\NNNLORR\sum_{\textrm{perms}}\dPS{m+2}{m+4}\nonumber\\
&&\times\sum_{i,k=1,2} \sum_j X_{ik,j}^0 |{\cal M}_{m+3}(\ldots,\hat{I},\hat{K},\ldots)|^2
J_{m}^{(m+1)}(\tilde{p}_3,\ldots,\tilde{p}_{m+4}).
\label{eq:sub3}
\end{eqnarray}
where as usual we denote momenta in the initial state with a hat. The radiators
$\hat{i}$ and $\hat{k}$ are replaced by new rescaled initial state partons $\hat{I}$ and $\hat{K}$ and all other spectator momenta
are Lorentz boosted to preserve momentum conservation as described in Appendix~\ref{sec:appendixAII}.

For the initial-initial configuration the phase space in $\eqref{eq:sub3}$ factorises into the convolution of an $(m+1)$-particle phase space, involving only redefined momenta, 
with the phase space of parton $j$ \cite{Daleo:2006xa} so that when 
$i=1$ and $k=2$,
\begin{eqnarray}
\label{eq:psx3}
\lefteqn{{\rm d}\Phi_{m+2}(p_3,\dots,p_{m+4};p_1,p_2)}\nonumber \\
&=&
{\rm d}\Phi_{m+1}(\tilde{p}_3,\ldots,\tilde{p}_{m+4};x_1 p_1, x_2 p_2)
\times\,x_1x_2\, \delta(x_1-\hat{x}_1)\,\delta(x_2-\hat{x}_2)\,[{\rm d} p_j] \,\frac{{\rm d} x_1}{x_1}\,\frac{{\rm d}x_2}{x_2},\nonumber \\
\end{eqnarray}
where the single particle phase space measure is $[{\rm d} p_j]={{\rm d}^{d-1}p_j}/{2E_j/(2\pi)^{d-1}}$ and $\hat{x}_i$ is defined in  Eq.~\eqref{eq:xiII}.

The only dependence on the original momenta lies in the antenna function $X_{ik,j}$ and the antenna phase space.  One can therefore carry out the integration over the unresolved phase space analytically, to find the integrated antenna function,
\begin{equation}
\label{eq:x3intii}
{\cal X}^0_{12,j}(s_{\bar{1}\bar{2}},x_1,x_2) = \frac{1}{C(\epsilon)}\int
[{\rm d} p_j]\,  x_1\,  x_2\,\delta(x_1-\hat{x}_1)\,\delta(x_2-\hat{x}_2)\,X_{12,j}^0\;,
\end{equation}
where $C(\epsilon)$ is given in 
Eq.~\eqref{eq:Cdef}.
Explicit forms for the integrated initial-initial three-parton
antennae are available in Ref.~\cite{Daleo:2006xa}.

We can therefore remove the explicit poles present in 
$\dsigma^{RV}_{NNLO}$ (defined in \eqref{eq:nnloonelalt}) associated with the colour-connected initial-state pair $\hat{i}$ and $\hat{k}$ with the subtraction term,
\begin{eqnarray}
\lefteqn{\int_1 \dsigma_{NNLO}^{S,a,(II)}
= \NNNLORV\, 
\sum_{\textrm{perms}}\dPSxx{m+1}{m+3}}\nonumber \\
&\times& \,  \sum_{i,k=1,2}\;   {\cal X}^0_{ik,j}(s_{\bar{i}\bar{k}},x_i,x_k) 
|{\cal M}_{m+3}(\ldots,\hat{I},\hat{K},\ldots)|^2\,
\JET_{m}^{(m+1)}(\{p\}_{m+3})\;.\nonumber \\
\label{eq:subv1aii}
\end{eqnarray}
As in the initial-final case, the redefined initial state momenta
$\hat{I}=\bar{i}$ and $\hat{K}=\bar{k}$ are not relabelled, neither in the functional
dependence of the integrated antenna $ {\cal  X}^0_{ik,j}(s_{\bar{i}\bar{k}},x_i,x_k) $ 
nor for the reduced matrix element squared $|{\cal M}_{m+3}(\ldots\hat{I},\hat{K},\ldots)|^2$.

Each of the integrated antenna have an explicit dependence on the
variables $x_1$ and $x_2$ as stated in eqs. 
(\ref{eq:x3intff}), (\ref{eq:x3intif})  and (\ref{eq:x3intii}).

Summing over the different final-final, initial-final and initial-initial
configurations, we find that the explicit poles in $\dsigma_{NNLO}^{RV}$ are
removed by $\int_1 \dsigma_{NNLO}^{S,a}$ to yield an integrand free from 
explicit infrared poles over the whole region of integration. This is of course
merely a consequence of the cancellation of infrared  poles between a virtual
contribution and integrated subtraction terms from real emission.  However, as
discussed later in Section~\ref{sec:sigb},  $\dsigma_{NNLO}^{RV}$ and $\int_1
\dsigma_{NNLO}^{S,a}$ develop further infrared singularities in singly
unresolved regions of the phase-space which  do not coincide. Therefore, we have
to  introduce a further subtraction term, $\dsigma_{NNLO}^{VS}$, to ensure an
integrand that has no explicit global $\eps$-poles, and that does not have 
implicit singularities  in single unresolved regions. 

\subsubsection{Integration of iterated antennae:  $\int_1 \dsigma_{NNLO}^{S,(b,c)}$}
\label{sec:RVbcff}

As discussed earlier in Sect.~\ref{subsec:RVS1}, contributions to the double real subtraction term due to colour-connected or almost colour-unconnected
hard radiators that have the generic form $X_3 \times X_3$ must also be added back integrated over the phase space of the first (outer) antenna. 
The structure of these terms is very similar to the single unresolved contributions of the previous sub-section; each term with an ``outer" final-final antenna present in $\dsigma_{NNLO}^{S,(b,c)}$ produces a contribution given by,
\begin{eqnarray}
\int_1 \dsigma_{NNLO}^{S,(b,c),(FF)}
\label{eq:bcintFF}
&=&   \NNNLORV\,\dPSxx{m+1}{m+3}
\nonumber \\
&\times& \,
{\cal X}^0_{ijk}(s_{ik},x_1,x_2)\,
X^0_3(\{p\}_{m+3}) \, |{\cal M}_{m+2}(\{p\}_{m+2})|^2\,
\JET_{m}^{(m)}(\{p\}_{m+2})\;.\nonumber\\
\end{eqnarray}
In this case, $i$ and $k$ represent the momenta in the set $\{p\}_{m+3}$.
The set of momenta denoted by $\{p\}_{m+2}$ is obtained from
$\{p\}_{m+3}$ set through a phase space mapping that is determined by the 
type of the unintegrated ``inner" antenna $X^0_3$ which is fixed by
the corresponding term in  $\dsigma_{NNLO}^{S}$.

When the outer antenna is of initial-final or initial-initial type, we find the integrated forms,
\begin{eqnarray}
\label{eq:bcintIF}
\int_1 \dsigma_{NNLO}^{S,(b,c),(IF)}
&=&   \NNNLORV\,\dPSxx{m+1}{m+3}\nonumber \\
& \times &   
    {\cal X}^0_{i,jk}(s_{\bar{i}k},x_1,x_2)\,
X^0_3(\{p\}_{m+3}) \, |{\cal M}_{m+2}(\{p\}_{m+2})|^2\,
\JET_{m}^{(m)}(\{p\}_{m+2})\;,  \nonumber \\
\label{eq:bcintII}
&&\\
\int_1 \dsigma_{NNLO}^{S,(b,c),(II)}
&=& \NNNLORV\,\dPSxx{m+1}{m+3}\nonumber \\
&\times& \,{\cal X}^0_{ik,j}(s_{\bar{i}\bar{k}},x_1,x_2) \,
X^0_3(\{p\}_{m+3})  \, |{\cal M}_{m+2}(\{p\}_{m+2})|^2\,
\JET_{m}^{(m)}(\{p\}_{m+2})\;.\nonumber \\
\end{eqnarray}

Note that the $(b,c)$-type  terms reflect different physical origins; $\dsigma^{S,b}_{NNLO}$
is designed to account for the unresolved contributions from two colour-connected unresolved partons, while $\dsigma^{S,c}_{NNLO}$ treats the singularities from two almost colour-connected unresolved partons.  

The general structure of $\dsigma^{S,b}_{NNLO}$ consists of groups of terms involving the difference between a four-parton antenna and products of three-parton antennae.   The latter removes the single unresolved limits present in the former.  There are two distinct types of four-parton antenna which, for sake of argument, we label $X_4^0(h_1,s_1,s_2,h_2)$ and $\tilde{X}_4^0(h_1,s_1,s_2,h_2)$, where $h_1$, $h_2$ represent the hard radiators and $s_1$, $s_2$ the unresolved particles.  By construction, there are no unresolved limits when $h_1$ or $h_2$ become unresolved. 
In both, the four partons are colour-connected.  The key difference between $X_4^0$ and $\tilde{X}_4^0$ is that in ${X}_4^0$, $s_1$ ($s_2$) is colour-connected to  $h_1$ ($h_2$) and $s_2$ ($s_1$), while in $\tilde{X}_4^0$, $s_1$ ($s_2$) is colour-connected to $h_1$ and $h_2$.

In the case of $X_4^0$, $\dsigma^{S,b}_{NNLO} $ has the form,
\begin{eqnarray}
\label{eq:dsigmaSbX}
\lefteqn{X_4^0(h_1,s_1,s_2,h_2) |{\cal M}_{n+2}(\ldots,\widetilde{(h_1s_1s_2)},\widetilde{(s_1s_2h_2)},\ldots)|^2 
}\nonumber \\
&&- X_3^0(h_1,s_1,s_2) X_3^0(\widetilde{(h_1s_1)},\widetilde{(s_1s_2)},h_2) 
|{\cal M}_n(\ldots,\widetilde{((h_1s_1)(s_1s_2))},\widetilde{((s_1s_2)h_2)},\ldots)|^2\nonumber \\
&&- X_3^0(s_1,s_2,h_2) X_3^0(h_1,\widetilde{(s_1s_2)},\widetilde{(s_2h_2)}) 
|{\cal M}_n(\ldots,\widetilde{(h_1(s_1s_2))},\widetilde{((s_1s_2)(s_2h_2))},\ldots)|^2
\end{eqnarray}
where the two $X_3^0 \times X_3^0$ terms subtract the single unresolved limits present in $X_4^0$. 
Most antennae fit into this class including, for example, the quark-antiquark antenna $A_4^0$, the quark-gluon subantenna $D_{4,a}^0$~\cite{GehrmannDeRidder:2007jk} and the gluon-gluon subantenna $F_{4,a}^0$~\cite{Glover:2010im}.
Upon integration, the  $X_3^0 \times X_3^0$ terms produce integrated antenna that cancel against the explicit poles present in $X_3^1$ that will be discussed in Section~\ref{sec:sigb}.

Similarly, one must also subtract the limits where $s_1$ and $s_2$ are unresolved from $\tilde{X}_4^0$,
\begin{eqnarray}
\label{eq:dsigmaSbXt}
\lefteqn{\tilde{X}_4^0(h_1,s_1,s_2,h_2) |{\cal M}_{n+2}(\ldots,\widetilde{h_1s_1s_2},\widetilde{s_1s_2h_2},\ldots)|^2 
}\nonumber \\
&&- X_3^0(h_1,s_1,h_2) X_3^0(\widetilde{(h_1s_1)},s_2,\widetilde{(s_1h_2)}) 
|{\cal M}_n(\ldots,\widetilde{(h_1s_1)s_2)},\widetilde{(s_2(s_1h_2))},\ldots)|^2\nonumber \\
&&- X_3^0(h_1,s_2,h_2) X_3^0(\widetilde{(h_1s_2)},s_1,\widetilde{(s_2h_2)}) 
|{\cal M}_n(\ldots,\widetilde{(h_1s_2)s_1)},\widetilde{(s_1(s_2h_2))},\ldots)|^2.
\end{eqnarray}
Antennae of this type include the quark-antiquark antenna $\tilde{A}_4^0$, the quark-gluon subantenna $D_{4,b}^0$~\cite{GehrmannDeRidder:2007jk} and the gluon-gluon subantenna $F_{4,b}^0$~\cite{Glover:2010im}.
As before, after integration the $X_3^0 \times X_3^0$ terms produce integrated antenna that cancel against the explicit poles present in $X_3^1$.
However, unless the single unresolved contribution $\dsigma^{S,a}_{NNLO}$ has the form,\footnote{This typically happens when both $s_1$ and $s_2$ are photon-like, see for example Eq.~(8.2) of Ref.~\cite{GehrmannDeRidder:2007jk}.}
\begin{eqnarray}
&&X_3^0(h_1,s_1,h_2)  
|{\cal M}_{n+1}(\ldots,\widetilde{(h_1s_1)},s_2,\widetilde{(s_1h_2)},\ldots)|^2,
\end{eqnarray}
it is clear that Eq.~\eqref{eq:dsigmaSbXt} oversubtracts the double unresolved limits. In this case, a correction term of the form 
\begin{eqnarray}
\label{eq:dsigmaSbXtfixup}
&&+\frac{1}{2}X_3^0(h_1,s_1,h_2) X_3^0(\widetilde{(h_1s_1)},s_2,\widetilde{(s_1h_2)}) 
|{\cal M}_n(\ldots,\widetilde{((h_1s_1)s_2)},\widetilde{(s_2(s_1h_2))},\ldots)|^2\nonumber \\
&&+\frac{1}{2}X_3^0(h_1,s_2,h_2) X_3^0(\widetilde{(h_1s_2)},s_1,\widetilde{(s_2h_2)}) 
|{\cal M}_n(\ldots,\widetilde{((h_1s_2)s_1)},\widetilde{(s_1(s_2h_2))},\ldots)|^2,
\end{eqnarray}
must be added to \eqref{eq:dsigmaSbXt} to compensate.  In the case of 
$\tilde{X}_4^0$, $\dsigma_{NNLO}^{S,b}$ is obtained by summing Eqs.~\eqref{eq:dsigmaSbXt} and \eqref{eq:dsigmaSbXtfixup}.
The terms in \eqref{eq:dsigmaSbXtfixup} are identified as having repeated hard radiators i.e.  the hard radiators of the outer antenna ($h_1$ and $h_2$) produce mapped momenta $\widetilde{(h_1s)}$ and $\widetilde{(sh_2)}$ that become the hard radiators of the inner antenna.  They can be integrated according to 
Eqs.~\eqref{eq:bcintFF}, \eqref{eq:bcintIF} and \eqref{eq:bcintII}, to give contributions of the form,
\begin{eqnarray}
&&+\frac{1}{2}{\cal X}_3^0(s_{h_1h_2},x_1,x_2) X_3^0(h_1,s_2,h_2) 
|{\cal M}_n(\ldots,\widetilde{(h_1s_2)},\widetilde{(s_2h_2)},\ldots)|^2. 
\end{eqnarray}
These terms produce poles that cancel against the integrated wide
angle soft terms discussed in Section~\ref{sec:RVeff}.
and the real-virtual subtraction terms discussed in Section~\ref{sec:sigc}.
The presence of $D_{4,b}^0$ and $F_{4,b}^0$ antennae in $\dsigma_{NNLO}^S$
is an indicator that there are contributions from wide angle soft radiation, see for example Refs.~\cite{GehrmannDeRidder:2007jk,Glover:2010im}.  

Similarly, the almost colour-connected contributions in $\dsigma^{S,c}_{NNLO}$ are produced by matrix elements of the type
$|{\cal M}_{n+2}(\ldots, a,s_1,h_1,s_2,h_2,\ldots)|^2$ and  
have the form~\cite{GehrmannDeRidder:2005cm,Glover:2010im},
\begin{eqnarray}
\label{eq:dsigmacex}
&&-\frac{1}{2}X_3^0(a,s_1,h_1) X_3^0(\widetilde{(h_1s_1)},s_2,h_2) 
|{\cal M}_n(\ldots,\widetilde{(as_1)},\widetilde{((h_1s_1)s_2)},\widetilde{(s_2h_2)},\ldots)|^2\nonumber \\
&&-\frac{1}{2}X_3^0(h_1,s_2,h_2) X_3^0(a,s_1,\widetilde{(h_1s_2)}) 
|{\cal M}_n(\ldots,\widetilde{(as_1)},\widetilde{((h_1s_2)s_1)},\widetilde{(s_2h_2)},\ldots)|^2.
\end{eqnarray}
Note that unlike the contributions described in Eq.~\eqref{eq:dsigmaSbXtfixup}, the radiators for the inner antenna are not {\em both} constructed from mapped momenta. 
Upon integration, these terms also produce poles that cancel against wide angle soft terms and will be further discussed in Section~\ref{sec:sigc}.  

\subsubsection{Integration of large angle soft terms: $\int_1 \dsigma_{NNLO}^{S,e}$}
\label{sec:RVeff}

For processes involving soft gluons the real-real channel
has an additional subtraction contribution denoted by $\dsigma_{NNLO}^{S,e}$ 
due to large angle soft radiation~\cite{Weinzierl:2008iv,Weinzierl:2009nz,GehrmannDeRidder:2007jk,Glover:2010im}. This term removes the
remnant soft gluon behaviour associated with the phase space
mappings of the iterated structures of the double-real subtraction
contributions $\dsigma_{NNLO}^{S,(b,c)}$.
In these contributions two successive mappings have
been applied and the gluon emission can occur before or after the
first mapping.
The large angle soft subtraction terms is 
constructed to account for that and the soft eikonal factors can involve {\it either} unmapped or mapped momenta.  

The wide angle soft terms are however not uniquely defined, only their soft
behaviour is determined.  They can be obtained after two successive mappings of
the same kind have been applied.  In other words, either two final-final, two
initial-final or two initial-initial mappings are used. Those are presented in
\cite{Glover:2010im}.

Alternatively, they can be constructed after the application of two successive
mappings where the first of those is {\em always} a final-final mapping,  $(i,j,k) \to (I,K)$. To
achieve this we modified the large angle soft contribution of the double-real
piece, given by Eqs.~(5.9), (5.13) and (5.17) of Ref.~\cite{Glover:2010im}, to the
equally valid forms given in \eqref{eq:LAST1}, \eqref{eq:LAST2} and
\eqref{eq:LAST3} respectively. Following this latter choice, the integrated
large angle soft subtraction terms can simply be obtained through integration
over the final-final antenna phase space ${\rm d}\Phi_{X_{ijk}}$ given
in eq. (\ref{eq:FFPSfact}).  

The unintegrated soft factor terms are eikonal factors of the form
$$S_{ajc}=\frac{2 s_{ac}}{s_{aj}s_{jc}}$$ where $a,c$ denotes initial or
final state external partons.
The hard radiators, $a$ and $c$, may be momenta present after the first mapping, or momenta produced by the second mapping, i.e.
\begin{equation}
p_a,~p_c \equiv p_a,~p_c \in \{p\}_{m+3}
\qquad
{\rm or}
\qquad
p_a,~p_c \equiv  p_{a^\prime},~p_{c^\prime} \in \{p\}_{m+2}.
\end{equation}
Because $j$ is eliminated in the first mapping, it is not a member of $\{p\}_{m+3}$ or $\{p\}_{m+2}$.
Nevertheless, in the latter case, the second mapping should also be applied to $p_j$ to produce a new soft momentum $p_{j^\prime}$.
When this is a final-final or initial-final mapping, it is a trivial mapping $p_j \to p_{j^\prime} \equiv p_{j}$.  However, when the second mapping is of the initial-initial form, then $p_j \to p_{j^\prime} \equiv \tilde{p}_j$ as given by Eq.~\eqref{3to2IImap}.

In general, and after factorisation of the phase space according to
Eq.~\eqref{eq:psx1},  
one always finds differences of terms of the form,
\begin{eqnarray}
\lefteqn{
\NNNLORR\, {\rm d}\Phi_{X_{ijk}}\, \dPS{m+1}{m+3}}\nonumber\\
&\times& \frac{1}{2}
\left [ S_{ajc}  - S_{a^\prime j^\prime c^\prime}  \right] \,
X^0_3(\{p\}_{m+3})  
\, |{\cal M}_{m+2}(\{p\}_{m+2})|^2 \JET_{m}^{(m)}(\{p\}_{m+2})\;,
\end{eqnarray}
where $X_3^0$ is the antenna corresponding to the second mapping, i.e. final-final, initial-final or initial-initial.  

For the final-final kinematical configuration,
where the second mapping transforms the momenta $(I, l, K)\to
(I^{'}, K^{'})$, we find a contribution of the form,
\begin{eqnarray}
\lefteqn{\int\dsigma_{NNLO}^{S,e,(FF)}
=\NNNLORR\,  \int {\rm d}\Phi_{X_{ijk}}\,\sum_{m+1}\dPS{m+1}{m+3}} \nonumber \\
&& \hspace{1.5cm}\times
 \frac{1}{2} \Big[S_{I^{'}jK^{'}}-S_{IjK}-S_{ajI^{'}}+S_{ajI}-S_{K^{'}jb}+S_{Kjb}\Big]\,
 X^0_{IlK}
\nonumber\\
&& \hspace{1.5cm} \times
|{\cal M}_{m+2}(\ldots,a,I^{'},K^{'},b,\ldots)|^2\,
\JET_{m}^{(m)}(p_{3},\ldots,p_{I^{'}},p_{K^{'}},\ldots,p_{m+3})\;.
\label{eq:Sff}
\end{eqnarray}
We see that there are three pairs of soft antennae which correspond to inserting the soft gluon around the radiator gluons, in this case, between $a$ and $I^\prime$, between $I^\prime$ and $K^\prime$ and between $K^\prime$ and $b$.

Similarly, when the second mapping is of the initial-final type corresponding to the mapping $(\hat{n}, l, K)\to
(\hat{N}, K^{\prime})$,  we have,
\begin{eqnarray}
\lefteqn{\int\dsigma_{NNLO}^{S,e,(IF)}
=\NNNLORR\, \int  {\rm d}\Phi_{X_{ijk}}\,\sum_{m+1}\dPS{m+1}{m+3}} \nonumber \\
&& \hspace{1.5cm}\times \sum_{n=1,2}
\frac{1}{2} \Big[S_{\hat{N}jK^{'}}-S_{\hat{n}jK}-S_{K^{\prime}jb}+S_{Kjb}
-S_{aj\hat{N}}+S_{aj\hat{n}} \Big]\,X^0_{n,lK}
\nonumber\\
&& \hspace{1.5cm} \times
|{\cal M}_{m+2}(\ldots,a,\hat{N},K^{'},b,\ldots)|^2\,
\JET_{m}^{(m)}(p_{3},\ldots,p_{K^{'}},\ldots,p_{m+3})\;.
\label{eq:Sif}
\end{eqnarray}
Once again, there are three pairs of soft terms, corresponding to inserting the soft gluon in the three positions into the colour-connected gluons, $\ldots,a,\hat{N},K^{'},b,\ldots$.   In this case, the last two terms simply cancel.  This is because the radiator momenta appears in both the numerator and denominator of the eikonal factor, coupled with the fact that the initial momentum is simply scaled in the initial-final mapping, $ S_{aj\hat{n}} \equiv S_{aj\hat{N}}$.

Finally, when the second mapping is of the initial-initial type:
$(\hat{n}, \hat{p}, l,\hdots,a,b,\hdots)\to (\hat{N},
\hat{P},\hdots,\tilde{a},\tilde{b},\hdots)$ we have,
\begin{eqnarray}
\lefteqn{\int\dsigma_{NNLO}^{S,e,(II)}
=\NNNLORR\,  \int {\rm d}\Phi_{X_{ijk}}\,\sum_{m+1}\dPS{m+1}{m+3}} \nonumber \\
&& \hspace{1.5cm}\times\sum_{n,p=1,2}
\frac{1}{2}\Big[-S_{\hat{N}\tilde{j}\hat{P}}+S_{\hat{n}j\hat{p}}+S_{\tilde{a}\tilde{j}\hat{N}}-S_{aj\hat{n}}
+S_{\hat{P}\tilde{j}\tilde{b}}-S_{\hat{p}jb}\Big]\, X^0_{np,l}
\nonumber\\
&& \hspace{1.5cm} \times
|{\cal M}_{m+2}(\ldots,\tilde{a},\hat{N},\hat{P},\tilde{b},\ldots)|^2\,
\JET_{m}^{(m)}(\tilde{p}_{3},\ldots,\tilde{p}_{m+3})\;.
\label{eq:Sii}
\end{eqnarray}
As anticipated, when the soft factor involves hard radiators produced by the initial-initial mapping, the soft momentum $p_j$ is replaced by the momentum $p_{\tilde{j}}$ obtained by applying the boost for the initial-initial mapping directly to momentum $p_j$.

When the unresolved momentum is unboosted and denoted by $p_{j}$, the integrated form
of the soft factor is given by, 
\begin{eqnarray}
\label{eq:ffSajc}
{\cal S}(s_{ac},s_{IK},x_{ac,IK}) &=& \frac{1}{C(\e)}\int \d \Phi_{X_{ijk}} S_{ajc},
\end{eqnarray}
where $a$ and $c$ are arbitrary hard radiator partons.
After relabelling the momenta such that $I,K \to i,k$, then \cite{GehrmannDeRidder:2007jk},
\begin{eqnarray}
\lefteqn{{\cal S}(s_{ac},s_{ik},x_{ac,ik})}\nonumber \\
&=& \left( \frac{s_{ik}}{\mu^2} \right)^{-\e}\;  
 \bigg[\frac{1}{\e^2} -\frac{1}{\e}\ln\left(x_{ac,ik}\right)  
-\Li_{2}\left(-\frac{1-x_{ac,ik}}{x_{ac,ik}}\right)-\frac{7\pi^2}{12}
 + {\cal O}(\e)\bigg]\;, 
\end{eqnarray}
where we have defined
\begin{equation}
x_{ac,ik} = \frac{s_{ac}s_{ik}}{(s_{ai}+s_{ak})(s_{ci}+s_{ck})}\;.
\end{equation}
Since the soft eikonal factor is invariant under crossing of one or two
partons from the final to the initial state the integrated
soft factor is independent of where the external partons are situated.   

The integrated final-final soft factor involving a boosted
unresolved momentum (denoted by $p_{\tilde{j}}$ in the soft eikonal factor is defined by,
\begin{eqnarray}
\tilde{\cal S}(s_{ac},s_{ik},x_{ac,ik})&=&\frac{1}{C(\e)}\int {\rm d}\Phi_{X_{ijk} } S_{a\tilde{j}c}.
\end{eqnarray}
Since $S_{abc}$ is composed of Lorentz invariants, we can simply invert the Lorentz
boost that mapped $j \to \tilde{j}$ \eqref{3to2IImap} such that
\begin{equation}
S_{a\tilde{j}c} = S_{\underline{a}j\underline{c}}
\end{equation}
where $\underline{a},~\underline{c}$ are also obtained by inverting the same Lorentz boost so that
\begin{eqnarray}
\tilde{\cal S}(s_{ac},s_{ik},x_{ac,ik})&=&\frac{1}{C(\e)}\int {\rm d}\Phi_{X_{ijk} } S_{\underline{a}j\underline{c}}.
\end{eqnarray}
The RHS has precisely the same form as Eq.~\eqref{eq:ffSajc} with $p_a \to p_{\underline{a}}$ and $p_c \to p_{\underline{c}}$, so that
\begin{eqnarray}
\tilde{\cal S}(s_{ac},s_{ik},x_{ac,ik}) \equiv {\cal S}(s_{\underline{a}\underline{c}},s_{ik},x_{\underline{a}\underline{c},ik}).
\end{eqnarray}
Furthermore, we can exploit the fact that ${\cal S}$ only depends on invariants and can apply the same Lorentz boost such that $$\{p_{\underline{a}}, p_{\underline{c}}, p_{i},p_{k}\} \to \{p_{a}, p_{c}, \tilde{p}_{i},\tilde{p}_{k}\}$$
so that,
\begin{eqnarray}
\tilde{\cal S}(s_{ac},s_{ik},x_{ac,ik}) \equiv {\cal S}(s_{ac},s_{\tilde{i}\tilde{k}},x_{ac,\tilde{i}\tilde{k}}).
\end{eqnarray}

The integrated soft factors only contribute to the soft region
($x_1=x_2=1$). Using the $(m+1)$ phase space factorisation as given in   
Eq.~\eqref{eq:psx1} and
inserting the integrated large angle subtraction factors into
Eqs.~\eqref{eq:Sff}, \eqref{eq:Sif} and \eqref{eq:Sii} while  relabelling
$I^\prime \to I$, $I \to i$, $K^\prime \to K$ and $K \to k$
we find,
\begin{eqnarray}
\lefteqn{\int\dsigma_{NNLO}^{S,e,(FF)}
= \NNNLORV\,\sum_{m+1}\, \dPSxx{m+1}{m+3}} \nonumber \\
&& \times
\frac{1}{2} X^0_{ilk} \,\delta(1-x_1)\,\delta(1-x_2)\bigg[
\S{I}{K}{i}{k}-\S{i}{k}{i}{k}-\S{a}{I}{i}{k}\nonumber \\
&& \hspace{3cm}+\S{a}{i}{i}{k}-\S{K}{b}{i}{k}+\S{k}{b}{i}{k}\bigg]
\nonumber\\
&& \hspace{1.5cm} \times
|{\cal M}_{m+2}(\ldots,a,I,K,b,\ldots)|^2\,
\JET_{m}^{(m)}(p_{3},\ldots,p_I,p_K,\ldots,p_{m+3})\;
\label{eq:Sff2},\\
&&
\lefteqn{\int\dsigma_{NNLO}^{S,e,(IF)}
=\NNNLORV\,\sum_{m+1}\, \dPSxx{m+1}{m+3}} \nonumber \\
&& \times\sum_{n=1,2}
\frac{1}{2} X^0_{n,lk} \delta(1-x_1)\,\delta(1-x_2)
\,\bigg[\S{\hat{N}}{K}{i}{k}-\S{\hat{n}}{k}{i}{k}\nonumber \\
&& \hspace{4cm}
-\S{K}{b}{i}{k}+\S{k}{b}{i}{k}\bigg]
\nonumber\\
&& \hspace{1.5cm} \times
|{\cal M}_{m+2}(\ldots,a,\hat{N},K,b,\ldots)|^2\,
\JET_{m}^{(m)}(p_{3},\ldots,p_{K},\ldots,p_{m+3})\;,
\label{eq:Sif2}\\
&&
\lefteqn{\int\dsigma_{NNLO}^{S,e,(II)}
=  \NNNLORV\,\sum_{m+1}\, \dPSxx{m+1}{m+3}} \nonumber \\
&& \times\sum_{n,p=1,2}
\frac{1}{2} X^0_{np,l}\delta(1-x_1)\,\delta(1-x_2) \,\bigg[-\S{\hat{N}}{\hat{P}}{\tilde{i}}{\tilde{k}}+
\S{\hat{n}}{\hat{p}}{i}{k}
\nonumber \\
&& \hspace{1.2cm}
+\S{\tilde{a}}{\hat{N}}{\tilde{i}}{\tilde{k}}-\S{a}{\hat{n}}{i}{k}
+\S{\hat{P}}{\tilde{b}}{\tilde{i}}{\tilde{k}}-\S{\hat{p}}{b}{i}{k}\bigg]
\nonumber\\
&& \hspace{1.5cm} \times
|{\cal M}_{m+2}(\ldots,\tilde{a},\hat{N},\hat{P},\tilde{b},\ldots)|^2\,
\JET_{m}^{(m)}(\tilde{p}_{3},\ldots,\tilde{p}_{m+3})\;.
\label{eq:Sii2}
\end{eqnarray}
Note that the double $\e$-poles cancel within the square brackets, so that,  for example, the leading pole for the square bracket in Eq.~\eqref{eq:Sff2} is
$$
\frac{1}{2\e}\log\left(\frac{s_{ai}s_{IK}s_{kb}}{s_{aI}s_{ik}s_{Kb}}\right)+ {\cal O}(1)
$$
and similarly for Eqs.~\eqref{eq:Sif2} and ~\eqref{eq:Sii2}.

Moreover, this combination of integrated soft factors does not have
any single unresolved limit and therefore does not require further subtraction.

\subsection{Subtraction terms for one-loop single-unresolved 
contributions: $\dsigma_{NNLO}^{VS}$}
\label{subsec:RVVS1}

As we have seen before, the contribution obtained by 
integrating the single unresolved subtraction terms present in $\dsigma_{NNLO}^{S,a}$ precisely cancels the pole structure of the one-loop $(m+1)$-parton contribution $\dsigma_{NNLO}^{RV}$. The real-virtual subtraction term  $\dsigma_{NNLO}^{VS}$ of Eq.~\eqref{eq:dsigmaTbreak} therefore has to subtract three types of contributions: 
\begin{itemize}
\item[(a)] Single unresolved limits of the  virtual one-loop
$(m+3)$-parton matrix element.\\
Each subtraction term takes
the form of a tree- or one-loop-antenna function multiplied by the
one-loop or tree- colour ordered matrix elements respectively which 
we denote by $\dsigma_{NNLO}^{VS,a}$.   
\item[(b)] Terms of the type ${\cal X}_3^0 X_3^0$ that cancel the explicit poles introduced by one-loop matrix elements and one-loop antenna functions present in $\dsigma_{NNLO}^{VS,a}$. This term is named $\dsigma_{NNLO}^{VS,b}$. 
\item[(c)] Terms of the type ${\cal X}_3^0 X_3^0$ that compensate for
any remaining poles, specifically those produced by $\int
\dsigma_{NNLO}^{S,(b,c)}$ and $\int \dsigma_{NNLO}^{S,e}$, and which are
denoted by  $\dsigma_{NNLO}^{VS,c}$,
\end{itemize}
so that \begin{equation}
 \dsigma_{NNLO}^{VS}=
 \dsigma_{NNLO}^{VS,a}
+ \dsigma_{NNLO}^{VS,b} 
+ \dsigma_{NNLO}^{VS,c}.
\end{equation}
The types of contributions present in each of these terms are summarised in Table~2.

Note that $\dsigma_{NNLO}^{VS}$ is a subtraction term, and it must therefore be added back in integrated form to the $m$-parton final state.  
There it combines with the twice integrated double real subtraction term
$\int_2\dsigma_{NNLO}^{S,2}$ and the mass-factorisation counterterm denoted by $\dsigma_{NNLO}^{MF,2}$ according to Eq.~\eqref{eq:Udef} to provide the singularity structure necessary to cancel the pole structure of the double virtual contribution $\dsigma_{NNLO}^{VV}$.

\begin{table}[t!]
\begin{center}
\begin{tabular}{|c|ccc|}
\hline
 &   $a$ & $a$  & $b,c$ \\\hline
$\dsigma_{NNLO}^{VS}$   & $ X_3^1 |{\cal M}^0_{m+2}|^2$    & $ X_3^0 |{\cal M}^1_{m+2}|^2$ & ${\cal X}_3^0 X_3^0 |{\cal M}^0_{m+2}|^2$   \\
$\int_1 \dsigma_{NNLO}^{VS}$ &  ${\cal X}_3^1 |{\cal M}^0_{m+2}|^2$& ${\cal X}_3^0 |{\cal M}^1_{m+2}|^2$ & ${\cal X}_3^0 {\cal X}_3^0 |{\cal M}^0_{m+2}|^2$ \\ \hline
\end{tabular}
\end{center}
\label{tab:VStypes}
\caption{Type of contribution to the real-virtual subtraction term ${\rm{d}}\hat\sigma_{NNLO}^{VS}$, together with the integrated form of each term.   The unintegrated antenna functions are denoted as $X_3^0$ and $X_3^1$ while their integrated forms are ${\cal X}_3^0$ and  ${\cal X}_3^1$ respectively.  $|{\cal M}^1_{n}|^2$ denotes the interference of the tree-level and one-loop $n$-particle colour ordered amplitude while $|{\cal M}^0_{n}|^2$ denotes the square of an $n$-particle tree-level colour ordered amplitude.  }
\end{table}

\subsubsection{One-loop single-unresolved 
contributions: $\dsigma_{NNLO}^{VS,a}$}
\label{sec:siga}

In single unresolved limits, the behaviour of $(m+3)$-parton 
one-loop amplitudes is described 
by the sum of two different contributions~\cite{Bern:1994zx,Bern:1998sc,
Kosower:1999xi,Kosower:1999rx,Bern:1999ry}: a single unresolved tree-level factor times a
$(m+2)$-parton one-loop amplitude and a single unresolved one-loop factor 
times a $(m+2)$-parton tree-level amplitude, as 
illustrated in Figure~1. Accordingly, we construct 
the one-loop single unresolved subtraction term from products of tree-
and one-loop antenna functions with one-loop and tree-amplitudes
respectively.

\begin{figure}[t!]{ 
\label{fig:subv}
\epsfig{file=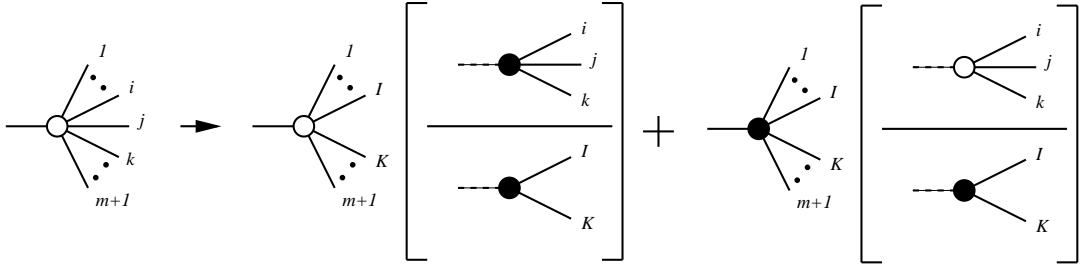,height=3.5cm}
\caption{Illustration of NNLO antenna factorisation representing the
factorisation of both the one-loop
``squared" matrix elements (represented by the white blob)
and the $(m+1)$-particle phase
space when the unresolved particles are colour-connected. 
The terms in square brackets
represent both the three-particle tree-level antenna function $X^0_{ijk}$ 
and the three-particle one-loop antenna function $X^1_{ijk}$
and the antenna phase space.}
}
\end{figure}
Analogously to the case of handling explicit infrared poles present in 
$\int_1 \dsigma_{NNLO}^{S,a}$ or to deal with single unresolved limits of
tree level amplitudes, we need to decompose this contribution into
three parts according to where the two hard radiators
are situated. We note that the momentum mappings which implement momentum conservation
away from the single unresolved limits of each configuration are the same 
as in the previous section, namely equations (\ref{3to2FFmap}), (\ref{3to2IFmap}) and (\ref{3to2IImap}).

In the final-final configuration, the subtraction term is given by,
\begin{eqnarray}
\dsigma_{NNLO}^{VS,a,(FF)}
&= & \NNNLORV\,\sum_{{\rm perms}}\dPSxx{m+1}{m+3} \nonumber \\
&\times& \,\sum_{j} \Bigg [X^0_{ijk}\,\delta(1-x_1)\delta(1-x_2)\,
|{\cal M}^1_{m+2}(\ldots,I,K,\ldots)|^2\,
\JET_{m}^{(m)}(\{p\}_{m+2})
\nonumber \\
&&\phantom{\sum_{j} }+\;X^1_{ijk}\,\delta(1-x_1)\delta(1-x_2)\,
|{\cal M}_{m+2}(\ldots,I,K,\ldots)|^2\,
\JET_{m}^{(m)}(\{p\}_{m+2})\;\Bigg
],\nonumber \\
\label{eq:subv2aff}
\end{eqnarray}
where the one-loop three-parton antenna
function $X^1_{ijk}$ depends only on the antenna momenta
$p_i,p_j,p_k$. $X^1_{ijk}$ correctly describes all simple unresolved limits of the difference between an $(m+2)$-parton one-loop corrected 
squared matrix element and the product of a tree-level antenna function 
with the  $m$-parton one-loop corrected squared matrix element.
It can therefore be constructed out of colour ordered and 
renormalised one-loop three-parton and two-parton 
matrix elements as
\begin{equation}
\label{eq:X1def}
X_{ijk}^1 = S_{ijk,IK}\, \frac{|{\cal M}^1_{ijk}|^2}{|{\cal M}^0_{IK}|^2} - 
X_{ijk}^0\, \frac{|{\cal M}^1_{IK}|^2}{|{\cal M}^0_{IK}|^2} \;,
\end{equation}
where $S_{ijk,IK}$ denotes the symmetry factor associated with the antenna, which accounts both for
potential identical particle symmetries and for the presence of more than one antenna in the basic two-parton process.

It should be noted that  $X^1_{ijk}$ is renormalised at a scale corresponding 
to the invariant mass of the antenna partons, $s_{ijk}$, while the one-loop
$(m+2)$-parton matrix element is renormalised at a scale $\mu^2$. To 
ensure correct subtraction of terms arising from renormalisation, 
we have to substitute 
\begin{equation}
X^1_{ijk} \to X^1_{ijk} + \frac{\beta_0}{\e}\,C(\e) X^0_{ijk} \left(
\left(s_{ijk}\right)^{-\e} - \left(\mu^2\right)^{-\e} \right)  
\end{equation}
in (\ref{eq:subv2aff}).
The terms arising from this substitution
 will in general be kept apart in the construction of the 
colour ordered subtraction terms, since they all share a  common 
colour structure $\beta_0$. 

Similar subtraction terms are appropriate in the initial-final and initial-initial configurations.
In the initial-final case, we have,
\begin{eqnarray}
\dsigma_{NNLO}^{VS,a,(IF)}
&= & \NNNLORV\,\sum_{{\rm perms}}\dPSxx{m+1}{m+3}\nonumber \\
&\times& \,\sum_{i=1,2}\sum_{j} \Bigg [X^0_{i,jk}\,\delta(1-x_1)\delta(1-x_2)\,
|{\cal M}^1_{m+2}(\ldots,\hat{I},K,\ldots)|^2\,
\JET_{m}^{(m)}(\{p\}_{m+2})\;
\nonumber \\
&&\phantom{\sum_{i=1,2}\sum_{j} }+ X^1_{i,jk}\,\delta(1-x_1)\delta(1-x_2)\,
|{\cal M}_{m+2}(\ldots,\hat{I},K,\ldots)|^2\,
\JET_{m}^{(m)}(\{p\}_{m+2})\;\Bigg]\;,\nonumber \\
\label{eq:subv2aif} 
\end{eqnarray}
where the one-loop antenna function $X^1_{i,jk}$ is obtained from $X^1_{ijk}$ by crossing parton $i$ from the final state to the initial state.

The initial-initial subtraction term is 
\begin{eqnarray}
\dsigma_{NNLO}^{VS,a,(II)}
&= & \NNNLORV\,\sum_{{\rm perms}}\dPSxx{m+1}{m+3} \nonumber \\
&\times& \,\sum_{i,k=1,2}\sum_{j} \Bigg [X^0_{ik,j}\,\delta(1-x_1)\delta(1-x_2)\,
|{\cal M}^1_{m+2}(\ldots,\hat{I},\hat{K},\ldots)|^2\,
\JET_{m}^{(m)}(\{p\}_{m+2})\;
\nonumber \\
&&\phantom{\sum_{i,k=1,2}\sum_{j} }+ X^1_{ik,j}\,\delta(1-x_1)\delta(1-x_2)\,
|{\cal M}_{m+2}(\ldots,\hat{I},\hat{K},\ldots)|^2\,
\JET_{m}^{(m)}(\{p\}_{m+2})\;\Bigg]\;,\nonumber \\
\label{eq:subv2aii} 
\end{eqnarray}
with $X^1_{ik,j}$ obtained by crossing partons $i$ and $k$ from $X^1_{ijk}$.
We note that in order to fulfil overall momentum conservation, the initial-initial momentum mapping requires all the momenta in the arguments of the reduced matrix elements and the jet functions to be redefined. 

The 
one-loop single unresolved subtraction terms, $X^1_{ijk}$,
$X^1_{i,jk}$ and $X^1_{ik,j}$ can never be related to integrals of
tree-level subtraction terms. 
Therefore, after integration over the three-parton antenna phase space, this 
component of the subtraction term must be added back in integrated
form to the terms yielding the $m$-parton final state contribution. 
This can be 
accomplished using the techniques described
in~\cite{GehrmannDeRidder:2003bm} 
and analytic expressions for all of the integrated three-parton one-loop antennae are available in Refs.~\cite{GehrmannDeRidder:2005cm,Daleo:2009yj,Gehrmann:2011wi}.

\subsubsection{Cancellation of explicit infrared divergences in $\dsigma_{NNLO}^{VS,a}$: $\dsigma_{NNLO}^{VS,b}$}
\label{sec:sigb}
By construction $\dsigma_{NNLO}^{VS,a}$
correctly approximates the one-loop $(m+3)$-parton 
contribution to $(m +1)$-jet final states in all single unresolved 
limits. However, the one-loop antenna functions
and one-loop reduced matrix elements present in
Eqs.~\eqref{eq:subv2aff}, \eqref{eq:subv2aif} and \eqref{eq:subv2aii} 
introduce explicit $\eps$-poles which do not correspond to poles 
of the real-virtual matrix element (\ref{eq:nnloonelalt}) and therefore
have to be removed. 

It is the purpose of this subsection to present a new term
 denoted by $\dsigma_{NNLO}^{VS,b}$ which removes the  
explicit $\eps$-poles present in  
$\dsigma_{NNLO}^{VS,a}$.
To achieve this, we introduce further partially integrated
subtraction terms built with products of
an unintegrated and an integrated three parton antenna 
function times a reduced matrix element squared. 
For a given configuration, the integrated antenna will be of
a given type (final-final, initial-final or initial-initial)
multiplying a sum of three unintegrated antennae of each configuration type. 

Let us consider first the explicit poles in the one-loop reduced matrix elements present in
Eqs.~\eqref{eq:subv2aff}, \eqref{eq:subv2aif} and \eqref{eq:subv2aii}.   
The pole structure of 
$|{\cal M}^1_{m+2}(\{p\}_{m+2})|^2$ is well understood as a sum of ${\bf I}^{(1)}$ operators~\cite{Catani:1998bh}. 
These $\eps$-poles can be simply subtracted using integrated ${\cal X}_3^0$ antennae.   To derive the relevant term, we simply add a contribution constructed from each ${\bf I}^{(1)}$ operator present in the pole structure of $|{\cal M}^1_{m+2}(\{p\}_{m+2})|^2$ according to the relation shown in Table~3,
\begin{equation}
2 {\bf I}^{(1)}_{xy}(s_{ab}) \to \frac{1}{S^X_{ab}}\, {\cal X}_3^0.
\end{equation}
The particle type $(x,y)$ fixes the flavour of antenna, while the momenta $(a,b)$ determines whether the integrated antenna is final-final, initial-final or initial-initial.
This is summarised by adding the contribution obtained by the replacement
\begin{equation}
|{\cal M}^1_{m+2}(\{p\}_{m+2})|^2\,\delta(1-x_1)\delta(1-x_2)
\to
\sum_{ab} \left( \frac{1}{S^X_{ab}}{\cal X}_3^0 (s_{ab},x_1,x_2) \right) |{\cal M}^0_{m+2}(\{p\}_{m+2})|^2,
\end{equation}
in Eqs.~\eqref{eq:subv2aff}, \eqref{eq:subv2aif} and \eqref{eq:subv2aii}.
Here $ab$ runs over the pairs of colour-connected particles, and ${\cal X}_3^0 (s_{ab})$ is an integrated antenna that depends on the particle type and on whether $a$ and $b$ are
in the initial or final states.
This contribution guarantees that the $\eps$-poles associated with soft and final state collinear singularities introduced through the one-loop matrix elements in 
Eqs.~\eqref{eq:subv2aff}, \eqref{eq:subv2aif} and \eqref{eq:subv2aii} 
are cancelled.\footnote{Any remaining initial state collinear singularities are removed by the mass factorisation counterterm.}   
Note that the momenta $a$ and $b$ lie in the set $\{p\}_{m+2}$ and that the invariant mass associated with each antenna is constructed from mapped momenta and cannot be related to terms coming from $\int_1\dsigma^{S,1}$.

\begin{table}
\begin{center}
\begin{tabular}{|l||c|c||c|c||c|c|}
\hline
& \multicolumn{2}{c||}{FF}& \multicolumn{2}{c||}{IF}& \multicolumn{2}{c|}{II} \\ \hline
& $S^{X}_{ij}$ & ${\cal X}_3^0$ & $S^{X}_{ij}$ & ${\cal X}_3^0$& $S^{X}_{ij}$ & ${\cal X}_3^0$ \\
\hline 
$2 {\bf I}^{(1)}_{qq}(s_{ij})$ & 1 & ${\cal A}_{qg\bar q}$ & 1 & ${\cal A}_{q,qg}$ & 1 & ${\cal A}_{q\bar q,g}$\\
$2 {\bf I}^{(1)}_{qg}(s_{ij})$ & 2 & ${\cal D}_{qg\bar q}$ & 2 & ${\cal D}_{q,gg}$, ${\cal D}_{g,gq}$ & 1 & ${\cal D}_{qg,g}$\\
$2 {\bf I}^{(1)}_{gg}(s_{ij})$ & 3 & ${\cal F}_{ggg}$ & 2 & ${\cal F}_{g,gg}$ & 1 & ${\cal F}_{gg,g}$\\
$2 {\bf I}^{(1)}_{qg,F}(s_{ij})$ & 2 & ${\cal E}_{q\bar q q^\prime}$ & 2 & ${\cal E}_{q,q^\prime \bar{q}^\prime}$ & & \\
$2 {\bf I}^{(1)}_{gg,F}(s_{ij})$ & 1 & ${\cal G}_{g q\bar q}$ & 1 & ${\cal G}_{g,q\bar{q}}$ & & \\\hline
\end{tabular}
\label{tab:I1map}
\caption{The relation between ${\bf I}^{(1)}$ operators and integrated antenna showing the dependence on whether the pair ($i,j$) is in the final-final, initial-final or initial-initial state.}
\end{center}
\end{table}

There are also explicit poles present in the one-loop antenna, $X^1_{ijk}$ which are produced by the physical matrix elements making up the antenna~\eqref{eq:X1def}.  These can also be described by integrated antenna, however now the relevant mass scale is constructed from momenta before the mapping i.e. the momenta lying in the set $\{p\}_{m+3}$.

The first type of poles are associated with $|{\cal M}^1_{ijk}|^2$ in Eq.~\eqref{eq:X1def}.   
The second contribution comes from the term proportional to $|{\cal M}^1_{IK}|^2$.  
For example, the explicit poles in $X^1_{ijk}$
are removed by adding the contribution obtained by the replacement
\begin{eqnarray}
\label{eq:X1rep}
\lefteqn{X^1_{ijk}\,\delta(1-x_1)\delta(1-x_2) \to }
\nonumber \\
&&
\left[\left(\sum_{cd} 
\frac{1}{S^X_{cd}}{\cal X}_3^0(s_{cd},x_1,x_2)
\right) 
- \frac{M_X}{S^X_{ik}} {\cal X}_3^0(s_{IK},x_1,x_2)\right] X^0_{ijk},\nonumber \\
\end{eqnarray}
where $cd$ runs over the $N_X$ pairs of colour-connected particles within the antenna.   
$N_X$ and $M_X$ depend on the type of antenna, and their values are given in Table~4.
Similar expressions are valid for initial-final and initial-initial antenna.
The terms in the sum in \eqref{eq:X1rep} are systematically provided by $\int_1
\dsigma_{NNLO}^{S,b,3\times 3}$, i.e. the $X_3^0 \times X_3^0$ type terms  coming from Eqs.~\eqref{eq:dsigmaSbX} and \eqref{eq:dsigmaSbXt}.  This
is another example of the cancellation of infrared 
poles between a virtual contribution and integrated subtraction terms
from real emission. On the other hand, the last term must be introduced as a new subtraction term in $ 
\dsigma_{NNLO}^{VS,b}$.  

\begin{table}
\label{tab:NXMX}
\begin{center}
\begin{tabular}{|l||c|c|c|c|c|c|c|c|c|c|c|c|c|}
\hline
$X_{3}^1$ & $A_3^1$ & $\tilde{A}_3^1$ & $\hat{A}_3^1$ & $D_3^1$ & $\hat{D}_3^1$ & $E_3^1$ & $\tilde{E}_3^1$ & $\hat{E}_3^1$  & $F_3^1$ & $\hat{F}_3^1$ & $G_3^1$ & $\tilde{G}_3^1$ & $\hat{G}_3^1$ \\\hline
$N_{X}$ & 2 & 1 & 2 & 3 & 3 & 2 & 1 & 0 & 3 & 3 & 2 & 1 & 0 \\
$M_{X}$ & 1 & 1 & 0 & 2 & 2 & 2 & 0 & 2 & 2 & 2 & 2 & 0 & 2 \\\hline
\end{tabular}
\caption{Number of colour-connected pairs $N_X$ in the one-loop antenna $X_3^1$, and the coefficient $M_X$.}
\end{center}
\end{table}

To be more explicit, in the final-final configuration, $\dsigma_{NNLO}^{VS,b}$ is given by,
\begin{eqnarray}
\lefteqn{\dsigma_{NNLO}^{VS,b,(FF)}
= \NNNLORV\,\sum_{{\rm perms}}\dPSxx{m+1}{m+3}} \nonumber \\
&&\hspace{1.5cm}\times  \sum_{j} \left[\left(\sum_{ab} \frac{1}{S^X_{ab}}{\cal X}^0_3(s_{ab},x_1,x_2)\right) - \frac{M_X}{S^X_{ik}} {\cal X}^0_3(s_{IK},x_1,x_2)\right]
\, X^0_{ijk}\,\nonumber \\
&&\hspace{3cm}\times
|{\cal M}^0_{m+2}(\{p\}_{m+2})|^2\,
\JET_{m}^{(m)}(\{p\}_{m+2}).\nonumber \\
\label{eq:subv2bff}
\end{eqnarray}
Here $p_I$ and $p_K$ are the momenta produced by the mapping for the $ijk$ antenna, while $ab$ are the pairs of colour-connected particles appearing in the matrix element ${\cal M}^0_{m+2}(\{p\}_{m+2})$.
Similar subtraction terms are appropriate in the initial-final and initial-initial configurations,
\begin{eqnarray}
\label{eq:subv2bif} 
\lefteqn{\dsigma_{NNLO}^{VS,b,(IF)}
= \NNNLORV\,\sum_{{\rm perms}}\dPSxx{m+1}{m+3}} \nonumber \\
&&\hspace{1.5cm}\times  \sum_{i=1,2}\sum_{j}  
\left[\left(\sum_{ab} \frac{1}{S^X_{ab}}{\cal X}^0_3(s_{ab},x_1,x_2)\right) -  \frac{M_X}{S^X_{ik}} {\cal X}^0_3(s_{\bar{i}K},x_1,x_2)\right]\, X^0_{i,jk}\nonumber \\
&&\hspace{3cm}\times
|{\cal M}_{m+2}(\{p\}_{m+2})|^2\,
\JET_{m}^{(m)}(\{p\}_{m+2})\;,  \\
\lefteqn{\dsigma_{NNLO}^{VS,b,(II)}
= \NNNLORV\,\sum_{{\rm perms}}\dPSxx{m+1}{m+3}} \nonumber \\
&&\hspace{1.5cm}\times  \sum_{i,k=1,2}\sum_{j} 
\left[\left(\sum_{ab} \frac{1}{S^X_{ab}}{\cal X}^0_3(s_{ab},x_1,x_2)\right) -  \frac{M_X}{S^X_{ik}}{\cal X}^0_3(s_{\bar{i}\bar{k}},x_1,x_2)\right]\, X^0_{ik,j}\,\nonumber \\
&&\hspace{3cm}\times
|{\cal M}_{m+2}(\{p\}_{m+2})|^2\,
\JET_{m}^{(m)}(\{p\}_{m+2})\;. 
\label{eq:subv2bii} 
\end{eqnarray}

\subsubsection{Compensation terms for remaining poles: $\dsigma_{NNLO}^{VS,c}$}
\label{sec:sigc}

In Sect.~\ref{sec:RVbcff}, we identified the integrated form of the double real radiation subtraction term that is due to almost colour-connected hard radiators.   In the final-final case, this has the form,
\begin{eqnarray}
\int_1 \dsigma_{NNLO}^{S,c,(FF)}
&=&   \NNNLORV\,\dPSxx{m+1}{m+3}
\nonumber \\
&\times& \,\frac{1}{2}\,
{\cal X}^0_{ijk}(s_{ik},x_1,x_2)\,
X^0_3(\{p\}_{m+3}) \, |{\cal M}_{m+2}(\{p\}_{m+2})|^2\,
\JET_{m}^{(m)}(\{p\}_{m+2})\;\nonumber\\
\end{eqnarray}
where the invariant mass $s_{ik}$ is constructed from momenta in the set 
$\{p\}_{m+3}$.    
For each term like this, we introduce a subtraction term (with the opposite sign) where the appropriate invariant mass is constructed out of momenta in the set 
$\{p\}_{m+2}$ i.e. the momenta produced by the mapping appropriate to the type of antenna $X^0_3(\{p\}_{m+3})$.
That is, 
\begin{eqnarray}
\dsigma_{NNLO}^{VS, c,(FF)}
&=&   -\NNNLORV\,\dPSxx{m+1}{m+3}
\nonumber \\
&\times& \,\frac{1}{2}\,
{\cal X}^0_{ijk}(s_{IK},x_1,x_2)\,
X^0_3(\{p\}_{m+3}) \, |{\cal M}_{m+2}(\{p\}_{m+2})|^2\,
\JET_{m}^{(m)}(\{p\}_{m+2})\;.\nonumber\\
\end{eqnarray}
Taken together, we find that
\begin{eqnarray}
\int_1 \dsigma_{NNLO}^{S,c,(FF)}+\dsigma_{NNLO}^{VS, c,(FF)}
&=&   \NNNLORV\,\dPSxx{m+1}{m+3}
\nonumber \\
&\times& \,\frac{1}{2}
\bigg[ {\cal X}^0_{ijk}(s_{ik},x_1,x_2) - {\cal X}^0_{ijk}(s_{IK},x_1,x_2)\bigg]\,
X^0_3(\{p\}_{m+3}) \nonumber \\
&\times& \,
|{\cal M}_{m+2}(\{p\}_{m+2})|^2\,
\JET_{m}^{(m)}(\{p\}_{m+2})\;, 
\end{eqnarray}
where the term in square brackets does not have a double pole in $\eps$, but has a leading singularity of the form 
$$\frac{1}{2\eps}\log\left(\frac{s_{ik}}{s_{IK}}\right)+ {\cal O}(1).$$
Combinations of terms like this, plus similar contributions from the  initial-final and initial-initial antenna,
\begin{eqnarray}
\int_1 \dsigma_{NNLO}^{S,c,(IF)}+\dsigma_{NNLO}^{VS, c,(IF)}
&=&  \NNNLORV\,\dPSxx{m+1}{m+3}
\nonumber \\
&\times& \,\frac{1}{2}\,
\bigg[ {\cal X}^0_{i,jk}(s_{ik},x_1,x_2) - {\cal X}^0_{i,jk}(s_{\bar{i}K},x_1,x_2)\bigg]\,
X^0_3(\{p\}_{m+3}) \nonumber \\
&\times& \,
|{\cal M}_{m+2}(\{p\}_{m+2})|^2\,
\JET_{m}^{(m)}(\{p\}_{m+2}) 
\;,\\
\int_1 \dsigma_{NNLO}^{S,c,(II)}+\dsigma_{NNLO}^{VS, c,(II)}
&=&   \NNNLORV\,\dPSxx{m+1}{m+3}
\nonumber \\
&\times& \,\frac{1}{2}\,
\bigg[ {\cal X}^0_{ik,j}(s_{ik},x_1,x_2) - {\cal X}^0_{ik,j}(s_{\bar{i}\bar{k}},x_1,x_2)\bigg]\,
X^0_3(\{p\}_{m+3}) \nonumber \\
&\times& \,
|{\cal M}_{m+2}(\{p\}_{m+2})|^2\,
\JET_{m}^{(m)}(\{p\}_{m+2})\;, 
\end{eqnarray}
together with correction terms coming from the oversubtraction of the single unresolved limits from the $\tilde{X}_4^0$ antenna (see Eq.~\ref{eq:dsigmaSbXtfixup})
ultimately cancel against the integrated wide angle soft radiation term 
$\int_1 \dsigma^{S,e}_{NNLO}$.

To make this cancellation more precise, consider the coefficients of terms proportional to 
\begin{equation}
X_{i\ell k}^0 |{\cal M}_{m+2}(\ldots,a,I,K,b,\ldots)|^2 J_m^{(m)}(\{p\}_{m+2}).
\end{equation}
Here we consider final-final radiation, but the argument is general.
The wide angle soft contribution is given by Eq.~\eqref{eq:Sff}.  As discussed in Section~\ref{sec:RVeff}, 
the leading singularity from the wide angle soft terms is 
proportional to
$$
\frac{1}{2\e} \log\left(\frac{s_{ai}s_{IK}s_{kb}}{s_{aI}s_{ik}s_{Kb}}\right)+ {\cal O}(1).
$$
This pole cancels against combinations of the form (where the dependence on $x_1,~x_2$ is suppressed),
\begin{eqnarray}
\label{eq:Xcomb}
\frac{1}{2}\left[ {\cal X}_{3}^0(s_{ik}) - {\cal X}_{3}^0(s_{ai}) - {\cal X}_{3}^0(s_{kb}) - {\cal X}_{3}^0(s_{IK}) + {\cal X}_{3}^0(s_{aI}) + {\cal X}_{3}^0(s_{Kb})\right].
\end{eqnarray}
The first three terms in Eq.~\eqref{eq:Xcomb} are obtained by integrating terms in the double real subtraction term $\int_1\dsigma^{S, (b,c)}_{NNLO}$.     
The first term is produced by terms coming from
$\int_1\dsigma_{NNLO}^{S,b,3\times 3}$ that compensate for the
oversubtraction of single unresolved poles associated with a
four-parton antenna of the $\tilde{X}_4^0$ type as in
Eq.~\eqref{eq:dsigmaSbXtfixup}.  This can be understood as being due
to repeated radiation where the hard radiators are particles $i$ and
$k$. The second and third terms have opposing signs (compared to the
first term) and come from the almost colour-connected term $\int_1\dsigma_{NNLO}^{S,c}$ and occur when the integrated antenna describes unresolved radiation which took place between hard radiators $a$ and $i$ ($b$ and $k$). For these three terms, the unresolved radiation is emitted from an outer $X_3^0$ antenna, while $X_{i\ell k}^0$ is the inner antenna.  
The final three terms form part of the real-virtual subtraction term $\dsigma_{NNLO}^{VS,c}$, which we can understand as
corresponding to emitting an unresolved momentum between pairs in the colour-connected set $\ldots,a,I,K,b,\ldots$ i.e. unresolved radiation from an inner antenna.  In this case, $X_{i\ell k}^0$ plays the role of outer antenna.   The signs are always fixed to be opposite of the partner term coming from $\int_1\dsigma^{S, (b,c)}_{NNLO}$.
As usual, the subtraction terms constituting $\dsigma_{NNLO}^{VS,c}$
must be integrated over the unresolved phase space and added back in integrated form to the double virtual contribution. 
Expanding \eqref{eq:Xcomb}, we find that the leading pole is proportional to
$$
-\frac{1}{2\e} \log\left(\frac{s_{ai}s_{IK}s_{kb}}{s_{aI}s_{ik}s_{Kb}}\right)+ {\cal O}(1)
$$
which cancels against the leading pole coming from the wide angle soft term.

\section{Renormalisation and mass factorisation}
\label{sec:massfac}
We are concerned with $m$-jet production in the collision of two hadrons  $h_1,h_2$ carrying momenta $H_1, H_2$
\begin{equation}
h_{1}(H_{1})+h_{2}(H_{2})\to j(p_{3})+\ldots+j(p_{m+2}).
\end{equation}
Within the framework of QCD factorisation the cross section for this process is  written in Eq.~\eqref{eq:totsig} as an integral over the infrared and ultraviolet finite hard partonic 
scattering cross section for $m$-jet production from quarks and gluons, multiplied by parton distribution functions (PDF's)
describing the momentum distribution of these partons inside the colliding hadrons.

However, after combining the real and virtual contributions together with the antenna subtraction terms, the partonic cross section ${\rm d}\tilde{\sigma}_{ij}$ contains both ultraviolet and initial-state
collinear singularities.  
We remove the UV singularities through coupling constant renormalisation in the $\overline{\rm MS}$ scheme and absorb the initial-state singularities into the PDFs using the $\overline{\rm MS}$ factorisation
scheme.	For simplicity, we first set the factorisation and renormalisation scales equal to a common scale,
$$
\mu^2 = \mu_R^2 = \mu_F^2.
$$
It is straightforward to restore the dependence of the partonic cross section on these scales using the requirement that the hadronic cross section is independent of them.

\subsection{Ultraviolet renormalisation}
In terms of the bare (and dimensionful) coupling $\alpha_s^b$, the unrenormalised cross section has the perturbative expansion,
\begin{equation}
\label{eq:sigpertunren}
{\rm d}\sigma^{un}_{ij} = {\rm d} \sigma_{ij}^{un,LO}
+\left(\frac{\alpha_s^b}{2\pi}\right){\rm d} \sigma_{ij}^{un,NLO}
+\left(\frac{\alpha_s^b}{2\pi}\right)^2{\rm d} \sigma_{ij}^{un,NNLO}
+{\cal O}\left((\alpha_s^{b})^{3}\right)
\end{equation}
bearing in mind that the LO cross section is ${\cal O}\left((\alpha_s^{b})^{m}\right)$.  Furthermore, each power of the bare coupling is accompanied by a factor of $\bar{C}(\e)$ given by Eq.~\eqref{eq:Cbar}.

The renormalisation is carried out by replacing 
the bare coupling $\alpha_s^{b}$ with the renormalised coupling 
$\alpha_s \equiv \alpha_s(\mu^2)$ evaluated at the renormalisation scale $\mu_R^2=\mu^2$,
\begin{equation}
\label{eq:aren}
\alpha_s^b = Z_{\alpha_s} \mu_R^{2\epsilon}  \alpha_s.
\end{equation}
In the $\overline{{\rm MS}}$ scheme \cite{msbar},
\begin{eqnarray}
\label{eq:Zdef}
Z_{\alpha_s} &=& \frac{1}{\bar{C}(\e)} \Bigg[  
1- \frac{\beta_0}{\e}\left(\frac{\alpha_s}{2\pi}\right) 
+\left(\frac{\beta_0^2}{\e^2}-\frac{\beta_1}{2\e}\right)
\left(\frac{\alpha_s}{2\pi}\right)^2+{\cal O}(\alpha_s^3) \Bigg] \; ,
\end{eqnarray}
where $\beta_0$ and $\beta_1$ are \cite{Gross:1973id,Politzer:1973fx,Caswell:1974gg,Jones:1974mm}
\begin{eqnarray}
\label{eq:bzero}
\beta_0 &=& 
\frac{11 C_A}{6}-\frac{2 N_F}{6},\\
\label{eq:bone}
\beta_1 &=& 
\frac{34 C_A^2}{12}-\frac{10 C_A N_F}{12}-\frac{C_F N_F}{2}.
\end{eqnarray}

Substituting Eqs.~\eqref{eq:aren} and \eqref{eq:Zdef}
into Eq.~\eqref{eq:sigpertunren} and collecting with respect to $\alpha_s$, gives the coefficients of the renormalised (but not infrared-finite) expansion,
\begin{equation}
\label{eq:sigpertren}
{\rm d}\tilde\sigma_{ij} = {\rm d}\tilde\sigma_{ij}^{LO}
+\left(\frac{\alpha_s}{2\pi}\right){\rm d}\tilde\sigma_{ij}^{NLO}
+\left(\frac{\alpha_s}{2\pi}\right)^2{\rm d}\tilde\sigma_{ij}^{NNLO}
+{\cal O}\left(\alpha_s^3 \right).
\end{equation}
Since the $m$-jet cross section has a leading order behaviour proportional to $\alpha_s^m$, we have,
\begin{eqnarray}
{\rm d}\tilde\sigma_{ij}^{LO}  &=& {\rm d} \sigma_{ij}^{un,LO} ,
\\
{\rm d}\tilde\sigma_{ij}^{NLO}  &=& 
\frac{1}{\bar{C}(\e)}  {\rm d} \sigma_{ij}^{un,NLO} 
-\frac{m\beta_0}{\e} {\rm d} \sigma_{ij}^{un,LO}  ,   \\
{\rm d}\tilde\sigma_{ij}^{NNLO} &=& 
\frac{1}{\bar{C}(\e)^2}  {\rm d} \sigma_{ij}^{un,NNLO} -\frac{(m+1)\beta_0}{\e} \frac{1}{\bar{C}(\e)} 
{\rm d} \sigma_{ij}^{un,NLO} \nonumber \\
&& -\left(\frac{m\beta_1}{2\e}-\frac{m(m+1)\beta_0^2}{2\e^2}\right)
{\rm d} \sigma_{ij}^{un,LO}.  
\end{eqnarray}
Note that for the double real radiation contribution $\dsigma_{NNLO}^{RR}$ and its associated subtraction term $\dsigma_{NNLO}^S$, renormalisation simply amounts to the replacement $\bar C(\epsilon)\alpha_s^b  \to \alpha_s$. Likewise, the inverse powers of $\bar{C}(\e)$ are immediately cancelled against the  additional factors of $\bar{C}(\e)$ present in  ${\rm d} \sigma_{ij}^{un, NLO}$ and 
${\rm d} \sigma_{ij}^{un, NNLO}$.

\subsection{Mass factorisation}

After renormalisation, the physical cross section is given by
\begin{eqnarray}
\label{eq:sigunfac}
{\rm d}\sigma(H_1,H_2) &=&\sum_{i,j} \int   
\frac{d\xi_1}{\xi_1} \frac{d\xi_2}{\xi_2} \tilde{f}_i(\xi_1) \tilde{f}_j(\xi_2) {\rm d}\tilde\sigma_{ij}(\xi_1 H_1,\xi_2 H_2)  
\end{eqnarray}
where the bare PDF for a parton of type $a$ carrying a momentum fraction $\xi$ is denoted by $\tilde{f}_a(\xi)$.
For clarity, we have made the dependence of the partonic cross section ${\rm d}\tilde\sigma_{ij}$ on the initial state parton momenta, $\xi_1 H_1$ and $\xi_2 H_2$, explicit. 

The initial-state singularities present in ${\rm d}\tilde\sigma_{ij}$  are removed using mass factorisation to produce the hadronic cross section 
where the finite partonic cross section multiplies 
the physical PDF $f_a(\xi,\mu_F^2)$ at the factorisation scale $\mu_F^2 =\mu_R^2=\mu^2$. The physical PDF are related to the bare 
PDF $\tilde{f}_a(\xi)$ by the convolution,
\begin{eqnarray}
\label{eq:fi}
f_a(\xi,\mu^2) &=&\int dx\,dy\;\tilde{f}_b(x)\Gamma_{ba}(y,\mu^2)\delta(\xi-xy)
= \left[ \tilde{f}_b \otimes \Gamma_{ba} \right] (\xi,\mu^2).
\end{eqnarray}   
The kernel $\Gamma_{ba}$ has the (renormalised) perturbative expansion, 
\begin{eqnarray}
\Gamma_{ba}(y,\mu^2)&=&\delta_{ba}\delta(1-y)
+\left(\frac{\alpha_s(\mu)}{2\pi}\right)\,\Gamma^{1}_{ba}(y)
+\left(\frac{\alpha_s(\mu)}{2\pi}\right)^2\,\Gamma^{2}_{ba}(y)
+{\cal O}(\alpha_{s}^{3})
\label{eq:pmodel}
\end{eqnarray}
where in the ${\overline{\rm MS}}$ scheme, 
\begin{eqnarray}
\Gamma^{1}_{ba}(y)&=&  
 -\frac{1}{\epsilon}\, p_{ba}^{0}(y) , \\
\Gamma^{2}_{ba}(y)&=& 
\frac{1}{2\epsilon^2}
\left[\sum_{c}\left[p_{bc}^{0}\otimes p_{ca}^{0}\right](y)
+2\beta_{0}\,p_{ba}^{0}(y)
\right]
-\frac{1}{2\epsilon}\,p_{ba}^{1}(y),
\end{eqnarray}
where the $p_{ba}^{(n)}$ are the standard four-dimensional LO and NLO Altarelli-Parisi kernels in the ${\overline{\rm MS}}$ scheme given in Refs.~\cite{Altarelli:1977zs,Curci:1980uw,Furmanski:1980cm,Floratos:1977au,Floratos:1978ny} and are collected in the Appendix~\ref{sec:appendixD} for the gluonic channel.

Eq.~\eqref{eq:fi} can be systematically inverted such that the bare PDF is given by,
\begin{eqnarray}
\label{eq:fxi}
\tilde{f}_a(\xi) &=&\int dx\,dy\;f_b(x,\mu^2)\Gamma^{-1}_{ba}(y,\mu^2)\delta(\xi-xy)
=\left[ f_b \otimes \Gamma^{-1}_{ba} \right] (\xi,\mu^2) 
\end{eqnarray}   
where,
\begin{eqnarray}
\label{eq:ginv}
\Gamma^{-1}_{ba}(y,\mu^2)&=&\delta_{ba}\delta(1-y)
-\left(\frac{\alpha_s(\mu)}{2\pi}\right)\,\Gamma^{1}_{ba}(y)-\left(\frac{\alpha_s(\mu)}{2\pi}\right)^2\,\left[
\Gamma^{2}_{ba}(y) -\sum_c \left[\Gamma^{1}_{bc} \otimes \Gamma^{1}_{ca}\right] (y)\right]\nonumber \\
&&\hspace{6cm}
+{\cal O}(\alpha_{s}^{3}).
\end{eqnarray}
Inserting Eq.~\eqref{eq:ginv} into \eqref{eq:sigunfac} and applying the rescaling $\xi_i \to x_i \xi_i$ we obtain
\begin{eqnarray}
\label{eq:sigrenfac}
{\rm d}\sigma(H_1,H_2)
&=&\sum_{i,j} \int   
\frac{d\xi_1}{\xi_1} \frac{d\xi_2}{\xi_2} f_i(\xi_1,\mu^2) f_j(\xi_2,\mu^2) \,\dsigma_{ij}(\xi_1 H_1,\xi_2 H_2)  
\end{eqnarray}
where   
the infrared-finite mass factorised partonic cross section is
\begin{eqnarray}
\dsigma_{ij}(\xi_1 H_1,\xi_2 H_2)
= 
\int \frac{{\rm d}x_1}{x_1}
\int \frac{{\rm d}x_2}{x_2}
\Gamma^{-1}_{ki}(x_1,\mu^2)
\Gamma^{-1}_{lj}(x_2,\mu^2)
\,{\rm d}\tilde\sigma_{kl}(x_1 \xi_1 H_1,x_2 \xi_2 H_2).
\end{eqnarray}
Expanding in the strong coupling as in Eq.~\eqref{eq:sigpert}, we find the connection between the infrared singular cross sections and the mass factorisation counterterms, 
\begin{eqnarray}
\dsigma_{ij,LO}(\xi_1 H_1,\xi_2 H_2) &=& {\rm d}\tilde\sigma_{ij,LO}(\xi_1 H_1,\xi_2 H_2),\\
\dsigma_{ij,NLO}(\xi_1 H_1,\xi_2 H_2) &=& {\rm d}\tilde\sigma_{ij,NLO}(\xi_1 H_1,\xi_2 H_2)
+ \dsigma_{ij,NLO}^{MF}(\xi_1 H_1,\xi_2 H_2),\\
\dsigma_{ij,NNLO}(\xi_1 H_1,\xi_2 H_2) &=& {\rm d}\tilde\sigma_{ij,NNLO}(\xi_1 H_1,\xi_2 H_2)
+\dsigma_{ij,NNLO}^{MF}(\xi_1 H_1,\xi_2 H_2),
\end{eqnarray}
where
\begin{eqnarray}
\label{eq:MF}
\dsigma_{ij,NLO}^{MF}(\xi_1 H_1,\xi_2 H_2) &=& 
-\int \frac{{\rm d}x_1}{x_1} \Gamma^{1}_{ki}(x_1) {\rm d}\hat\sigma_{kj,LO}(x_1 \xi_1 H_1,\xi_2 H_2)\nonumber \\
&&  -\int \frac{{\rm d}x_2}{x_2} \Gamma^{1}_{lj}(x_2) {\rm d}\hat\sigma_{il,LO}(\xi_1 H_1,x_2\xi_2 H_2), \\
\dsigma_{ij,NNLO}^{MF}(\xi_1 H_1,\xi_2 H_2) &=& 
-\int \frac{{\rm d}x_1}{x_1} \Gamma^{2}_{ki}(x_1) {\rm d}\hat\sigma_{kj,LO}(x_1 \xi_1 H_1,\xi_2 H_2)\nonumber \\ 
&&-\int \frac{{\rm d}x_1}{x_1} \Gamma^{1}_{ki}(x_1) {\rm d}\hat\sigma_{kj,NLO}(x_1 \xi_1 H_1,\xi_2 H_2)\nonumber \\
&&-\int \frac{{\rm d}x_2}{x_2} \Gamma^{2}_{lj}(x_2) {\rm d}\hat\sigma_{il,LO}(\xi_1 H_1,x_2\xi_2 H_2)\nonumber \\
&&-\int \frac{{\rm d}x_2}{x_2} \Gamma^{1}_{lj}(x_2) {\rm d}\hat\sigma_{il,NLO}(\xi_1 H_1,x_2\xi_2 H_2)\nonumber \\
&&-\int \frac{{\rm d}x_1}{x_1}\int \frac{{\rm d}x_2}{x_2} \Gamma^{1}_{ki}(x_1) \Gamma^{1}_{lj}(x_2) {\rm d}\hat\sigma^{LO}_{kl}(x_1 \xi_1 H_1,x_2\xi_2 H_2).\nonumber \\
\end{eqnarray}
Recalling that
\begin{eqnarray}
\label{eq:NLO3jet}
\dsigma_{ij,NLO}&=&\int_{{\rm d}\Phi_{m+1}}\left(\dsigma_{ij,NLO}^{R}-\dsigma_{ij,NLO}^{S}\right)
+\int_{{\rm d}\Phi_m}\left(\int_1\dsigma_{ij,NLO}^{S}+\dsigma_{ij,NLO}^{V}+\dsigma_{ij,NLO}^{MF}\right),\nonumber\\
\end{eqnarray}
we find that the NNLO 
mass factorisation counterterm for the $(m+1)$-parton phase space $\dsigma_{ij,NNLO}^{MF,1}$ in equation (\ref{eq:Tdef}) is given by,
\begin{eqnarray}
\label{eq:MFone}
\dsigma_{ij,NNLO}^{MF,1}(\xi_1 H_1,\xi_2 H_2)&=&\nonumber \\
-\int\frac{{\rm d}x_{1}}{x_{1}} \frac{{\rm d}x_{2}}{x_{2}} 
&&\hspace{-0.5cm}\delta(1-x_{2})\,\Gamma_{ki}^{1}(x_{1})\,
\bigg[\dsigma_{kj,NLO}^{R}
-\dsigma_{kj,NLO}^S\bigg](x_1 \xi_1 H_1,x_2 \xi_2 H_2) \nonumber \\
-\int\frac{{\rm d}x_{1}}{x_{1}} \frac{{\rm d}x_{2}}{x_{2}} 
&&\hspace{-0.5cm}\delta(1-x_{1})\,\Gamma_{lj}^{1}(x_{2})\,
\bigg[\dsigma_{il,NLO}^{R}
-\dsigma_{il,NLO}^S\bigg] (x_1\xi_1 H_1,x_2\xi_2 H_2),\nonumber \\
\end{eqnarray}
while $\dsigma_{ij,NNLO}^{MF,2}$ in equation (\ref{eq:Udef}) contributes to the $m$-parton phase space counterterm,
\begin{eqnarray}
\label{eq:MFtwo}
\dsigma_{ij,NNLO}^{MF,2}(\xi_1 H_1,\xi_2 H_2)&=&\nonumber \\
-\int\frac{{\rm d}x_{1}}{x_{1}} \frac{{\rm d}x_{2}}{x_{2}}&&\hspace{-0.5cm} \bigg \{
\,\delta(1-x_{2})\,\delta_{lj}\,\left(\Gamma_{ki}^{2}(x_{1})-
\sum_{a} \left[\Gamma_{a i}^{1}\otimes \Gamma_{k a}^{1}\right](x_{1})\right)\nonumber \\
&&\hspace{-0.5cm}+
\,\delta(1-x_{1})\,\delta_{ki}\,\left(\Gamma_{lj}^{2}(x_{2})-
\sum_{a} \left[\Gamma_{a j}^{1}\otimes \Gamma_{l a}^{1}\right](x_{2})\right)\nonumber \\
&&\hspace{-0.5cm}-
\Gamma_{ki}^{1}(x_{1})\Gamma_{lj}^{1}(x_2) \bigg \}\,
\dsigma_{kl,LO}(x_1\xi_1 H_1,x_2\xi_2 H_2)
\nonumber\\
-\int\frac{{\rm d}x_{1}}{x_{1}} \frac{{\rm d}x_{2}}{x_{2}} 
&&\hspace{-0.5cm}
\delta(1-x_{2})\,\Gamma_{ki}^{1}(x_{1}) 
\left[\int_1\dsigma_{kj,NLO}^{S}
+\dsigma_{kj,NLO}^V\right](x_1 \xi_1 H_1,x_2 \xi_2 H_2) \nonumber \\
-\int\frac{{\rm d}x_{1}}{x_{1}} \frac{{\rm d}x_{2}}{x_{2}} 
&&\hspace{-0.5cm}\delta(1-x_{1})\,\Gamma_{lj}^{1}(x_{2}) \left[\int_1\dsigma_{il,NLO}^{S}
+\dsigma_{il,NLO}^V\right] (x_1\xi_1 H_1,x_2\xi_2 H_2),\nonumber \\
\end{eqnarray}
and will be discussed elsewhere.

\subsection{Scale dependence of the partonic cross section}
We start from the hadronic cross section that depends on $\mu_F=\mu_R=\mu$ through the strong coupling and the PDF's, 
\begin{equation}
{\rm d}\sigma =\sum_{i,j} \int   
\frac{d\xi_1}{\xi_1} \frac{d\xi_2}{\xi_2} f_i(\xi_1,\mu^2) f_j(\xi_2,\mu^2) \dsigma_{ij}(\alpha_s(\mu),\mu).
\label{eq:Hxsec}
\end{equation}
The scale variation of the coupling constant $\alpha_s(\mu)$ is given by,
\begin{eqnarray}
\mu^2\frac{\partial}{\partial \mu^2} \left(\frac{\alpha_s(\mu)}{2\pi}\right) &=& -\beta(\alpha_s(\mu)),
\end{eqnarray}
where,
\begin{eqnarray}
\beta(\alpha_s(\mu)) &=&  
\beta_0 \left(\frac{\alpha_s(\mu)}{2\pi}\right)^2
+\beta_1 \left(\frac{\alpha_s(\mu)}{2\pi}\right)^3 + {\cal O}(\alpha_s^4),
\end{eqnarray}
with $\beta_0, \beta_1$ given in (\ref{eq:bzero}), (\ref{eq:bone}).
Similarly, the scale variation of the parton distribution function $f_{i}(x,\mu^2)$ is determined by the DGLAP evolution equation,
\begin{eqnarray}
\mu^2\frac{\partial}{\partial \mu^2} f_i(x,\mu^2) &=& \left(\frac{\alpha_s(\mu)}{2\pi}\right)\left[ p_{ij} \otimes f_{j}(\mu^2) \right] (x),
\end{eqnarray}
where, 
\begin{eqnarray}
p_{ij}(x) &=& p^0_{ij}(x) + \left(\frac{\alpha_s(\mu)}{2\pi}\right) p^1_{ij}(x) + {\cal O} (\alpha_s^2).
\end{eqnarray}

Demanding the independence of the physical cross section in~\eqref{eq:Hxsec} on the unphysical scale $\mu^2$,
\begin{equation}
\mu^2\frac{\partial}{\partial \mu^2} {\rm d}\sigma=0,
\end{equation}
we obtain the following scale variation equation for the partonic cross section,
\begin{eqnarray}
\label{eq:scaledep}
\mu^2\frac{\partial}{\partial \mu^2}{\rm d}\hat{\sigma}_{ij}(\alpha_s(\mu),\mu)=-\left(\frac{\alpha_s(\mu)}{2\pi}\right)\Big[
p_{ik}\otimes{\rm d}\hat{\sigma}_{kj}(\alpha_s(\mu),\mu)+
{\rm d}\hat{\sigma}_{ik}(\alpha_s(\mu),\mu)\otimes p_{kj}
\Big].\nonumber \\
\end{eqnarray}

Bearing in mind that the leading order $m$-jet cross section depends on $\alpha_s^m$, and solving Eq.~\eqref{eq:scaledep} order by order in $\alpha_s$,
we see that the partonic cross section at scale $\mu_2$ is related to that at scale $\mu_1$ via,

\begin{eqnarray}
\dsigma_{ij} (\alpha_s(\mu_2),\mu_2) &=& \dsigma_{ij,LO} (\alpha_s(\mu_2),\mu_1)\nonumber \\
&&
+ \left(\frac{\alpha_s(\mu_2)}{2\pi}\right)
\Bigg(
\dsigma_{ij,NLO} (\alpha_s(\mu_2),\mu_1)
+ m\; \beta_0\; L \dsigma_{ij,LO}(\alpha_s(\mu_2),\mu_1) \nonumber \\
&&  \hspace{1cm}
- L\; p^0_{ki}\otimes \dsigma_{kj,LO}(\alpha_s(\mu_2),\mu_1)
- L\; p^0_{kj}\otimes  \dsigma_{ik,LO} (\alpha_s(\mu_2),\mu_1)
\Bigg)\nonumber \\
&&+\left(\frac{\alpha_s(\mu_2)}{2\pi}\right)^2
\Bigg(\dsigma_{ij,NNLO} (\alpha_s(\mu_2),\mu_1)\nonumber \\
&& \hspace{1cm}
+(m+1)\; \beta_0\; L\; \dsigma_{ij,NLO}(\alpha_s(\mu_2),\mu_1)\nonumber \\
&&
\hspace{1cm}
+\left(m\; \beta_1\; L + \frac{m(m+1)}{2}\; \beta_0^2\;L^2 \right)
\dsigma_{ij,LO}(\alpha_s(\mu_2),\mu_1)\nonumber \\
&&\hspace{1cm}
- L\; p^0_{ki}\otimes \dsigma_{kj,NLO}(\alpha_s(\mu_2),\mu_1)
- L\; p^0_{kj}\otimes  \dsigma_{ik,NLO}(\alpha_s(\mu_2),\mu_1) \nonumber \\
&&\hspace{1cm}
- \left( L\; p^1_{ki}+\frac{(m+1)}{2}\; \beta_0 \;L^2 \; p^0_{ki}\right) \otimes \dsigma_{kj,LO}(\alpha_s(\mu_2),\mu_1)\nonumber \\
&&\hspace{1cm}
- \left( L\; p^1_{kj}+\frac{(m+1)}{2}\; \beta_0\; L^2 \; p^0_{kj}\right) \otimes  \dsigma_{ik,LO} (\alpha_s(\mu_2),\mu_1)
\Bigg),\nonumber \\
\end{eqnarray}
where $L = \ln(\mu_2^2/\mu_1^2)$.

\section{Real-Virtual corrections for gluon scattering at NNLO}
\label{sec:RVgluon}
In this section we discuss the amplitudes that enter in the implementation of the mixed real-virtual correction. We focus on the pure gluon
channel and describe the colour decomposition of the gluonic matrix elements at tree and loop-level. The remaining contributions that we use 
for subtraction, namely the three parton tree-level unintegrated and integrated antennae as
well as one-loop three parton antennae have been derived in previous publications. For convenience and completeness we collect
them in appendix \ref{sec:appANT}.

\subsection{Gluonic amplitudes}
\label{subsec:gluon}

The leading colour contribution to the 
$m$-gluon $n$-loop amplitude can be written as~\cite{Berends:1987cv,Kosower:1987ic,Mangano:1987xk,Mangano:1990by,Bern:1994zx},
\begin{eqnarray}
\lefteqn{{\bf A}^n_m(\{p_i,\lambda_i,a_i\})}\nonumber \\
&=&2^{m/2} g^{m-2} 
\left(\frac{g^2 N C(\epsilon)}{2}\right)^n \sum_{\sigma\in S_m/Z_m}\textrm{Tr}(T^{a_{\sigma(1)}}\cdots T^{a_{\sigma(m)}})
{\cal A}^n_m(\sigma(1),\cdots,\sigma(m)).\nonumber \\
\label{eq:cdecomploop}
\end{eqnarray}
where the  permutation sum, $S_m/Z_m$ is the group of non-cyclic permutations of $m$ symbols.  
We systematically extract a loop factor of $C(\epsilon)/2$ per loop with
$C(\epsilon)$ defined in \eqref{eq:Cdef}.
The helicity information is not relevant to the discussion of the subtraction terms and from now on, we will systematically suppress the helicity labels. The $T^a$ are fundamental representation $SU(N)$ colour matrices, normalised such that ${\rm Tr}(T^aT^b)= \delta^{ab}/2$.
${\cal A}^n_m(1,\cdots,n)$ denotes the $n$-loop colour ordered \textit{partial amplitude}.
It is gauge invariant, as well as being invariant under cyclic permutations of the gluons. For simplicity, we will frequently denote the momentum $p_j$ of gluon $j$ by $j$.

At leading colour,  the tree-level $(m+2)$-gluon contribution to the $M$-jet cross section is given by,
\begin{eqnarray}
\label{eq:treemaster}
\dsigma&=&{\cal N}_{m+2}^0 
{\rm d}\Phi_{m}(p_3,\dots,p_{m+2};p_1,p_2)\frac{1}{m!}\nonumber \\
&& \times
\sum_{\sigma\in S_{m+2
}/Z_{m+2}}A_{m+2}^0(\sigma(1),\dots,\sigma(m+2))\JET_M^{(m)}(p_3,...,p_{m+2}) 
\end{eqnarray}
where the sum runs over the group of non-cyclic permutations of $n$ symbols and where   
\begin{equation}
\label{eq:A0m}
{A}^0_m(\sigma(1),\ldots,\sigma(m)) = \sum_{\rm helicities}  
{\cal A}_m^{0\dagger}(\sigma(1),\ldots,\sigma(m))
{\cal A}_m^{0}(\sigma(1),\ldots,\sigma(m)) 
+ {\cal O}\left(\frac{1}{N^2}\right).
\end{equation}
For tree-processes involving four- and five-gluons, there are no sub-leading colour contributions. 
The normalisation factor ${\cal N}_{m+2}^0$ includes the average over initial spins and colours and is given by,
\begin{eqnarray}
{\cal N}_{m+2}^0 &=& {\cal N}_{LO} \times \left(\frac{\alpha_s N}{2\pi}\right)^{m-2} \frac{\bar{C}(\epsilon)^{m-2}}{C(\epsilon)^{m-2}},
\label{eq:NNLORR}
\end{eqnarray}
with
\begin{eqnarray}
{\cal N}_{LO} &=& \frac{1}{2s} \times \frac{1}{4(N^2-1)^2}\times \left(g^2 N\right)^{2} (N^2-1),
\end{eqnarray}
and where we have absorbed the factors of $g^2$ using the useful factors $C(\epsilon)$ \eqref{eq:Cdef} and $\bar{C}(\epsilon)$ \eqref{eq:Cbar},
$$
g^2 N C(\epsilon) = \left(\frac{\alpha_s N}{2\pi}\right) \bar{C}(\epsilon).
$$

The leading colour six-gluon real-real contribution to the NNLO dijet cross section is obtained by setting $m=4$ and $M=2$ in~\eqref{eq:NNLORR} such
that in this case $\NNNLORR$ appearing in Section~\ref{sec:RVsub} in~\eqref{eq:RRcross} is given by $\NNNLORR={\cal N}_{6}^{0}$.

For convenience, we introduce the additional notation for the one-loop ``squared" matrix elements
\begin{equation}
\label{eq:A1m}
{A}^1_m(\sigma(1),\ldots,\sigma(m)) = \sum_{\rm helicities} \sum_{i=0,1}
{\cal A}_m^{i\dagger}(\sigma(1),\ldots,\sigma(m))
{\cal A}_m^{1-i}(\sigma(1),\ldots,\sigma(m)) 
+ {\cal O}\left(\frac{1}{N^2}\right)
\end{equation}
so that the one-loop, $(m+2)$-gluon contribution to the $M$-jet cross section is given by,
\begin{eqnarray}
\label{eq:RVmaster}
\dsigma&=&{\cal N}_{m+2}^1 
{\rm d}\Phi_{m}(p_3,\dots,p_{m+2};p_1,p_2)\frac{1}{m!}\nonumber \\
&& \times
\sum_{\sigma\in S_{m+2
}/Z_{m+2}}A_{m+2}^1(\sigma(1),\dots,\sigma(m+2))\JET_M^{(m)}(p_3,...,p_{m+2}).
\end{eqnarray}
As before, the normalisation factor ${\cal N}_{m+2}^1$ includes the average over initial spins and colours and is given by,
\begin{eqnarray}
{\cal N}_{m+2}^1 &=& {\cal N}_{LO} \times \left(\frac{\alpha_s N}{2\pi}\right)^{m-1} \frac{\bar{C}(\epsilon)^{m-1}}{C(\epsilon)^{m-2}}.
\label{eq:NNLORV}
\end{eqnarray}
We will encounter both ${A}_5^1$ and ${A}_4^1$ when computing the real-virtual corrections relevant for the NNLO dijet cross section.

\subsection{The five-gluon real-virtual contribution ${\rm d}\hat\sigma_{NNLO}^{RV}$}
\label{subsec:five}
The leading colour five-gluon real-virtual contribution to the NNLO dijet cross section is obtained by setting $m=3$ and $M=2$ in (${\ref{eq:NNLORV}}$) such that
in this case $\NNNLORV={\cal N}_{5}^{1}$ and we have,

\begin{eqnarray}
\dsigma_{NNLO}^{RV}&=&\NNNLORV 
{\rm d}\Phi_3(p_3,\dots,p_5;p_1,p_2)\frac{1}{3!}
\sum_{\sigma\in S_3/Z_3}A_5^1(\sigma(1),\dots,\sigma(5))\JET_2^{(3)}(p_3,...,p_5)\nonumber\\
&=&{\cal N}_{LO} \left(\frac{\alpha_sN}{2\pi}\right)^2 \frac{\bar{C}(\epsilon)^2}{C(\epsilon)}
{\rm d}\Phi_3(p_3,\dots,p_5;p_1,p_2)\frac{2}{3!}\nonumber\\
&\times&\sum_{P(i,j,k)\in(3,4,5)} \,
\Bigg(
{ A}^{1}_{5}(\hat{1}_g,\hat{2}_g,i_g,j_g,k_g)+{ A}^{1}_{5}(\hat{1}_g,i_g,\hat{2}_g,j_g,k_g)\, \Bigg) J_{2}^{(3)}(p_i,p_j,p_k)\nonumber\\
\label{eq:RVNLO}
\end{eqnarray}
where the sums runs over the 3! permutations of the final state gluons.
 
Therefore, depending on the position of the initial state gluons in the colour ordered matrix elements, we have two different topologies. These are labelled by the the colour ordering of initial and final state gluons.   We denote the configurations where the two initial state gluons are colour-connected (i.e. adjacent) as IIFFF and those where the colour ordering allows one final state gluon to be sandwiched between the initial state gluons is denoted by IFIFF, 
\begin{equation}
{\rm d}\hat\sigma_{NNLO}^{RV}
=
{\rm d}\hat\sigma_{NNLO}^{RV,IIFFF}+
{\rm d}\hat\sigma_{NNLO}^{RV,IFIFF},
\end{equation}
where,
\begin{eqnarray}
{\rm d}\hat\sigma_{NNLO}^{RV,IIFFF}&=& {\cal N}_{LO} \left(\frac{\alpha_sN}{2\pi}\right)^2 \frac{\bar{C}(\epsilon)^2}{C(\epsilon)}
{\rm d}\Phi_{3}(p_{3},\ldots,p_{5};p_1,p_2)
 \, \nonumber\\
&\times &
 \frac{2}{3!}\,
\sum_{P(i,j,k)\in(3,4,5)}
A_{5}^{1} (\hat{1}_g,\hat{2}_g,i_g,j_g,k_g)\, \JET_{2}^{(3)}(p_{3},\ldots,p_{5}),
\label{eq:RVIIFFF}\\
{\rm d}\hat\sigma_{NNLO}^{RV,IFIFF}&=& {\cal N}_{LO} \left(\frac{\alpha_sN}{2\pi}\right)^2 \frac{\bar{C}(\epsilon)^2}{C(\epsilon)}
{\rm d}\Phi_{3}(p_{3},\ldots,p_{5};p_1,p_2)
 \, \nonumber\\
&\times &
 \frac{2}{3!}\,
\sum_{P(i,j,k)\in(3,4,5)}
A_{5}^1 (\hat{1}_g,i_g,\hat{2}_g,j_g,k_g)\, \JET_{2}^{(3)}(p_{3},\ldots,p_{5}),
\label{eq:RVIFIFF}
\end{eqnarray}

The one-loop helicity amplitudes for $gg\to ggg$ have been available for some time~\cite{Bern:1993mq}. 
We have cross checked our implementation of the one-loop helicity amplitudes of Ref.~\cite{Bern:1993mq} against the numerical package {\tt NGluon}~\cite{Badger:2010nx}.

We note that the renormalised singularity structure of the contribution in (\ref{eq:RVNLO}) can be easily written in terms of the tree-level
squared matrix elements multiplied by combinations of the colour ordered infrared singularity operator~\cite{Catani:1998bh} 
\begin{eqnarray}
{\bom I}_{gg}^{(1)}(s_{gg})=-\frac{e^{\epsilon\gamma}}{2\Gamma(1-\epsilon)}\left[\frac{1}{\epsilon^2}+
\frac{11}{6\epsilon}\right]\Re\left(-\frac{s_{gg}}{\mu^2}\right)^{-\epsilon}.
\label{eq:Ione}
\end{eqnarray}
Therefore the real-virtual correction contains only ${\bom I}^{(1)}$ type of operators and in the gluonic approximation ${\bom I}_{gg}^{(1)}$
is the only operator that appears. The singular part of the renormalised colour ordered gluonic amplitude takes the form,
\begin{eqnarray}
&&{ A}^{1}_{5}(\hat{1}_g,\hat{2}_g,i_g,j_g,k_g)= \nonumber\\
&& 2\left({\bf I}_{gg}^{(1)}(\e,s_{12})+{\bf I}_{gg}^{(1)}(\e,s_{2i})+{\bf I}_{gg}^{(1)}(\e,s_{ij})
+{\bf I}_{gg}^{(1)}(\e,s_{jk})+{\bf I}_{gg}^{(1)}(\e,s_{k1})\right)A_5^0(\hat{1}_g,\hat{2}_g,i_g,j_g,k_g)+{\cal O}(\eps^0),\nonumber\\
&&{ A}^{1}_{5}(\hat{1}_g,i_g,\hat{2}_g,j_g,k_g)=\nonumber \\
&& 2\left({\bf I}_{gg}^{(1)}(\e,s_{1i})+{\bf I}_{gg}^{(1)}(\e,s_{i2})+{\bf I}_{gg}^{(1)}(\e,s_{2j})
+{\bf I}_{gg}^{(1)}(\e,s_{jk})+{\bf I}_{gg}^{(1)}(\e,s_{k1})\right)A_5^0(\hat{1}_g,i_g,\hat{2}_g,j_g,k_g)+{\cal O}(\eps^0).\nonumber\\
\label{eq:RVepoles}
\end{eqnarray}

In Section~\ref{sec:DsigmaVSNNLO} we will explicitly write down the counterterm that regularises 
the infrared divergences of the real-virtual correction for topology 
(\ref{eq:RVIIFFF}) and (\ref{eq:RVIFIFF}) separately.

\section{Construction of the NNLO real-virtual subtraction term}
\label{sec:DsigmaVSNNLO}

As stated in the introduction, the aim of this paper is to construct the
subtraction term for the real-virtual contribution such that the $(m+1)$-parton
contribution to the $m$-jet rate is free from explicit $\e$-poles over the whole of phase space and the
subtracted integrand is well behaved in the single unresolved regions of phase
space. As discussed earlier, this is achieved with the help of the antenna functions and,  as will be explained here and in the subsequent section,
the limit $\e\to0$ can be safely taken and the finite remainders evaluated
numerically in four dimensions.

We start this section by recovering the general formula for the real
virtual channel which has to be integrated over the $(m+1)$-parton
final state phase space numerically. It reads,
\begin{eqnarray}
\int_{{\rm d}\Phi_{m+1}}\left(
\dsigma_{NNLO}^{RV}-\dsigma_{NNLO}^{VS}+\int_{1}\dsigma_{NNLO}^{S,1}
+\dsigma_{NNLO}^{MF,1}\right),
\label{eq:RVimp}
\end{eqnarray}
where $\int_{1}\dsigma_{NNLO}^{S,1}$ contains the once integrated part of the integrated double real subtraction term discussed in Section~\ref{subsec:RVS1},
$\dsigma_{NNLO}^{VS}$ is the the real-virtual subtraction term discussed in Section~\ref{subsec:RVVS1} and the mass factorisation term $\dsigma_{NNLO}^{MF,1}$ is given in Eq.~\eqref{eq:MFone}.

The remaining contribution from the double real subtraction term denoted by $\int_2\dsigma_{NNLO}^{S,2}$ must
be integrated over two unresolved particles and contributes directly to the $m$-jet
final states. It must therefore
be added to the integrated real-virtual subtraction term  $\int_1
\dsigma_{NNLO}^{VS}$, the two-loop matrix elements
$\dsigma_{NNLO}^{VV}$ together with the mass-factorisation counterterm denoted by $\dsigma_{NNLO}^{MF,2}$. These all contribute to the $m$-parton final state
and will be treated elsewhere. At the end of this section, however, we present the contribution from ${\dsigma_{NNLO}^{VS}}$ which must be added back
in integrated form to the $m$-parton final state.

As discussed in Section~\ref{sec:RVsub}, the phase space for the $(m+1)$-parton phase space can be written as an integral over the longitudinal momentum fractions $x_{1},x_{2}$ in the form
$$
\dPSxx{m+1}{m+3}
$$
to account for initial state radiation.

In the next subsections we present separately the subtraction terms
required for gluon-gluon scattering divided for the IIFFF and IFIFF topologies. Subtraction terms related to each topology
are denoted with an superscript  $X_{5}$ and  $Y_{5}$ respectively such that,\footnote{This notation follows naturally from that used for the double real subtraction term ~\cite{Glover:2010im} where the IIFFFF, IFIFFF and IFFIFF topologies were labelled $X_6$, $Y_6$ and $Z_6$ respectively, and anticipates a similar labelling in the double virtual contribution where the IIFF and IFIF topologies will be labelled $X_4$ and $Y_4$.  We note that $X_5$ receives contributions from singly integrated contributions from $X_6$ and $Y_6$, while $Y_5$ receives contributions from $Y_6$ and $Z_6$.} 
\begin{equation}
\dsigma_{NNLO}^{T}=\dsigma_{NNLO}^{T,X_{5}}+\dsigma_{NNLO}^{T,Y_{5}}.
\end{equation}

When presenting the subtraction term we group the terms which are free of explicit $\e$-poles in square brackets. For conciseness we suppress the explicit
$x_{1}, x_{2}$ dependence of the integrated antennae appearing in the formulae below.

\subsection{IIFFF topology}
The one-loop single unresolved subtraction term for the IIFFF topology is:
\begin{eqnarray}
&&\dsigma_{NNLO}^{T,X_{5}}={\cal N}_{LO} \left(\frac{\alpha_sN}{2\pi}\right)^2 \frac{\bar{C}(\epsilon)^2}{C(\epsilon)}
\,\frac{2}{3!}\sum_{P(i,j,k)\in(3,4,5)}\PSh\Bigg\{\nonumber\\
&&-\Bigg(\frac{1}{2}{\cal F}_{3}^{0}(s_{\bar{2}i})+\frac{1}{3}{\cal F}_{3}^{0}(s_{ij})
+\frac{1}{3}{\cal F}_{3}^{0}(s_{jk})+\frac{1}{2}{\cal F}_{3}^{0}(s_{k\bar{1}})
+{\cal F}_{3}^{0}(s_{\bar{1}\bar{2}})-\Gamma_{gg}^{1}(x_{1})\delta(1-x_{2})\nonumber\\
&&\hspace{1.0cm}-\Gamma_{gg}^{1}(x_{2})\delta(1-x_{1})\Bigg) 
A_5^0(\hat{\bar{1}}_g,\hat{\bar{2}}_g,i_g,j_g,k_g)\,\JET_{2}^{(3)}(p_i,p_j,p_k)\nonumber\\
&&+f_{3}^{0}(\hat{\bar{2}}_g,i_g,j_g)\Bigg[\delta(1-x_{1})\delta(1-x_{2}) A_{4}^{1}(\hat{\bar{1}}_g,\hat{\bar{\bar{2}}}_g,(\widetilde{ij})_g,k_g)
+\left(\calF(s_{\bar{1}\bar{\bar{2}}})+\frac{1}{2}\calF(s_{\bar{\bar{2}}\wt{(ij)}})\nonumber\right.\\
&&\left.\hspace{1.0cm}+\frac{1}{3}\calF(s_{\wt{(ij)}k})+\frac{1}{2}\calF(s_{\bar{1}k})-\Gamma_{gg}^{1}(x_{1})\delta(1-x_{2})
-\Gamma_{gg}^{1}(x_{2})\delta(1-x_{1})\right)\nonumber\\
&&\hspace{1.0cm}\times A_{4}^{0}(\hat{\bar{1}}_g,\hat{\bar{\bar{2}}}_g,(\widetilde{ij})_g,k_g)\Bigg]\JET_{2}^{(2)}(\tilde{p}_{ij},p_k)\nonumber\\
&&+f_{3}^{0}(i_g,j_g,k_g)\Bigg[\delta(1-x_{1})\delta(1-x_{2})A_{4}^{1}(\hat{\bar{1}}_g,\hat{\bar{2}}_g,(\widetilde{ij})_g,(\widetilde{jk})_g)
+\left(\calF(s_{\bar{1}\bar{2}})+\frac{1}{2}\calF(s_{\bar{2}\wt{(ij)}})\right.\nonumber\\
&&\left.\hspace{1.0cm}+\frac{1}{3}\calF(s_{\wt{(ij)}\wt{(jk)}})+\frac{1}{2}\calF(s_{\wt{(jk)}\bar{1}})-\Gamma_{gg}^{1}(x_{1})\delta(1-x_{2})
-\Gamma_{gg}^{1}(x_{2})\delta(1-x_{1})\right)\nonumber\\
&&\hspace{1.0cm}\times A_{4}^{0}(\hat{\bar{1}}_g,\hat{\bar{2}}_g,(\widetilde{ij})_g,(\widetilde{jk})_g)\Bigg]
\,\JET_{2}^{(2)}(\tilde{p}_{ij},\tilde{p}_{jk})\nonumber\\
&&+f_{3}^{0}(j_g,k_g,\hat{\bar{1}}_g)\Bigg[\delta(1-x_{1})\delta(1-x_{2})A_{4}^{1}(\hat{\bar{\bar{1}}}_g,\hat{\bar{2}}_g,i_g,(\widetilde{kj})_g)
+\left(\calF(s_{\bar{\bar{1}}\bar{2}})+\frac{1}{2}\calF(s_{\bar{2}i})\right.\nonumber\\
&&\left.\hspace{1.0cm}+\frac{1}{3}\calF(s_{i\wt{(kj)}})+\frac{1}{2}\calF(s_{\bar{\bar{1}}\wt{(kj)}})-\Gamma_{gg}^{1}(x_{1})\delta(1-x_{2})
-\Gamma_{gg}^{1}(x_{2})\delta(1-x_{1})\right)\nonumber\\
&&\hspace{1.0cm}\times A_{4}^{0}(\hat{\bar{\bar{1}}}_g,\hat{\bar{2}}_g,i_g,(\widetilde{kj})_g)\Bigg]\JET_{2}^{(2)}(p_i,\tilde{p}_{kj})\nonumber\\
&&+\Bigg[f_{3}^{1}(\hat{\bar{2}}_g,i_g,j_g)\delta(1-x_{1})\delta(1-x_{2})+\left(\frac{1}{2}\calF(s_{\bar{2}i})+\frac{1}{2}\calF(s_{\bar{2}j})+\frac{1}{3}\calF(s_{ij})
-\calF(s_{\bar{\bar{2}}\wt{(ij)}})\right)\nonumber\\
&&\hspace{1.0cm}\times
f_{3}^{0}(\hat{\bar{2}}_g,i_g,j_g)\Bigg]\, A_{4}^{0}(\hat{\bar{1}}_g,\hat{\bar{\bar{2}}}_g,(\widetilde{ij})_g,k_g)\,\JET_{2}^{(2)}(\tilde{p}_{ij},p_k)\nonumber\\
&&+\Bigg[f_{3}^{1}(i_g,j_g,k_g)\delta(1-x_{1})\delta(1-x_{2})+\left(\frac{1}{3}\calF(s_{ij})+\frac{1}{3}\calF(s_{ik})+\frac{1}{3}\calF(s_{jk})-\frac{2}{3}\calF(s_{\wt{(ij)}\wt{(jk)}})\right)
\nonumber\\
&&\hspace{1.0cm}\times f_{3}^{0}(i_g,j_g,k_g)\Bigg]\, A_{4}^{0}(\hat{\bar{1}}_g,\hat{\bar{2}}_g,(\widetilde{ij})_g,(\widetilde{jk})_g)
\,\JET_{2}^{(2)}(\tilde{p}_{ij},\tilde{p}_{jk})\nonumber\\
&&+\Bigg[f_{3}^{1}(j_g,k_g,\hat{\bar{1}}_g)\delta(1-x_{1})\delta(1-x_{2})+\left(\frac{1}{3}\calF(s_{jk})+\frac{1}{2}\calF(s_{\bar{1}j})+\frac{1}{2}\calF(s_{\bar{1}k})-\calF(s_{\bar{\bar{1}}\wt{(kj)}})\right)
\nonumber\\
&&\hspace{1.0cm}\times f_{3}^{0}(j_g,k_g,\hat{\bar{1}}_g)\Bigg]\, A_{4}^{0}(\hat{\bar{\bar{1}}}_g,\hat{\bar{2}}_g,i_g,(\widetilde{kj})_g)
\,\JET_{2}^{(2)}(p_i,\tilde{p}_{kj})\nonumber\\
&&+\frac{11}{6}\log\left(\frac{\mu^2}{|s_{\bar{2}ij}|}\right)f_{3}^{0}(\hat{\bar{2}}_g,i_g,j_g)\delta(1-x_{1})\delta(1-x_{2})\, A_{4}^{0}(\hat{\bar{1}}_g,\hat{\bar{\bar{2}}}_g,(\widetilde{ij})_g,k_g)
\,\JET_{2}^{(2)}(\tilde{p}_{ij},p_k)\nonumber\\
&&+\frac{11}{6}\log\left(\frac{\mu^2}{s_{ijk}}\right)f_{3}^{0}(i_g,j_g,k_g)\delta(1-x_{1})\delta(1-x_{2})\, A_{4}^{0}(\hat{\bar{1}}_g,\hat{\bar{2}}_g,(\widetilde{ij})_g,(\widetilde{jk})_g)
\,\JET_{2}^{(2)}(\tilde{p}_{ij},\tilde{p}_{jk})\nonumber\\
&&+\frac{11}{6}\log\left(\frac{\mu^2}{|s_{\bar{1}jk}|}\right)f_{3}^{0}(j_g,k_g,\hat{\bar{1}}_g)\delta(1-x_{1})\delta(1-x_{2})\, A_{4}^{0}(\hat{\bar{\bar{1}}}_g,\hat{\bar{2}}_g,i_g,(\widetilde{kj})_g)
\,\JET_{2}^{(2)}(p_i,\tilde{p}_{kj})\nonumber\\
&&+\frac{1}{2}\Bigg[-\frac{1}{2}\calF(s_{\bar{2}\wt{(ij)}})+\frac{1}{2}\calF(s_{\bar{2}i})-\frac{1}{2}\calF(s_{\bar{1}\wt{(jk)}})+\frac{1}{2}\calF(s_{\bar{1}k})
+\frac{1}{3}\calF(s_{\wt{(ij)}\wt{(jk)}})-\frac{1}{3}\calF(s_{ik})
\nonumber \\
&&\hspace{1cm}+\delta(1-x_{1})\delta(1-x_{2})\Big(\calS(s_{\bar{2}\wt{(ij)}},s_{ik},x_{\bar{2}\wt{(ij)},ik})-\calS(s_{\bar{2}i},s_{ik},x_{\bar{2}i,ik})
+\calS(s_{\bar{1}\wt{(jk)}},s_{ik},x_{\bar{1}\wt{(jk)},ik})\nonumber\\
&&\hspace{1cm}-\calS(s_{\bar{1}k},s_{ik},x_{\bar{1}k,ik})-\calS(s_{\wt{(ij)}\wt{(jk)}},s_{ik},x_{\wt{(ij)}\wt{(jk)},ik})+\calS(s_{ik},s_{ik},1)\Big)\Bigg]\nonumber\\
&&\hspace{1cm}\times f_{3}^{0}(i_g,j_g,k_g)\,
A_{4}^{0}(\hat{\bar{1}}_g,\hat{\bar{2}}_g,(\widetilde{ij})_g,(\widetilde{jk})_g)\,\JET_{2}^{(2)}(\tilde{p}_{ij},\tilde{p}_{jk})\nonumber\\
&&+\frac{1}{2}\Bigg[-\frac{1}{3}\calF(s_{i\wt{(kj)}})+\frac{1}{3}\calF(s_{ij})+\frac{1}{2}\calF(s_{\bar{\bar{1}}\wt{(kj)}})-\frac{1}{2}\calF(s_{\bar{1}j})
-\calF(s_{\bar{\bar{1}}\bar{2}})+\calF(s_{\bar{1}\bar{2}})\nonumber\\
&&\hspace{0.5cm}+\delta(1-x_{1})\delta(1-x_{2})\Big(\calS(s_{i\wt{(kj)}},s_{ij},x_{i\wt{(kj)},ij})-\calS(s_{ij},s_{ij},1)-\calS(s_{\bar{\bar{1}}\wt{(kj)}},s_{ij},x_{\bar{\bar{1}}\wt{(kj)},ij})
\nonumber\\
&&\hspace{0.5cm}+\calS(s_{\bar{1}j},s_{ij},x_{\bar{1}j,ij})\Big)\Bigg]f_{3}^{0}(\hat{\bar{1}}_g,k_g,j_g)
\,A_{4}^{0}(\hat{\bar{\bar{1}}}_g,\hat{\bar{2}}_g,i_g,(\widetilde{kj})_g)\,\JET_{2}^{(2)}(p_i,\tilde{p}_{kj})\nonumber\\
&&+\frac{1}{2}\Bigg[-\frac{1}{3}\calF(s_{k\wt{(ij)}})+\frac{1}{3}\calF(s_{kj})+\frac{1}{2}\calF(s_{\bar{\bar{2}}\wt{(ij)}})-\frac{1}{2}\calF(s_{\bar{2}j})
-\calF(s_{\bar{1}\bar{\bar{2}}})+\calF(s_{\bar{1}\bar{2}})\nonumber\\
&&\hspace{0.5cm}+\delta(1-x_{1})\delta(1-x_{2})\Big(\calS(s_{k\wt{(ij)}},s_{kj},x_{k\wt{(ij)},kj})-\calS(s_{kj},s_{kj},1)-\calS(s_{\bar{\bar{2}}\wt{(ij)}},s_{kj},x_{\bar{\bar{2}}\wt{(ij)},kj})
\nonumber\\
&&\hspace{0.5cm}+\calS(s_{\bar{2}j},s_{kj},x_{\bar{2}j,kj})\Big)\Bigg] f_{3}^{0}(\hat{\bar{2}}_g,i_g,j_g)
\, A_{4}^{0}(\hat{\bar{1}}_g,\hat{\bar{\bar{2}}}_g,(\widetilde{ij})_g,k_g)\,\JET_{2}^{(2)}(\tilde{p}_{ij},p_k)\nonumber\\
&&+\frac{1}{2}\Bigg[-\calF(s_{\bar{\bar{1}}\bar{\bar{2}}})+\calF(s_{\bar{1}\bar{2}})-\frac{1}{2}\calF(s_{\bar{2}j})+\frac{1}{2}\calF(s_{\bar{\bar{2}}\tilde{j}})
-\frac{1}{2}\calF(s_{\bar{1}k})+\frac{1}{2}\calF(s_{\bar{\bar{1}}\tilde{k}})\nonumber\\
&&\hspace{0.5cm}+\delta(1-x_{1})\delta(1-x_{2})\Big(
 \calS(s_{\bar{\bar{1}}\bar{\bar{2}}},s_{\tilde{j}\tilde{k}},x_{{\bar{\bar{1}}}{\bar{\bar{2}}},\tilde{j}\tilde{k}})
-\calS(s_{\bar{1}\bar{2}},s_{jk},x_{\bar{1}\bar{2},jk})
+\calS(s_{\bar{2}j},s_{jk},x_{\bar{2}j,jk})\nonumber\\
&&\hspace{0.5cm}
-{\calS}(s_{\bar{\bar{2}}\t{j}},s_{\tilde{j}\tilde{k}},x_{{\bar{\bar{2}}}j,\tilde{j}\tilde{k}})
+\calS(s_{\bar{1}k},s_{jk},x_{\bar{1}k,jk})
-{\calS}(s_{\bar{\bar{1}}\t{k}},s_{\tilde{j}\tilde{k}},x_{{\bar{\bar{1}}}k,\tilde{j}\tilde{k}})\Big)\Bigg]\nonumber\\
&&\hspace{0.5cm}\times F_{3}^{0}(\hat{\bar{1}}_g,i_g,\hat{\bar{2}}_g)\, 
A_{4}^{0}(\hat{\bar{\bar{1}}}_g,\hat{\bar{\bar{2}}}_g,\tilde{j}_g,\tilde{k}_g)\,\JET_{2}^{(2)}(\tilde{p}_j,\tilde{p}_{k})\Bigg\}.\nonumber\\
\label{eq:RVsofttop1}
\end{eqnarray}

\subsection{IFIFF topology}
The one-loop single unresolved subtraction term for the IFIFF topology is:
\begin{eqnarray}
&&\dsigma_{NNLO}^{T,Y_{5}}={\cal N}_{LO} \left(\frac{\alpha_sN}{2\pi}\right)^2 \frac{\bar{C}(\epsilon)^2}{C(\epsilon)}
\,\frac{2}{3!}\sum_{P(i,j,k)\in(3,4,5)}\PSh\Bigg\{\nonumber\\
&&-\Bigg(\frac{1}{2}{\cal F}_{3}^{0}(s_{\bar{1}i})+\frac{1}{2}{\cal F}_{3}^{0}(s_{i\bar{2}})
+\frac{1}{2}{\cal F}_{3}^{0}(s_{\bar{2}j})+\frac{1}{3}{\cal F}_{3}^{0}(s_{jk})
+\frac{1}{2}{\cal F}_{3}^{0}(s_{k\bar{1}})-\Gamma_{gg}^{1}(x_{1})\delta(1-x_{2})\nonumber\\
&&\hspace{1.0cm}-\Gamma_{gg}^{1}(x_{2})\delta(1-x_{1})\Bigg)
A_5^0(\hat{\bar{1}}_g,i_g,\hat{\bar{2}}_g,j_g,k_g)\,\JET_2^{(3)}(p_i,p_j,p_k)\nonumber\\
&&+F_{3}^{0}(\hat{\bar{1}}_g,i_g,\hat{\bar{2}}_g)\Bigg[\delta(1-x_{1})\delta(1-x_{2}) A_{4}^{1}(\hat{\bar{\bar{1}}}_g,\hat{\bar{\bar{2}}}_g,\tilde{j}_g,\tilde{k}_g)+
\left(\calF(s_{\bar{\bar{1}}\bar{\bar{2}}})+\frac{1}{2}\calF(s_{\bar{\bar{2}}\tilde{j}})
\nonumber\right.\\
&&\left.\hspace{1.0cm}+\frac{1}{3}\calF(s_{\tilde{j}\tilde{k}})+\frac{1}{2}\calF(s_{\bar{\bar{1}}\tilde{k}})-\Gamma_{gg}^{1}(x_{1})\delta(1-x_{2})
-\Gamma_{gg}^{1}(x_{2})\delta(1-x_{1})\right)\nonumber\\
&&\hspace{1.0cm}\times A_{4}^{0}(\hat{\bar{\bar{1}}}_g,\hat{\bar{\bar{2}}}_g,\tilde{j}_g,\tilde{k}_g)
\Bigg]\,\JET_{2}^{(2)}(\tilde{p}_j,\tilde{p}_k)\nonumber\\
&&+f_{3}^{0}(\hat{\bar{2}}_g,j_g,k_g)\Bigg[\delta(1-x_{1})\delta(1-x_{2}) A_{4}^{1}(\hat{\bar{1}}_g,i_g,\hat{\bar{\bar{2}}}_g,(\widetilde{jk})_g)
+\left(\frac{1}{2}\calF(s_{\bar{1}i})+\frac{1}{2}\calF(s_{\bar{\bar{2}}i})\nonumber\right.\\
&&\left.\hspace{1.0cm}+\frac{1}{2}\calF(s_{\bar{\bar{2}}\wt{(jk)}})+\frac{1}{2}\calF(s_{\bar{1}\wt{(jk)}})-\Gamma_{gg}^{1}(x_{1})\delta(1-x_{2})
-\Gamma_{gg}^{1}(x_{2})\delta(1-x_{1})\right)\nonumber\\
&&\hspace{1.0cm}\times A_{4}^{0}(\hat{\bar{1}}_g,i_g,\hat{\bar{\bar{2}}}_g,(\widetilde{jk})_g)\Bigg]
\,\JET_{2}^{(2)}(p_i,\tilde{p}_{jk})\nonumber\\
&&+f_{3}^{0}(j_g,k_g,\hat{\bar{1}}_g)\Bigg[\delta(1-x_{1})\delta(1-x_{2}) A_{4}^{1}(\hat{\bar{\bar{1}}}_g,i_g,\hat{\bar{2}}_g,(\widetilde{kj})_g)
+\left(\frac{1}{2}\calF(s_{\bar{\bar{1}}i})+\frac{1}{2}\calF(s_{\bar{2}i})\nonumber\right.\\
&&\left.\hspace{1.0cm}+\frac{1}{2}\calF(s_{\bar{2}\wt{(kj)}})+\frac{1}{2}\calF(s_{\bar{\bar{1}}\wt{(kj)}})-\Gamma_{gg}^{1}(x_{1})\delta(1-x_{2})
-\Gamma_{gg}^{1}(x_{2})\delta(1-x_{1})\right)\nonumber\\
&&\hspace{1.0cm}\times A_{4}^{0}(\hat{\bar{\bar{1}}}_g,i_g,\hat{\bar{2}}_g,(\widetilde{kj})_g)\Bigg]
\JET_{2}^{(2)}(p_i,\tilde{p}_{kj})\nonumber\\
&&+\Bigg[F_{3}^{1}(\hat{\bar{1}}_g,i_g,\hat{\bar{2}}_g)\delta(1-x_{1})\delta(1-x_{2})+\left(\frac{1}{2}\calF(s_{\bar{1}i})+\calF(s_{\bar{1}\bar{2}})+\frac{1}{2}\calF(s_{\bar{2}i})
-2\calF(s_{\bar{\bar{1}}\bar{\bar{2}}})\right)\nonumber\\
&&\hspace{1.0cm}\times F_{3}^{0}(\hat{\bar{1}}_g,i_g,\hat{\bar{2}}_g)\Bigg]\, 
A_{4}^{0}(\hat{\bar{\bar{1}}}_g,\hat{\bar{\bar{2}}}_g,\tilde{j}_g,\tilde{k}_g)\,\JET_{2}^{(2)}(\tilde{p}_j,\tilde{p}_k)
\nonumber\\
&&+\Bigg[f_{3}^{1}(\hat{\bar{2}}_g,j_g,k_g)\delta(1-x_{1})\delta(1-x_{2})+\left(\frac{1}{2}\calF(s_{\bar{2}j})+\frac{1}{2}\calF(s_{\bar{2}k})+\frac{1}{3}\calF(s_{jk})
-\calF(s_{\bar{\bar{2}}\wt{(jk)}})\right)\nonumber\\
&&\hspace{1.0cm}\times f_{3}^{0}(\hat{\bar{2}}_g,j_g,k_g)\Bigg]\, A_{4}^{0}(\hat{\bar{1}}_g,i_g,\hat{\bar{\bar{2}}}_g,(\widetilde{jk})_g)
\,\JET_{2}^{(2)}(p_i,\tilde{p}_{jk})\nonumber\\
&&+\Bigg[f_{3}^{1}(j_g,k_g,\hat{\bar{1}}_g)\delta(1-x_{1})\delta(1-x_{2})+\left(\frac{1}{3}\calF(s_{jk})+\frac{1}{2}\calF(s_{\bar{1}j})+\frac{1}{2}\calF(s_{\bar{1}k})
-\calF(s_{\bar{\bar{1}}\wt{(kj)}})\right)\nonumber\\
&&\hspace{1.0cm}\times f_{3}^{0}(j_g,k_g,\hat{\bar{1}}_g)\Bigg]\, A_{4}^{0}(\hat{\bar{\bar{1}}}_g,i_g,\hat{\bar{2}}_g,(\widetilde{kj})_g)
\,\JET_{2}^{(2)}(p_i,\tilde{p}_{kj})\nonumber\\
&&+\frac{11}{6}\log\left(\frac{\mu^2}{|s_{\bar{1}\bar{2}i}|}\right)F_{3}^{0}(\hat{\bar{1}}_g,i_g,\hat{\bar{2}}_g)\delta(1-x_{1})\delta(1-x_{2})
\, A_{4}^{0}(\hat{\bar{\bar{1}}}_g,\hat{\bar{\bar{2}}}_g,\tilde{j}_g,\tilde{k}_g)
\,\JET_{2}^{(2)}(\tilde{p}_j,\tilde{p}_k)\nonumber\\
&&+\frac{11}{6}\log\left(\frac{\mu^2}{|s_{\bar{2}jk}|}\right)f_{3}^{0}(\hat{\bar{2}}_g,j_g,k_g)
\delta(1-x_{1})\delta(1-x_{2})\, A_{4}^{0}(\hat{\bar{1}}_g,i_g,\hat{\bar{\bar{2}}}_g,(\widetilde{jk})_g)
\,\JET_{2}^{(2)}(p_i,\tilde{p}_{jk})\nonumber\\
&&+\frac{11}{6}\log\left(\frac{\mu^2}{|s_{\bar{1}jk}|}\right)f_{3}^{0}(j_g,k_g,\hat{\bar{1}}_g)
\delta(1-x_{1})\delta(1-x_{2})\, A_{4}^{0}(\hat{\bar{\bar{1}}}_g,i_g,\hat{\bar{2}}_g,(\widetilde{kj})_g)
\,\JET_{2}^{(2)}(p_i,\tilde{p}_{kj})\nonumber\\
&&+\Bigg[\calF(s_{\bar{\bar{1}}\bar{\bar{2}}})-\calF(s_{\bar{1}\bar{2}})+\frac{1}{2}\calF(s_{\bar{2}j})
-\frac{1}{2}\calF(s_{\bar{\bar{2}}\tilde{j}})+\frac{1}{2}\calF(s_{\bar{1}k})-\frac{1}{2}\calF(s_{\bar{\bar{1}}\tilde{k}})\nonumber\\
&&\hspace{0.5cm}+\delta(1-x_{1})\delta(1-x_{2})\Big(-\calS(s_{\bar{\bar{1}}\bar{\bar{2}}},s_{\tilde{j}\tilde{k}},x_{\bar{\bar{1}}\bar{\bar{2}},\tilde{j}\tilde{k}})
+\calS(s_{\bar{1}\bar{2}},s_{jk},x_{\bar{1}\bar{2},jk})
-\calS(s_{\bar{2}j},s_{jk},x_{\bar{2}j,jk})\nonumber\\
&&\hspace{0.5cm}+\calS(s_{\bar{\bar{2}}\t{j}},s_{\tilde{j}\tilde{k}},x_{\bar{2}\tilde{j},\tilde{j}\tilde{k}})-\calS(s_{\bar{1}k},s_{jk},x_{\bar{1}k,jk})
+\calS(s_{\bar{\bar{1}}\t{k}},s_{\tilde{j}\tilde{k}},x_{\bar{\bar{1}}\tilde{k},\tilde{j}\tilde{k}})\Big)\Bigg]\nonumber\\
&&\hspace{0.5cm}\times
F_{3}^{0}(\hat{\bar{1}}_g,i_g,\hat{\bar{2}}_g)\,A_{4}^{0}(\hat{\bar{\bar{1}}}_g,\hat{\bar{\bar{2}}}_g,\tilde{j}_g,\tilde{k}_g)\,\JET_{2}^{(2)}(\tilde{p}_j,\tilde{p}_k)\nonumber\\
&&+\frac{1}{2}\Bigg[-\frac{1}{2}\calF(s_{\bar{2}\wt{(kj)}})+\frac{1}{2}\calF(s_{\bar{2}j})-\frac{1}{2}\calF(s_{\bar{1}j})
+\frac{1}{2}\calF(s_{\bar{\bar{1}}\wt{(jk)}})-\frac{1}{2}\calF(s_{\bar{\bar{1}}i})+\frac{1}{2}\calF(s_{\bar{1}i})\nonumber\\
&&\hspace{0.5cm}+\delta(1-x_{1})\delta(1-x_{2})\Big(\calS(s_{\bar{2}\wt{(kj)}},s_{ij},x_{\bar{2}\wt{(kj)},ij})-\calS(s_{\bar{2}j},s_{ij},x_{\bar{2}j,ij})
+\calS(s_{\bar{1}j},s_{ij},x_{\bar{1}j,ij})\nonumber\\
&&\hspace{0.5cm}-\calS(s_{\bar{\bar{1}}\wt{(kj)}},s_{ij},x_{\bar{\bar{1}}\wt{(kj)},ij})\Big)\Bigg]
f_{3}^{0}(\hat{\bar{1}}_g,k_g,j_g)\, A_{4}^{0}(\hat{\bar{\bar{1}}}_g,i_g,\hat{\bar{2}}_g,\wt{(kj)}_g)\,\JET_{2}^{(2)}(p_i,\tilde{p}_{kj})\nonumber\\
&&+\frac{1}{2}\Bigg[-\frac{1}{2}\calF(s_{\bar{1}\wt{(jk)}})+\frac{1}{2}\calF(s_{\bar{1}k})
-\frac{1}{2}\calF(s_{\bar{2}k})+\frac{1}{2}\calF(s_{\bar{\bar{2}}\wt{(kj)}})-\frac{1}{2}\calF(s_{\bar{\bar{2}}i})+\frac{1}{2}\calF(s_{\bar{2}i})\nonumber\\
&&\hspace{0.5cm}+\delta(1-x_{1})\delta(1-x_{2})\Big(\calS(s_{\bar{1}\wt{(kj)}},s_{ik},x_{\bar{1}\wt{(kj)},ik})-\calS(s_{\bar{1}k},s_{ik},x_{\bar{1}k,ik})
+\calS(s_{\bar{2}k},s_{ik},x_{\bar{2}k,ik})\nonumber\\
&&\hspace{0.5cm}-\calS(s_{\bar{\bar{2}}\wt{(kj)}},s_{ik},x_{\bar{\bar{2}}\wt{(kj)},ik})\Big)\Bigg]
f_{3}^{0}(\hat{\bar{2}}_g,j_g,k_g)\,A_{4}^{0}(\hat{\bar{1}}_g,i_g,\hat{\bar{\bar{2}}}_g,\wt{(jk)}_g)\,\JET_{2}^{(2)}(p_i,\tilde{p}_{jk})\Bigg\}.
\nonumber \\
\label{eq:RVsofttop2}
\end{eqnarray}

For this topology, the contribution to the hard region ($x_{1},x_{2}\ne1$) turns out to be
identically zero. This is because the only terms that contribute in the hard region are 
integrated initial-initial antennae.  This configuration
(\ref{eq:RVIFIFF}) does not involve colour-connected initial state gluons 
and therefore does not contain any integrated initial-initial antennae.\footnote{Due to the grouping of terms in Eqs.~\eqref{eq:RVsofttop2} 
to demonstrate the cancellation of $\e$-poles, this type of initial-initial antenna does appears.  However, they always cancel pairwise amongst themselves.}

\subsection{Infrared structure}

With the explicit expressions for the integrated  antenna functions given in
Appendix~\ref{sec:appANT},  the integrated large angle soft terms given in
Section~\ref{sec:RVeff} and the pole structure of the one-loop matrix elements,
it is straightforward to check that the explicit $\e$ poles analytically cancel in each and every one of
the groups of terms in square brackets in
Eqs.~\eqref{eq:RVsofttop1}-\eqref{eq:RVsofttop2}.  The only remaining poles lie in the second lines of
Eqs.~\eqref{eq:RVsofttop1} and \eqref{eq:RVsofttop2} which precisely reproduce
those of the real-virtual matrix elements.   The limit $\e\to0$ can therefore
be safely taken. 

In summary, we have shown that, within the antenna subtraction formalism, the
real-virtual corrections to gluon-gluon scattering are locally free of explicit $\e$-poles
providing us with a stringent check on the construction of the necessary
subtraction terms. This is in direct contradiction to the statement made in Ref.~\cite{Bolzoni:2010bt}.

Furthermore, Eqs.~\eqref{eq:RVsofttop1} and \eqref{eq:RVsofttop2} are free of implicit
kinematical singularities corresponding to the single unresolved regions of the
phase space as will be demonstrated in Section~\ref{sec:numerical}.

\subsection{Contributions to the $m$-jet final state}
In this subsection we identify the contributions from the real-virtual channel that we have subtracted in unintegrated form and which therefore must be added back
in integrated form in the double virtual $(m+2)$-parton channel. As expected contributions from the $X_{5}$ topology collapse in integrated form to the $X_{4}$ topology of
the virtual-virtual contribution while contributions from the $Y_{5}$ topology contribute to both $X_{4}$ and $Y_{4}$.
The contributions of ${\rm d}\sigma_{NNLO}^{VS}$, which when integrated over the antenna phase space become proportional to the $X_{4}$ topology are denoted
by $\dsigma_{NNLO}^{VS}|_{X_{4}^{0}}$ and are given by,
\begin{eqnarray}
&&\dsigma_{NNLO}^{VS}|_{X_{4}^{0}}={\cal N}_{LO} \left(\frac{\alpha_sN}{2\pi}\right)^2 \frac{\bar{C}(\epsilon)^2}{C(\epsilon)}
\,\frac{2}{3!}\sum_{P(i,j,k)\in(3,4,5)}\PSs\Bigg\{\nonumber\\
&&\phantom{+}F_{3}^{0}(\hat{1}_g,i_g,\hat{2}_g)A_{4}^{1}(\hat{1}_{g},\hat{\bar{2}}_g,\tilde{j}_g,\tilde{k}_{g})\, J_{2}^{(2)}(\tilde{p}_{j},\tilde{p}_{k})\nonumber\\
&&+f_{3}^{0}(\hat{2}_g,i_g,j_g)A_{4}^{1}(\hat{1}_{g},\hat{\bar{2}}_g,\widetilde{(ij)}_g,k_{g})\,J_{2}^{(2)}(\tilde{p}_{ij},p_{k})\nonumber\\
&&+f_{3}^{0}(i_g,j_g,k_g)A_{4}^{1}(\hat{1}_g,\hat{2}_g,\widetilde{(ij)}_g,\widetilde{(jk)}_g)\,J_{2}^{(2)}(\tilde{p}_{ij},\tilde{p}_{jk})\nonumber\\
&&+f_{3}^{0}(j_{g},k_{g},\hat{1}_g)A_{4}^{1}(\hat{\bar{1}}_g,\hat{2}_g,i_{h},\widetilde{(kj)}_g)\,J_{2}^{(2)}(\tilde{p}_i,\tilde{p}_{kj})\nonumber\\
&&+\left(F_{3}^{1}(\hat{1}_g,i_g,\hat{2}_g)+\frac{11}{6\epsilon}F_{3}^{0}(\hat{1}_g,i_g,\hat{2}_g)
\left((s_{12i})^{-\e}-(\mu^2)^{-\e}\right)\right)A_{4}^{0}(\hat{1}_{g},\hat{\bar{2}}_g,\tilde{j}_g,\tilde{k}_{g})\, J_{2}^{(2)}(\tilde{p}_{j},\tilde{p}_{k})\nonumber\\
&&+\left(f_{3}^{1}(\hat{2}_g,i_g,j_g)+\frac{11}{6\epsilon}f_{3}^{0}(\hat{2}_g,i_g,j_g)
\left((|s_{2ij}|)^{-\e}-(\mu^2)^{-\e}\right)\right)
A_{4}^{0}(\hat{1}_{g},\hat{\bar{2}}_g,\widetilde{(ij)}_g,k_{g})\,J_{2}^{(2)}(\tilde{p}_{ij},p_{k})\nonumber\\
&&+\left(f_{3}^{1}(i_g,j_g,k_g)+\frac{11}{6\epsilon}f_{3}^{0}(i_g,j_g,k_g)
\left((s_{ijk})^{-\e}-(\mu^2)^{-\e}\right)\right)A_{4}^{0}(\hat{1}_g,\hat{2}_g,\widetilde{(ij)}_g,\widetilde{(jk)}_g)\,J_{2}^{(2)}(\tilde{p}_{ij},\tilde{p}_{jk})\nonumber\\
&&+\left(f_{3}^{1}(j_{g},k_{g},\hat{1}_g)+\frac{11}{6\epsilon}f_{3}^{0}(j_g,k_g,\hat{\bar{1}}_g)
\left((|s_{1kj}|)^{-\e}-(\mu^2)^{-\e}\right)\right)
A_{4}^{0}(\hat{\bar{1}}_g,\hat{2}_g,i_{g},\widetilde{(kj)}_g)\,J_{2}^{(2)}(\tilde{p}_i,\tilde{p}_{kj})\Bigg\}\nonumber\\
&&\nonumber\\
&&+{\cal N}_{LO} \left(\frac{\alpha_sN}{2\pi}\right)^2\frac{\bar{C}(\epsilon)^2}{C(\epsilon)}\,\frac{2}{3!}\sum_{P(i,j,k)\in(3,4,5)}\PSh\Bigg\{\nonumber\\
&&+F_{3}^{0}(\hat{\bar{1}}_g,i_g,\hat{\bar{2}}_g)\Bigg[-\frac{1}{2}\calF(s_{\bar{\bar{1}}\bar{\bar{2}}})+\frac{1}{4}\calF(s_{\bar{\bar{2}}\tilde{j}})
+\frac{1}{3}\calF(s_{\tilde{j}\tilde{k}})+\frac{1}{4}\calF(s_{\bar{\bar{1}}\tilde{k}})\Bigg]\nonumber\\
&&\hspace{1.0cm}\times A_{4}^{0}(\hat{\bar{\bar{1}}}_g,\hat{\bar{\bar{2}}}_g,\tilde{j}_g,\tilde{k}_g)\,J_{2}^{(2)}(\tilde{p}_{j},\tilde{p}_{k})\nonumber\\
&&+f_{3}^{0}(\hat{\bar{2}}_g,i_g,j_g)\Bigg[\frac{1}{2}\calF(s_{\bar{1}\bar{\bar{2}}})
-\frac{1}{4}\calF(s_{\bar{\bar{2}}\widetilde{(ij)}})+\frac{1}{6}\calF(s_{k\widetilde{(ij)}})+\frac{1}{2}\calF(s_{\bar{1}k})\Bigg]\nonumber\\
&&\hspace{1.0cm}\times A_{4}^{0}(\hat{\bar{1}}_g,\hat{\bar{\bar{2}}}_g,\widetilde{(ij)}_g,k_g)\,J_{2}^{(2)}(\tilde{p}_{ij},p_{k})\nonumber\\
&&+f_{3}^{0}(i_g,j_g,k_g)\Bigg[\calF(s_{\bar{1}\bar{2}})+\frac{1}{4}\calF(s_{\bar{2}\widetilde{(ij)}})
-\frac{1}{6}\calF(s_{\wt{(ij)}\wt{(jk)}})+\frac{1}{4}(s_{\bar{1}\wt{(jk)}})\Bigg]\nonumber\\
&&\hspace{1.0cm}\times A_{4}^{0}(\hat{\bar{1}}_g,\hat{\bar{2}}_g,\wt{(ij)}_g,\wt{(jk)}_g)\, J_{2}^{(2)}(\tilde{p}_{ij},\tilde{p}_{jk})\nonumber\\
&&+f_{3}^{0}(j_g,k_g,\hat{\bar{1}}_g)\Bigg[\frac{1}{2}\calF(s_{\bar{\bar{1}}\bar{2}})+\frac{1}{2}\calF(s_{\bar{2}i})
+\frac{1}{6}\calF(s_{i\wt{(kj)}})-\frac{1}{4}\calF(s_{\bar{\bar{1}}\wt{(kj)}})\Bigg]\nonumber\\
&&\hspace{1.0cm}\times A_{4}^{0}(\hat{\bar{\bar{1}}}_g,\hat{\bar{2}}_g,i_{g},\wt{(kj)}_g)\, J_{2}^{(2)}(p_i,\tilde{p}_{kj}).
\end{eqnarray}

Similarly, the contributions which become proportional to the $Y_{4}$ topology are,
\begin{eqnarray}
&&\dsigma_{NNLO}^{VS}|_{Y_{4}^{0}}={\cal N}_{LO} \left(\frac{\alpha_sN}{2\pi}\right)^2 \frac{\bar{C}(\epsilon)^2}{C(\epsilon)}
\,\frac{2}{3!}\sum_{P(i,j,k)\in(3,4,5)}\PSs\Bigg\{\nonumber\\
&&\phantom{+}f_{3}^{0}(\hat{2}_g,j_g,k_g)A_{4}^{1}(\hat{1}_{g},i_g,\hat{\bar{2}}_g,\wt{(jk)}_g)\, J_{2}^{(2)}(p_i,\tilde{p}_{(jk)})\nonumber\\
&&+f_{3}^{0}(j_g,k_g,\hat{1}_g)A_{4}^{1}(\hat{\bar{1}}_g,i_{g},\hat{2}_g,\wt{(kj)}_g)\, J_{2}^{(2)}(p_i,\tilde{p}_{kj})\nonumber\\
&&+\left(f_{3}^{1}(\hat{2}_g,j_g,k_g)+\frac{11}{6\epsilon}f_{3}^{0}(\hat{2}_g,j_g,k_g)\left((|s_{2jk}|)^{-\e}-(\mu^2)^{-\e}\right)\right)
A_{4}^{0}(\hat{1}_{g},i_g,\hat{\bar{2}}_g,\wt{(jk)}_g)\, J_{2}^{(2)}(p_i,\tilde{p}_{(jk)})\nonumber\\
&&+\left(f_{3}^{1}(j_g,k_g,\hat{1}_g)+\frac{11}{6\epsilon}f_{3}^{0}(j_g,k_g,\hat{1}_g)\left((|s_{1kj}|)^{-\e}-(\mu^2)^{-\e}\right)\right)
A_{4}^{0}(\hat{\bar{1}}_{g},i_g,\hat{2}_g,\wt{(kj)}_g)\, J_{2}^{(2)}(p_i,\tilde{p}_{(kj)})\Bigg\}\nonumber\\
&&\nonumber\\
&&+{\cal N}_{LO} \left(\frac{\alpha_sN}{2\pi}\right)^2\frac{\bar{C}(\epsilon)^2}{C(\epsilon)}\,\frac{2}{3!}\sum_{P(i,j,k)\in(3,4,5)}\PSh\Bigg\{\nonumber\\
&&+f_{3}^{0}(\hat{2}_g,j_g,k_g)\Bigg[\frac{1}{2}\calF(s_{\bar{1}i})+\frac{1}{4}\calF(s_{\bar{\bar{2}}i})
-\frac{1}{4}\calF(s_{\bar{\bar{2}}\wt{(jk)}})+\frac{1}{4}\calF(s_{\bar{1}\wt{(jk)}})\Bigg]\nonumber\\
&&\hspace{1.0cm}\times A_{4}^{0}(\hat{\bar{1}}_g,i_g,\hat{\bar{\bar{2}}}_g,\wt{(jk)}_g)\, J_{2}^{2}(p_{i},\tilde{p}_{jk})\nonumber\\
&&+f_{3}^{0}(j_g,k_g,\hat{\bar{1}}_g)\Bigg[\frac{1}{4}\calF(s_{\bar{\bar{1}}i})+\frac{1}{2}\calF(s_{\bar{2}i})
+\frac{1}{4}\calF(s_{\bar{2}\wt{(kj)}})-\frac{1}{4}\calF(s_{\bar{\bar{1}}\wt{(kj)}})\Bigg]\nonumber\\
&&\hspace{1.0cm}\times A_{4}^{0}(\hat{\bar{\bar{1}}}_g,i_g,\hat{\bar{2}}_g,\wt{(kj)}_g)\, J_{2}^{2}(p_{i},\tilde{p}_{kj}).
\end{eqnarray}

\section{Numerical results}
\label{sec:numerical}

In this section we will test how well the real virtual subtraction term $\dsigma_{NNLO}^{T}$ derived in the previous section approaches
the real virtual contribution $\dsigma_{NNLO}^{RV}$ in
all single unresolved regions of the phase space so that their difference can
be integrated numerically in four dimensions. We will do this by generating a series of phase space points using
 {\tt RAMBO}~\cite{Kleiss:1985gy} that approach a given single unresolved limit. For each generated point we compute the ratio of the finite parts of $\dsigma^{RV}_{NNLO}$ and $\dsigma^{T}_{NNLO}$,
\begin{equation}
R=\frac{\Finite(\dsigma^{RV}_{NNLO})}{\Finite(\dsigma^{T}_{NNLO})}.
\end{equation}
Here $\dsigma^{RV}_{NNLO}$ is the interference between the one-loop and tree-level five gluon matrix elements given by Eq.~(\ref{eq:RVNLO}),
and $\dsigma^T_{NNLO}$ is the real virtual subtraction term given by Eqs.~(\ref{eq:RVsofttop1}) and (\ref{eq:RVsofttop2}). The ratio $R$ should approach unity as we get closer to any singularity showing that the subtraction captures the infrared singularity structure of the real virtual contribution.

For each unresolved configuration, we will define a variable that controls how we approach the singularity subject to the requirement that there are at least
two jets in the final state with $p_{T}>50$ GeV where the jets have been clustered with the anti-$k_t$ jet algorithm~\cite{Cacciari:2005hq,Cacciari:2008gp}
with radius R=0.4. The partonic center-of-mass energy $\sqrt{s}$ is fixed to be 1000~GeV.

\subsection{Soft limit}
\begin{figure}[htp!]
\begin{minipage}[b]{0.3\linewidth}
\centering
\includegraphics[width=4cm]{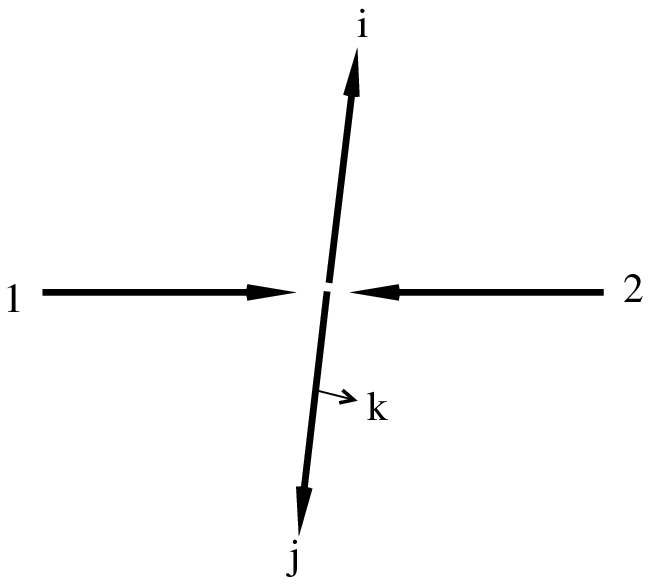}\\
\vspace{0.2cm}
(a)
\end{minipage}
\hspace{0.5cm}
\begin{minipage}[b]{0.7\linewidth}
\centering
\includegraphics[width=5cm,angle=270]{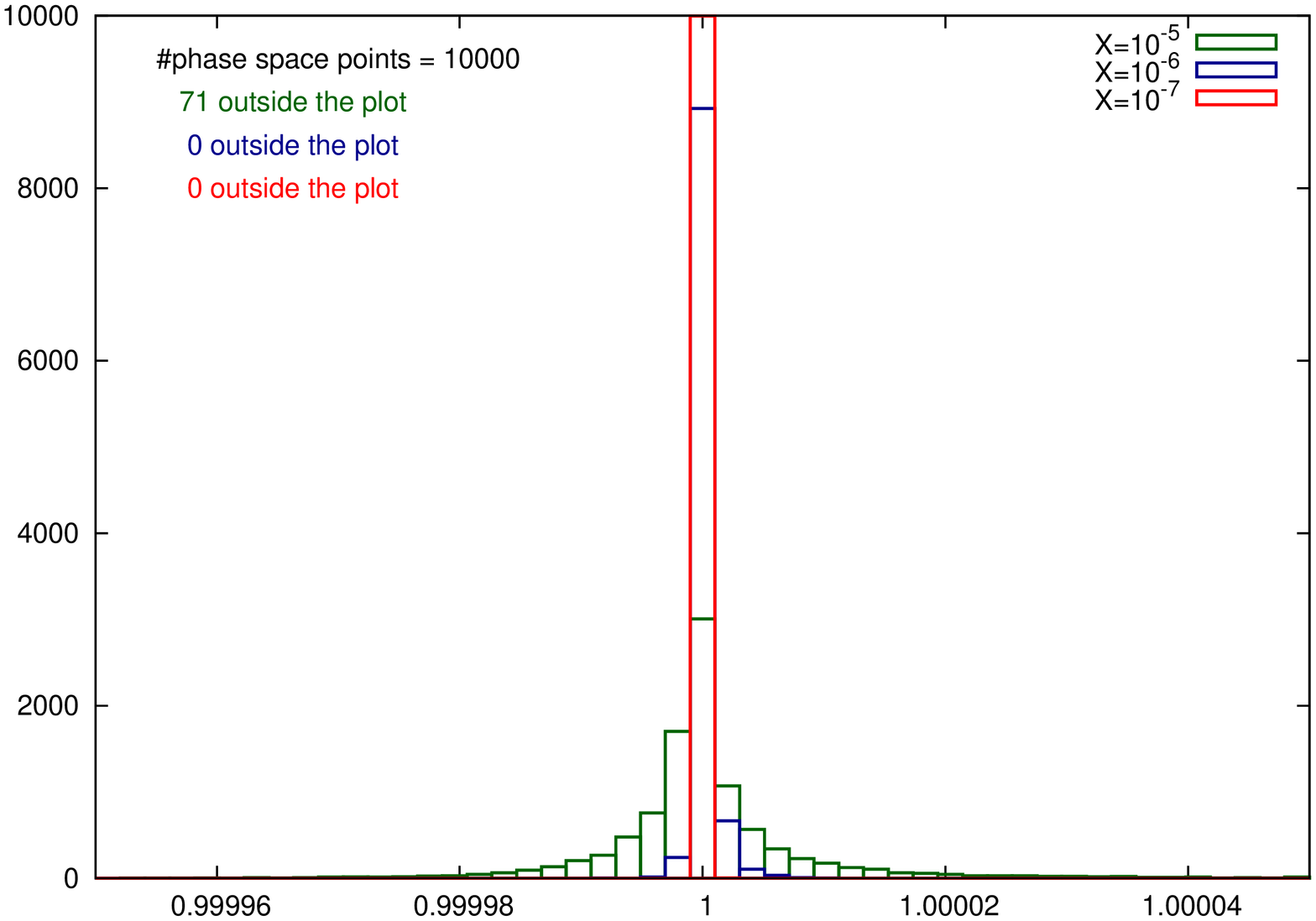}\\
\vspace{0.2cm}
(b)
\end{minipage}
\caption[Single soft singularity]{(a) Example configuration of a single soft event with $s_{ij}\approx s_{12}=s$.
(b) Distribution of $R$ for 10000 single soft phase space points.}
\label{fig:ssoft}
\end{figure}

To probe the soft regions of the phase space, we generate an event configuration with a soft final state gluon $k$ by making    invariant $s_{ij}$ close to the full center of mass energy $s_{12}$. This kinematic configuration is pictured in Fig.~\ref{fig:ssoft}(a).  We define the small parameter $x=(s-s_{ij})/s$ and show 
the distributions of the ratio between the real-virtual matrix element and the subtraction term for $x=10^{-5}$ (green), $x=10^{-6}$ (blue) and $x=10^{-7}$ (red)
in Fig.~\ref{fig:ssoft}(b) using 10000 phase space points. The plot also shows the number of points that lie outside the range of the histogram.
We see that the subtraction term rapidly converges to the matrix element as we approach the single soft limit. 

\subsection{Collinear limit}
\label{sec:colllim}
\begin{figure}[htp!]
\begin{minipage}[b]{0.3\linewidth}
\centering
\includegraphics[width=4cm]{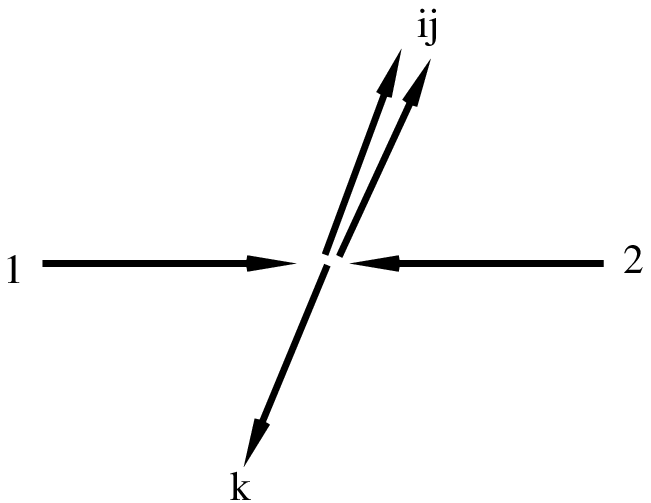}\\
\vspace{0.2cm}
(a)
\end{minipage}
\hspace{0.5cm}
\begin{minipage}[b]{0.7\linewidth}
\centering
\includegraphics[width=5cm,angle=270]{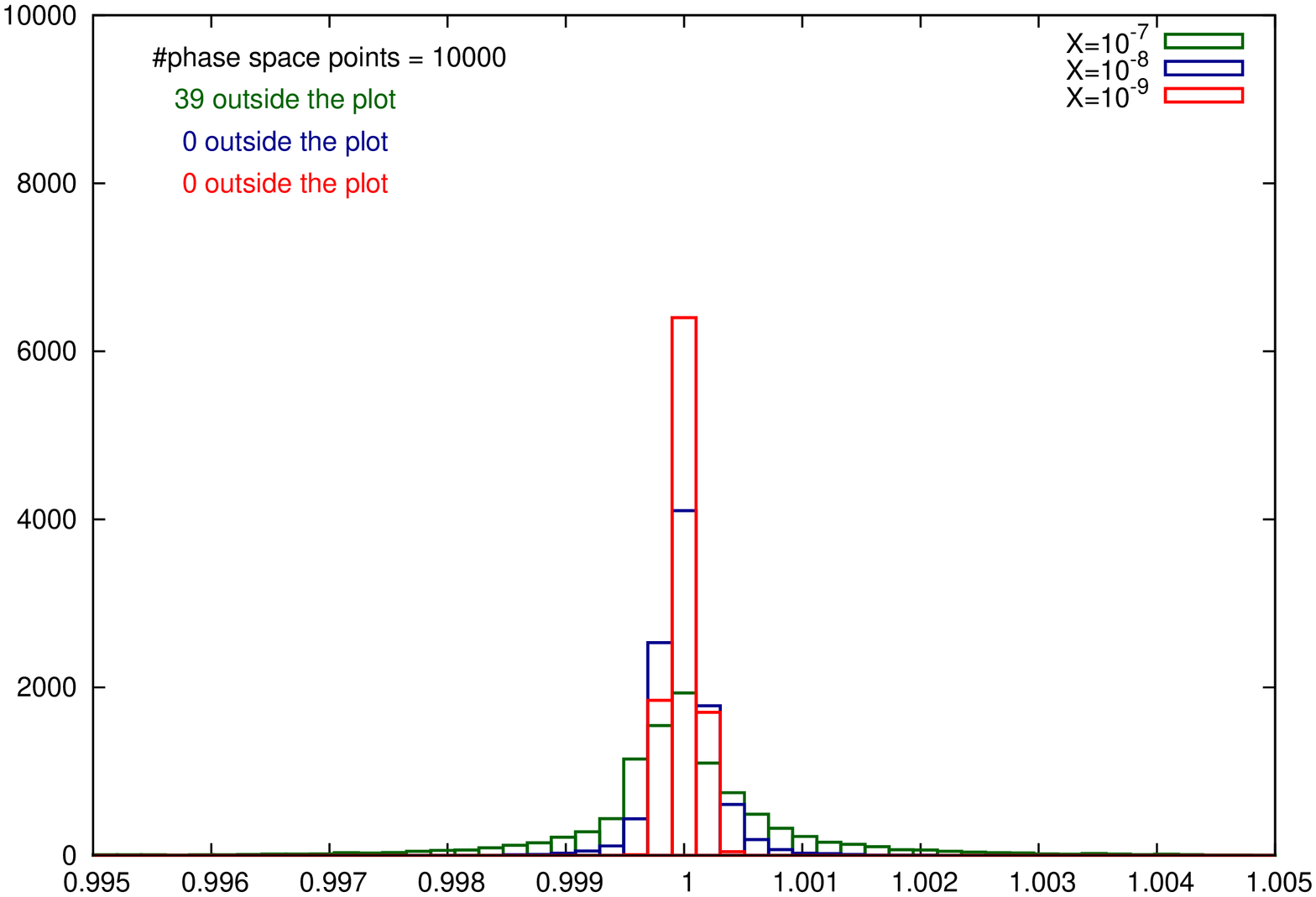}\\
\vspace{0.2cm}
(b)
\end{minipage}
\caption[Single collinear limit final state singularity]{(a) Example configuration of a single collinear event with $s_{ij}\to 0$.
(b) Distribution of $R$ for 10000 single collinear phase space points.}
\label{fig:scollff}
\end{figure}
\begin{figure}[htp!]
\begin{minipage}[b]{0.3\linewidth}
\centering
\includegraphics[width=4cm]{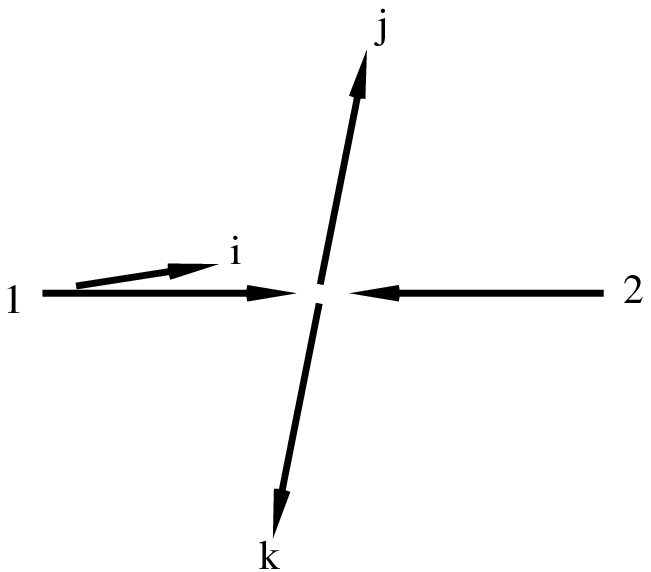}\\
\vspace{0.2cm}
(a)
\end{minipage}
\hspace{0.5cm}
\begin{minipage}[b]{0.7\linewidth}
\centering
\includegraphics[width=5cm,angle=270]{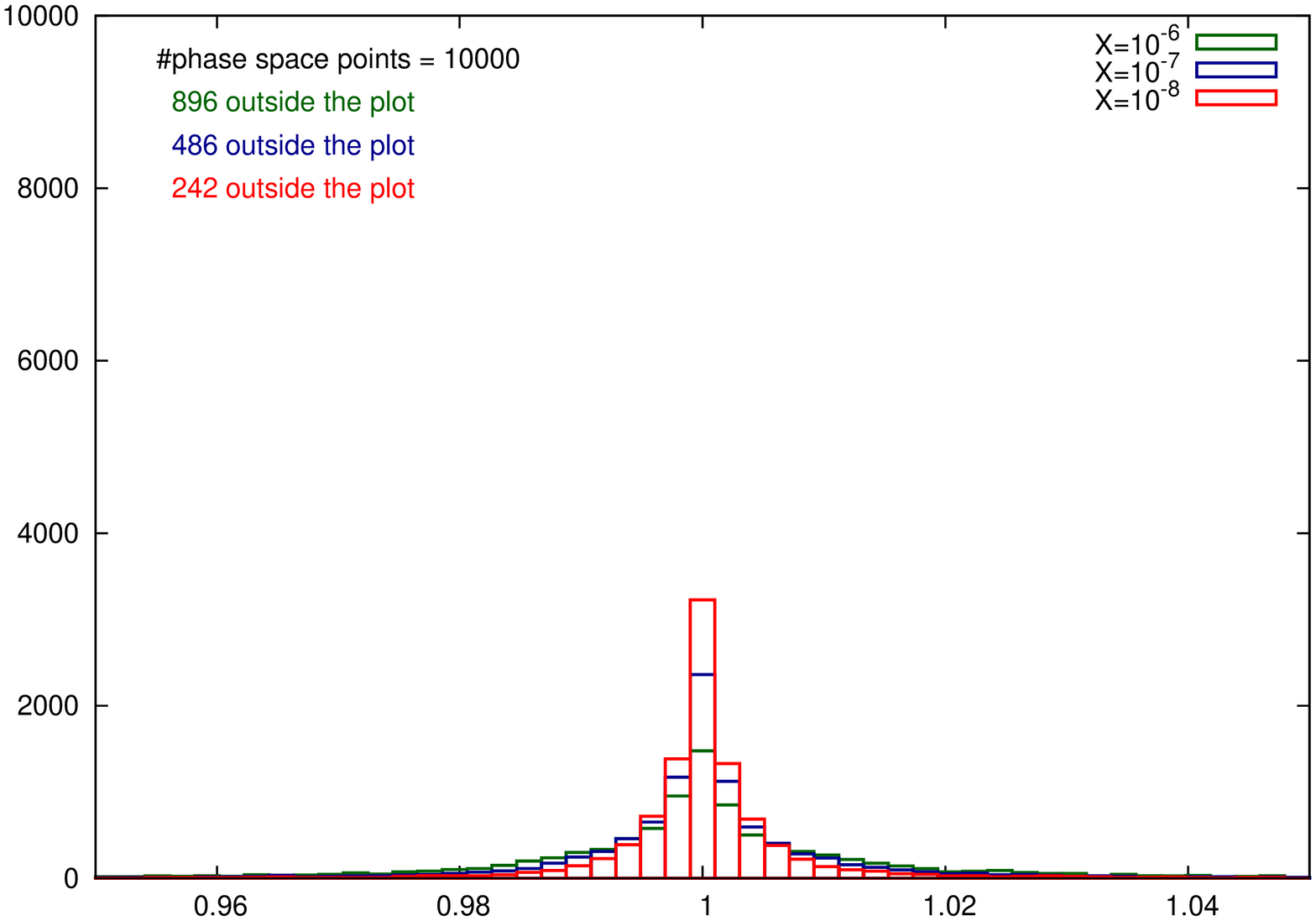}\\
\vspace{0.2cm}
(b)
\end{minipage}
\caption[Single collinear limit initial state singularity]{(a) Example configuration of a single collinear event with $s_{1i}\to 0$.
(b) Distribution of $R$ for 10000 single collinear phase space points.}
\label{fig:scollif}
\end{figure}

Next we probe the final and initial state single collinear regions of the phase space. These event topologies are depicted in Figs.~\ref{fig:scollff}(a) when gluons $i$ and $j$ become collinear, and \ref{fig:scollif}(a) where gluon $i$ becomes collinear with the incoming gluon $1$. 

For the final-final collinear singularity, we introduce the small parameter $x = s_{ij}/s_{12}$. Fig.~\ref{fig:scollff}(b) shows the distribution in $R$ obtained for 10000 phase space points for
$x=10^{-7}$ (green),   $x=10^{-8}$ (blue) and  $x=10^{-9}$ (red). 

Similarly in the initial-final collinear limit, the small parameter is $x = s_{1i}/s_{12}$ and Fig.~\ref{fig:scollif}(b) shows the distributions of $R$ for the same $x$-values of $x=10^{-7}$ (green),   $x=10^{-8}$ (blue) and  $x=10^{-9}$ (red).

As the small parameter $x$ gets smaller, we see a systematic improvement in the convergence of the real-virtual matrix elements and the subtraction term. 
This is in contrast with the collinear limit of the double real all gluon subtraction term \cite{Glover:2010im,Pires:2010jv}, but not surprising due to the simplicity
of the final state where the partons are fixed to be in back-to-back pairs as shown in Figs.~\ref{fig:scollff}(a) and \ref{fig:scollif}(a).
Nevertheless, Figs.~\ref{fig:scollff}(b) and \ref{fig:scollif}(b) show that the subtraction term does not approximate the real-virtual matrix element as well
as in the soft limit (Fig.~\ref{fig:ssoft}(b)).  This is due to the presence of angular correlations in the matrix elements stemming from gluon splittings $g\to gg$.
The collinear limits of tree and one-loop matrix elements are controlled by the unpolarised Altarelli-Parisi splitting functions which explicitly depend on the
transverse momentum $k_{\perp}$ of the collinear gluons with respect to the collinear direction and on the helicity of the parent
parton. As a result of this, the   splitting functions produce spin correlations with respect to the directions of other momenta in the
matrix element besides the momenta becoming collinear.  These azimuthal terms coming from
the single collinear limits vanish after integration over the azimuthal angle of the collinear system. This occurs globally after an
azimuthal integration over the unresolved phase space. Here we are performing a point-by-point analysis on the integrand defined by the
real-virtual matrix element and the subtraction term and because we use spin-averaged antenna functions to subtract the collinear singularities, 
the azimuthal angular terms produced by
the spin correlations are simply not accounted for in the antenna subtraction procedure. 

To improve on this, several approaches have been discussed in the past.  One possible strategy discussed in
\cite{GehrmannDeRidder:2005cm} is to proceed with a tensorial reconstruction of the angular terms within the antenna subtraction terms. 
A second approach is to cancel the angular terms by combining phase space points which are related by rotating the collinear partons by
an angle of $\pi/2$ around the collinear parton direction \cite{Weinzierl:2006wi,GehrmannDeRidder:2007jk}.  In this case, the azimuthal
correlations present in the matrix element at the rotated point cancel precisely the azimuthal correlations of the un-rotated point.  

This second procedure was demonstrated to be extremely powerful in improving the convergence of the double-real radiation subtraction
contribution to dijet production in \cite{Glover:2010im,Pires:2010jv} in the pure gluonic channel. The strategy of combining pairs of
phase space points related by a $\pi/2$ rotation eliminated the correlations from both:  the real-radiation and its subtraction term. In
the latter case, the four-parton antennae are responsible for angular correlations. In the real-virtual contribution discussed in this
paper, the correlations can arise in the real-virtual matrix elements ${\rm d}\hat\sigma_{NNLO}^{V,1}$ and in the tree-level five gluon
matrix elements present in the subtraction term. There is no contribution from the three-parton antennae. Therefore, the azimuthal
effect is expected to be smaller than in the double-real case~\cite{Glover:2010im}. Looking at the distributions shown in 
Fig.~\ref{fig:scollff}(b) for $x=10^{-7}$ (green),   $x=10^{-8}$ (blue) and  $x=10^{-9}$ (red) and Fig.~\ref{fig:scollif}(b)  for
$x=10^{-6}$ (green),   $x=10^{-7}$ (blue) and  $x=10^{-8}$ (red), we see that the correlations are clearly visible, but are indeed
relatively small.  

\begin{figure}[t]
\begin{minipage}[b]{0.3\linewidth}
\centering
\includegraphics[width=5cm,angle=270]{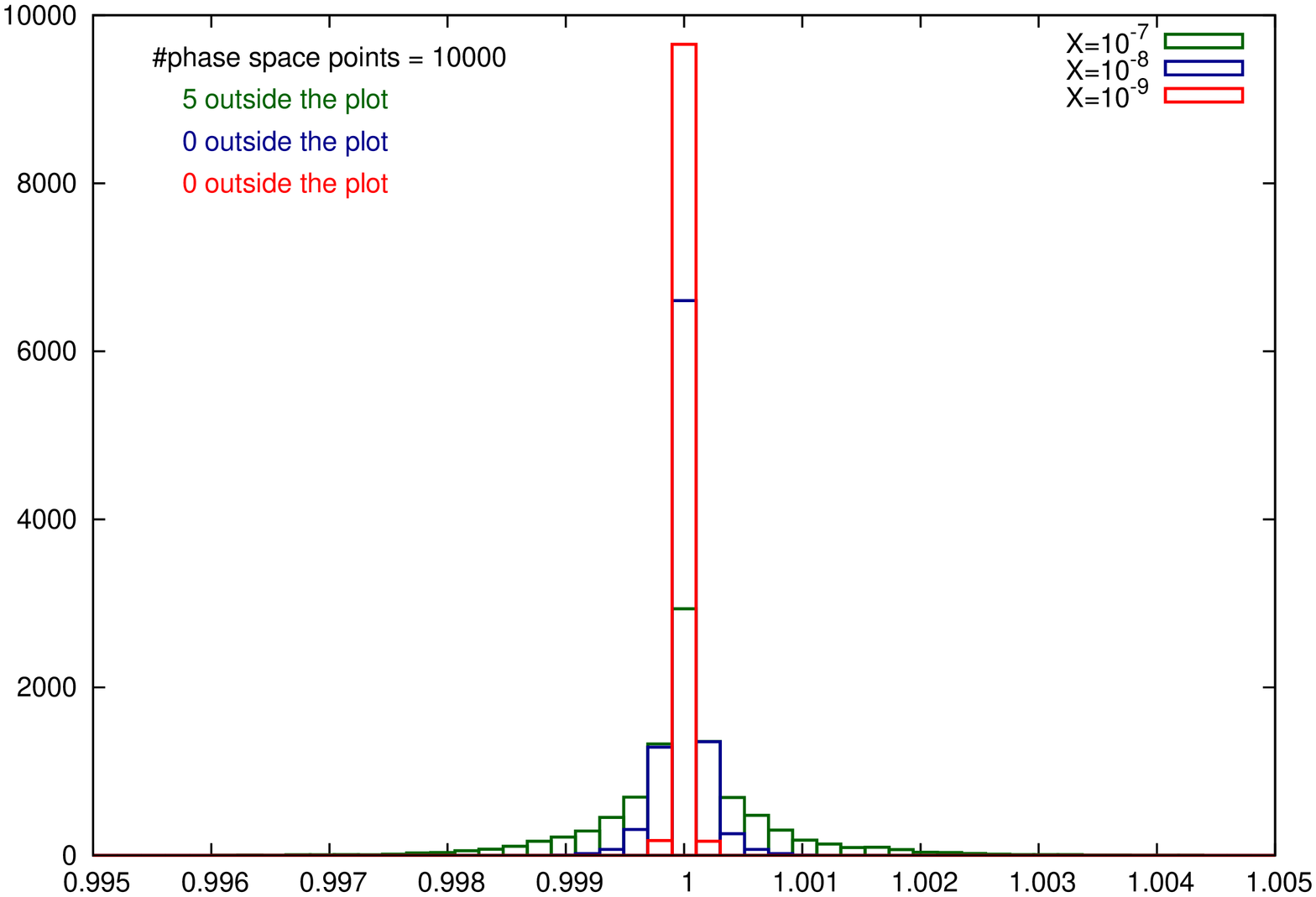}\\
\vspace{0.2cm}
(a)
\end{minipage}
\hspace{0.5cm}
\begin{minipage}[b]{0.7\linewidth}
\centering
\includegraphics[width=5cm,angle=270]{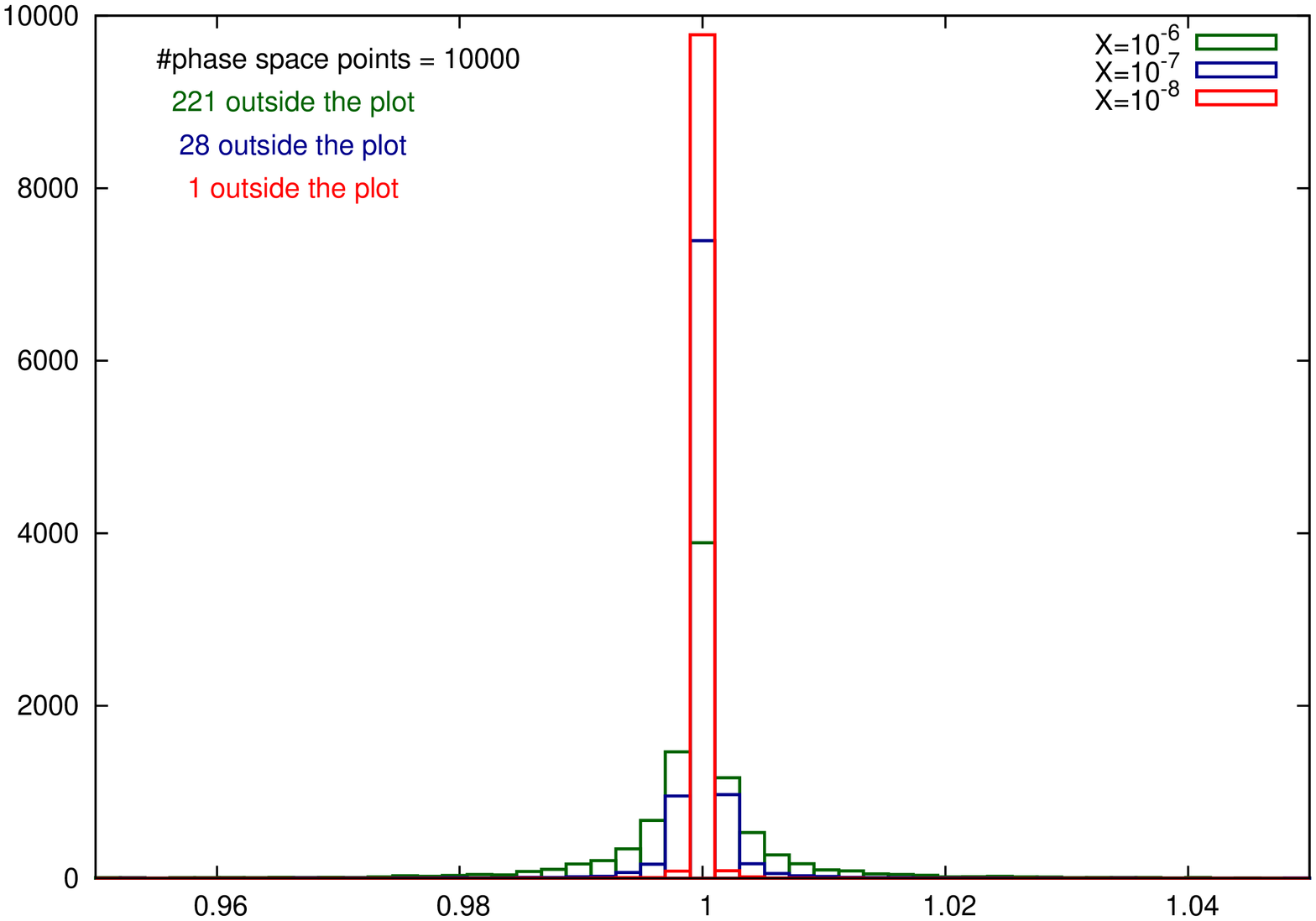}\\
\vspace{0.2cm}
(b)
\end{minipage}
\caption[Single collinear limit final state singularity]{Distribution of R for 10000 single collinear phase space point pairs where the pair of phase space points is related by an azimuthal rotation of $\pi/2$ about the collinear direction. (a) Final state collinear singularity 
(b) Initial state collinear singularity}
\label{fig:col-rot}
\end{figure}

Nevertheless, to eliminate the remaining azimuthal correlations, we show the effect of combining related phase space points discussed
above in Fig.~\ref{fig:col-rot}(a) for the final-state collinear singularity and Fig.~\ref{fig:col-rot}(b) for the initial-state 
collinear singularity for the same values of the small parameter as in  Figs.~\ref{fig:scollff}(b) and \ref{fig:scollif}(b)
respectively. We observe a significant improvement in the convergence of the subtraction term, particularly in the  case of the
initial-final collinear limit. The conclusion is that by combining azimuthally correlated phase space points, the antenna subtraction
term correctly subtracts the azimuthally enhanced terms in a point-by-point manner.

\section{Conclusions} 
\label{sec:conc}

In this paper, we have generalised the antenna subtraction method for the calculation of higher order QCD corrections to derive the
real-virtual subtraction term for exclusive collider observables for situations with partons in the initial state to NNLO.  We focussed particular attention on the application of the antenna subtraction formalism to construct the subtraction term relevant for
the gluonic real-virtual contribution to dijet production. The gluon scattering channel is expected
to be the dominant contribution at NNLO. The
subtraction term includes a mixture of  integrated and unintegrated tree- and one-loop three-parton antennae functions in final-final,
initial-final and initial-initial configurations. We note that the subtraction terms for processes involving quarks, as required for
dijet or vector boson plus jet processes, will make use of the same types of antenna building blocks as those discussed here.

By construction the counterterm removes the explicit infrared poles present
on the one-loop amplitude, as well as the implicit singularities that occur in the soft and collinear limits. 
The $\eps$-poles present in the real-virtual contribution are analytically cancelled by the $\eps$-poles in the subtraction term rendering the real-virtual contribution locally finite over the whole of phase space.

We tested that our numerical implementation of the antenna subtraction term behaves in the expected way by comparing the behaviour of the finite parts of the one-loop
real-virtual contribution $\dsigma^{RV}_{NNLO}$ with the finite part of the real-virtual subtraction term $\dsigma^{T}_{NNLO}$ for the five-gluon process in the regions of phase space where one particle is unresolved.     The numerical convergence of these terms  gives confidence that the infrared structure of the real-virtual matrix element is captured by antenna subtraction method in a systematic and accurate manner.

The real-virtual subtraction terms presented here
provide a major step towards the NNLO evaluation of the dijet observables at
hadron colliders. 
Future steps include; 
\begin{itemize}
\item[(i)] completion of the analytic integration of the initial-initial four-parton antennae. 
\item[(ii)] analytic cancellation of infrared poles between the analytically integrated antennae present in the subtraction terms and the two-loop four-gluon matrix elements.
\item[(iii)] full parton-level Monte Carlo implementation of the finite four-, five- and six-gluon channels.
\item[(iv)] the construction of similar subtraction terms, etc., for processes
involving quarks.
\end{itemize}
The final goal is the construction of a numerical program to compute the 
NNLO QCD corrections to dijet production in hadron-hadron collisions.

\acknowledgments 

We thank James Currie and Thomas Gehrmann for stimulating discussions. We thank Simon Badger for assistance in comparing with the {\tt NGluon} package and
Stephan Buehler and Claude Duhr for help in using the {\tt CHAPLIN} package.  This
research was supported in part by the UK Science and Technology Facilities
Council,  in part by the Swiss National Science Foundation (SNF) under contract
PP0022-118864, in part by an ETH-Fellowship program under grant No. FEL-15 10-2,
in part by the Research Executive Agency (REA) of the European Union under the Grant Agreement number PITN-GA-2010-264564 (LHCPhenoNet) and in part by the US National Science Foundation
under Grant No. NSF PHY05-51164. EWNG gratefully acknowledges the support of the
Wolfson Foundation and the Royal Society and thanks the Institute for
Theoretical Physics at the ETH for its kind hospitality during the completion of
this work. AG and EWNG also thank the Kavi Institute for Theoretical Physics,
Santa Barbara for its kind hospitality during the early stages of this work.

\appendix
\section{Momentum mappings}
\label{sec:appendixA}

The NNLO corrections to an $m$-jet final state receive contributions from processes with different numbers of final state particles. In the antenna subtraction scheme, one is replacing antennae consisting of 
two hard radiators plus unresolved particles with two new hard radiators.
A key element of the antenna subtraction scheme is the factorisation of the matrix elements and phase space in the singular limits where one or more particles are unresolved.   This factorisation is guaranteed by the momentum mapping.

In this section we denote the set of momenta for the $M$-particle process by $\{p\}_M$.
In order to subtract a particular singular configuration in a given process, 
we derive subtraction terms which reproduce the exact singular behaviour of the matrix element in the unresolved configuration and employ a 
momentum mapping to implement momentum conservation away from the unresolved limit. This has the consequence of mapping a singular configuration
in an $M$-particle process to an $({M} -1)$ or $({M} -2)$-particle process, depending
whether the given singular configuration involves 
a single or a double unresolved limit. In integrated form
these subtraction terms have explicit $\e$-poles and contribute
to final sates with fewer particles. The consistent momentum maps we require are,
\begin{eqnarray}
\label{eq:map4to3}
\{p\}_{m+4} &\to &\{p\}_{m+3},\\
\label{eq:map4to2}
\{p\}_{m+4} &\to &\{p\}_{m+2},\\
\label{eq:map3to2}
\{p\}_{m+3} &\to &\{p\}_{m+2}.
\end{eqnarray} 

Let us consider the single unresolved emission that is relevant in this paper - either as part of the integrated single unresolved subtraction term that cascades down from the double real emission process \eqref{eq:map4to3} or 
in the single unresolved limit of the real-virtual contribution \eqref{eq:map3to2}.
If the antenna consists of an unresolved particle $j$ colour linked to two hard radiators
$i$ and $k$, then the mapping must produce two new hard radiators $I$ and $K$.  
Each mapping must conserve four-momentum and maintain the on-shellness of the particles involved.
There are three distinct cases,
\begin{align}
&{\rm final-final~configuration}   &  i j k \to I K \nonumber\\
&{\rm initial-final~configuration} &  \hat i j k \to \hat I K\nonumber \\ 
&{\rm initial-initial~configuration} &  \hat i j \hat k \to \hat I \hat K\nonumber
\end{align}
where, as usual, initial state particles are denoted by a hat.
In principle, the momenta not involved in the antenna are also affected by the mapping.   For the final-final and initial-final maps, this is trivial.  Only in the initial-initial case are the spectator momenta actually modified.
The momentum transformations for these three mappings are described in
refs.~\cite{Kosower:1997zr,Kosower:2003bh,Daleo:2006xa} and will be
recalled below.

\subsection{Final-Final mapping}
\label{sec:appendixAFF}

The final-final mapping is given in 
\cite{Kosower:1997zr} and reads
\begin{eqnarray}
p_I^{\mu} \equiv p_{\widetilde{(ij)}}^{\mu}&=&x\,p_i^{\mu}+r\,p_j^{\mu}+z\,p_k^{\mu}\nonumber\\
p_K^{\mu} \equiv p_{\widetilde{(jk)}}^{\mu}&=&(1-x)\,p_i^{\mu}+(1-r)\,p_j^{\mu}+(1-z)\,p_k^{\mu}\;
\label{3to2FFmap}
\end{eqnarray}
where, 
\begin{eqnarray}
x&=&\frac{1}{2(s_{ij}+s_{ik})}\Big[(1+\rho)\,s_{ijk} -2\,r\,s_{jk}     \Big],\nonumber\\
z&=&\frac{1}{2(s_{jk}+s_{ik})}\Big[(1-\rho)\,s_{ijk} -2\,r\,s_{ij}     \Big],\nonumber\\
&&\nonumber\\
\rho^2&=&1+\frac{4\,r(1-r)\,s_{ij}s_{jk}}{s_{ijk}s_{ik}}.\;
\end{eqnarray}
The parameter $r$ can be chosen conveniently~\cite{Kosower:1997zr,Kosower:2003bh} and we use $ r=s_{jk}/(s_{ij}+s_{jk}).$

The mapping \eqref{3to2FFmap} implements momentum conservation $p_{\widetilde{(ij)}}+p_{\widetilde{(jk)}}=p_i+p_j+p_k$ and satisfies the following properties:
\begin{eqnarray}
p_{\widetilde{(ij)}}^2=0,\qquad&&\qquad p_{\widetilde{(jk)}}^2=0,\nonumber\\
p_{\widetilde{(ij)}}\to p_i,\qquad&&\qquad p_{\widetilde{(jk)}}\to p_k \qquad\qquad\textrm{when \textit{j} is soft},\nonumber\\
p_{\widetilde{(ij)}}\to p_i+p_j,\qquad&&\qquad p_{\widetilde{(jk)}}\to p_k \qquad\qquad\textrm{when \textit{i} becomes collinear with \textit{j}},\nonumber\\
p_{\widetilde{(ij)}}\to p_i,\qquad&&\qquad p_{\widetilde{(jk)}}\to p_j+p_k \qquad\textrm{when \textit{j} becomes collinear with \textit{k}}.\nonumber
\end{eqnarray}

\subsection{Initial-Final mapping}
\label{sec:appendixAIF}

The initial-final mapping is given in 
 \cite{Daleo:2006xa} and reads
\begin{eqnarray} 
p_{\hat I}^\mu &\equiv& \bar{p}_{i}^{\mu}=\hat{x}_i\,p_i^{\mu},\nonumber\\
p_{K}^\mu &\equiv& p_{\widetilde{(jk)}}^{\mu}=p_j^{\mu}+\,p_k^{\mu}-(1-\hat{x}_i)\,p_i^{\mu}\;,
\label{3to2IFmap}
\end{eqnarray}
with $p_{\hat I}^2=p_{ K}^2=0$ and where $\hat{x}_i$ is given by,
\begin{eqnarray}
\label{eq:xiIF}
\hat{x}_i&=&\frac{s_{ij}+s_{ik}+s_{jk}}{s_{ij}+s_{ik}}.
\end{eqnarray}
Proper subtraction of infrared singularities requires that the momenta mapping satisfies,
\begin{eqnarray}
&&\hat{x}_i p_i\rightarrow p_i\,,\qquad\qquad p_{\widetilde{(jk)}}\rightarrow p_k\qquad\qquad\mbox{when $j$ becomes soft}\,,\nonumber\\
&&\hat{x}_i p_i\rightarrow p_i\,,\qquad\qquad p_{\widetilde{(jk)}}\rightarrow p_j+p_k
\qquad\mbox{when $j$ becomes collinear with $k$}\,,\nonumber\\
&&\hat{x}_i p_i\rightarrow p_i-p_j\,,\qquad p_{\widetilde{(jk)}}\rightarrow p_k
\qquad\qquad\mbox{when $j$ becomes collinear with $\hat{i}$}\,.\nonumber
\end{eqnarray}

\subsection{Initial-Initial mapping}
\label{sec:appendixAII}

The initial-initial mapping for $ik,j \to IK$ is given in Ref.~\cite{Daleo:2006xa} and reads
\begin{eqnarray}
p_{\hat I}^\mu \equiv \bar{p}_i^{\mu}&=&\hat{x}_i\,p_i^{\mu},\nonumber\\
p_{\hat K}^\mu \equiv \bar{p}_k^{\mu}&=&\hat{x}_k\,p_k^{\mu},\nonumber\\
\tilde{p}_\ell^{\mu}&=&p_\ell^{\mu}-\frac{2p_\ell\cdot(q+\tilde{q})}{(q+\tilde{q})^2}\;(q^{\mu}+\tilde{q}^{\mu})+\frac{2p_\ell\cdot q}{q^2}\;\tilde{q}^{\mu},
\label{3to2IImap}
\end{eqnarray}
where $p_{\hat I}^2=p_{\hat K}^2=\tilde{p}_\ell^2=0$ and  
$$
q^{\mu}=p_i^{\mu}+p_k^{\mu}-p_j^{\mu},\qquad\qquad
\tilde{q}^{\mu}=\bar{p}_i^{\mu}+\bar{p}_k^{\mu}\quad.
$$
The rescaling of the initial state momenta are given by the fractions $\hat{x}_i$ and $\hat{x}_k$ which read \cite{Daleo:2006xa},
\begin{eqnarray}
\label{eq:xiII}
\hat{x}_i&=&\sqrt{\frac{s_{ik}+s_{jk}}{s_{ik}+s_{ij}}}\sqrt{\frac{s_{ik}+s_{ij}+s_{jk}}{s_{ik}}},\nonumber\\
\hat{x}_k&=&\sqrt{\frac{s_{ik}+s_{ij}}{s_{ik}+s_{jk}}}\sqrt{\frac{s_{ik}+s_{jk}+s_{ij}}{s_{ik}}}.
\end{eqnarray}
With these definitions it is straightforward to check that the momenta mapping satisfies the following limits required
for proper subtraction of infrared singularities,
\begin{eqnarray}
\bar{p}_i\rightarrow p_i\,,&&\qquad\bar{p}_k\rightarrow p_k\qquad\mbox{when $j$ becomes soft}\,,\nonumber\\
\qquad  \bar{p}_i\rightarrow (1-z_i)p_i\,,&&
\qquad\bar{p}_k\rightarrow p_k\qquad\mbox{when $j$ becomes collinear with $\hat{i}$}\,,\nonumber\\
\bar{p}_k\rightarrow (1-z_k)p_k\,,&&
\qquad\bar{p}_i\rightarrow p_i\qquad~\mbox{when $j$ becomes collinear with $\hat{k}$}\,.\nonumber
\end{eqnarray}

\section{Gluonic antennae}

\label{sec:appANT}

In this section we collect the three-parton tree-level unintegrated and integrated antennae as
well as one-loop three-parton antennae used in the implementation of the real virtual subtraction
term of Section \ref{sec:DsigmaVSNNLO}.

\subsection{Tree-level three-parton antennae}
\label{subsec:unintegratedthree}
The
tree-level antenna functions are obtained by normalising the 
colour-ordered three-parton tree-level 
squared matrix elements to the squared matrix element 
for the basic two-parton process,
\begin{eqnarray}
X_{ijk}^0 = S_{ijk,IK}\, \frac{|{\cal M}^0_{ijk}|^2}{|{\cal M}^0_{IK}|^2}\;,\nonumber
\end{eqnarray}
where $S$ denotes the symmetry factor associated with the antenna, which accounts
both for potential identical particle symmetries and for the presence 
of more than one antenna in the basic two-parton process.
For the purposes of this paper the relevant antennae are, those containing the pure gluon final state. They are obtained from the 
radiative corrections to Higgs boson decay into gluons~\cite{GehrmannDeRidder:2005aw}.

\subsubsection{Final-Final emitters}
\label{sec:ff}
The tree-level three-parton antenna corresponding to the gluon-gluon-gluon final state is \cite{GehrmannDeRidder:2005cm}:
\begin{eqnarray}
F_3^0(1_g,2_g,3_g)&=&\frac{2}{s_{123}^2}\Bigg(\frac{s_{123}^2s_{12}}{s_{13}s_{23}}
+\frac{s_{123}^2s_{13}}{s_{12}s_{23}}+\frac{s_{123}^2s_{23}}{s_{12}s_{13}}
+\frac{s_{12}s_{13}}{s_{23}}+\frac{s_{12}s_{23}}{s_{13}}+\frac{s_{13}s_{23}}{s_{12}}\nonumber\\
&&+4s_{123}+ {\cal O}(\epsilon)\ \Bigg),
\label{eq:bigFff}
\end{eqnarray}
where $s_{ij}=(p_i+p_j)^2$.
As can be seen from the pole structure, this tree-level antenna function contains three antenna configurations, corresponding 
to the three possible configurations of emitting a gluon in between a gluon pair. We make the decomposition \cite{GehrmannDeRidder:2005cm} 
\begin{equation}
F_3^0(1,2,3)=f_3^0(1,3,2)+f_3^0(3,2,1)+f_3^0(2,1,3),
\label{eq:decomp3FF}
\end{equation}
where
\begin{eqnarray}
f_3^0(1,3,2)&=&\frac{1}{s_{123}^2}\Bigg(2\frac{s_{123}^2s_{12}}{s_{13}s_{23}}+\frac{s_{12}s_{13}}{s_{23}}
+\frac{s_{12}s_{23}}{s_{13}}+\frac{8}{3}s_{123}+ {\cal O}(\epsilon)\ \Big).
\label{eq:smallFff}
\end{eqnarray}
The subantenna $f_3^0(i,j,k)$ has the full $j$ soft limit and part of the $i\parallel j$ and $j\parallel k$ limits of the full antenna (\ref{eq:bigFff}), such that $i$ and $k$ can be identified as hard radiators.  Each
subantenna is associated with a unique \{3$\to$2\} momenta mapping,  $(i,j,k)\to(I,K)$~\cite{Kosower:1997zr,Kosower:2003bh} given by Eq.~(\ref{3to2FFmap}).

\subsubsection{Initial-Final emitters}
\label{sec:if}
The initial-final three-gluon antenna function can be obtained from its final-final counterpart \eqref{eq:bigFff} by the appropriate crossing of one of the particles from the final to the initial state, i.e. by making the replacements,
$s_{23}\to(p_2+p_3)^2$,
$s_{12}\to(p_1-p_2)^2$,
$s_{13}\to(p_1-p_3)^2$ and $Q^2=s_{12}+s_{13}+s_{23}$.
It reads \cite{Daleo:2006xa},
\begin{eqnarray}
&&F_3^0(\hat{1}_g,2_g,3_g)=\frac{1}{2(Q^2)^2}\Bigg(\frac{8s_{12}^2}{s_{13}}+\frac{8s_{12}^2}{s_{23}}
+\frac{8s_{13}^2}{s_{12}}+\frac{8s_{13}^2}{s_{23}}+\frac{8s_{23}^2}{s_{12}}+\frac{8s_{23}^2}{s_{13}}\nonumber\\
&&
+\frac{12s_{12}s_{13}}{s_{23}}+\frac{12s_{23}s_{13}}{s_{12}}
+\frac{12s_{12}s_{23}}{s_{13}}
+\frac{4s_{12}^3}{s_{23}s_{13}}+\frac{4s_{13}^3}{s_{23}s_{12}}+\frac{4s_{23}^3}{s_{12}s_{13}}\nonumber\\
&&
+24s_{23}+24s_{12}+24s_{13} + {\cal O}(\epsilon)\ \Bigg),
\label{eq:bigFif}
\end{eqnarray}
where the hat identifies the gluon crossed to the initial state.
Because the initial gluon can never be soft, it is convenient to decompose this antenna into two contributions, that each contain the soft singularities of one of the final state gluons,
\begin{equation}
F_3^0(\hat{1},2,3)=f_3^0(\hat{1},2,3)+f_3^0(\hat{1},3,2).
\end{equation}
Here,
\begin{eqnarray}
f_3^0(\hat{1},2,3)&=&\frac{1}{2(Q^2)^2}\Big(\frac{8s_{13}^2}{s_{12}}+\frac{8s_{23}^2}{s_{12}}
+\frac{12s_{23}s_{13}}{s_{12}}+\frac{4s_{13}^3}{s_{23}s_{12}}+\frac{4s_{23}^3}{s_{12}(s_{12}+s_{13})}\nonumber\\
&&+\frac{8s_{13}^2}{s_{23}}+\frac{6s_{12}s_{13}}{s_{23}}
+12s_{23}+12s_{12}+12s_{13}+ {\cal O}(\epsilon)\ \Big). 
\end{eqnarray}
The subantenna $f_3^0(\hat{1},j,k)$ has the full $j$ soft limit, the full $1\parallel j$ limit and part of the $j\parallel k$ limit of the full antenna (\ref{eq:bigFif}), such that we can identify $\hat{1}$ as the initial state radiator and $k$ the final state radiator. To numerically implement this antenna we use the \{3$\to$2\} mapping, $(\hat{i},j,k)\to(\hat{I},K)$~\cite{Daleo:2006xa} given by (\ref{3to2IFmap}).

\subsubsection{Initial-Initial emitters}
\label{sec:ii}
The initial-initial gluon-gluon-gluon antenna is obtained by crossing symmetry from the corresponding initial-final antenna function (\ref{eq:bigFif}), with the replacements 
$s_{12}\to(p_1+p_2)^2$, 
$s_{13}\to(p_1-p_3)^2$, $s_{23}\to(p_2-p_3)^2$ and $Q^2=s_{12}+s_{13}+s_{23}$. It reads \cite{Daleo:2006xa},
\begin{eqnarray}
&&F_3^0(\hat{1}_g,3_g,\hat{2}_g)=\frac{1}{2(Q^2)^2}\Bigg(\frac{8s_{13}^2}{s_{23}}+\frac{8s_{13}^2}{s_{12}}
+\frac{8s_{23}^2}{s_{13}}+\frac{8s_{23}^2}{s_{12}}+\frac{8s_{12}^2}{s_{13}}
+\frac{8s_{12}^2}{s_{23}}\nonumber\\
&&+\frac{12s_{13}s_{23}}{s_{12}}+\frac{12s_{12}s_{23}}{s_{13}}
+\frac{12s_{13}s_{12}}{s_{23}}
+\frac{4s_{13}^3}{s_{12}s_{23}}+\frac{4s_{23}^3}{s_{12}s_{13}}
+\frac{4s_{12}^3}{s_{13}s_{23}}\nonumber\\
&&+24s_{12}+24s_{13}+24s_{23}+ {\cal O}(\epsilon)\ \Bigg),
\label{eq:bigFii}
\end{eqnarray}
where the hat identifies the gluons crossed to the initial state. Only the final state gluon $j$ may be soft, and it can also be collinear with the initial state gluons $\hat{i}$ or $\hat{k}$.
Having well defined hard radiators, $F_3^0(\hat{i},j,\hat{k})$
does not need to be further decomposed. The full antenna can be used with a single
initial-initial mapping, $(\hat{i},j,\hat{k})\to(\hat{I},\hat{K})$ \cite{Daleo:2006xa} of the type given in Eq.~(\ref{3to2IImap}).

\subsection{Integrated tree-level three-parton antennae}
\label{subsec:integratedthree}
In this subsection we give the expressions for the integrated forms of the antennae in (\ref{eq:bigFff}), (\ref{eq:bigFif}) and (\ref{eq:bigFii}).
The integrated three-parton antennae contain explicit $\e$-poles from the integration over the antenna phase space for one unresolved emission and
finite remainders. They appear in the real virtual channel as single analytic integration of subtractions in the $(m+2)$-parton channel, corresponding to the double real emission~\cite{Glover:2010im},
as well as in the form of genuine subtraction terms to compensate for oversubtracted poles as discussed in Section \ref{sec:sigb}. Note that the full $\e$ dependence in $F_3^0$ is retained during integration over the antenna phase space.

\subsubsection{Final-Final emitters}
For final-final kinematics the integrated antenna was computed in~\cite{GehrmannDeRidder:2005cm} and reads,
\begin{eqnarray}
{\cal F}_3^0(s_{123}) = - 6 {\bf I}^{(1)}_{gg} \left(\e,s_{123}  \right)+\frac{73}{4} + {\cal O}(\e)\;,
\end{eqnarray}
where the colour-ordered infrared singularity operator ${\bom I}_{gg}^{(1)}$ was defined in Eq.~(\ref{eq:Ione}).

\subsubsection{Initial-Final emitters}
The full set of integrated initial-final three-parton tree-level antennae were computed in~\cite{Daleo:2006xa}. The pure gluon antenna reads,
\begin{eqnarray}
{\cal F}^0_{1,jk}(s_{\bar{1}k},x_{1},x_{2})&=&\delta(1-x_{2})\Bigg(\delta(1-x_{1})\l(-4{\bf I}_{gg}^{(1)}(\e,s_{\bar{1}k})+\frac{67}{18}-\frac{1}{3}\pi^2\r)\nonumber\\
&+&\left(\frac{-s_{\bar{1}k}}{\mu^2}\right)^{-\e}\Bigg[
-\frac{1}{\eps}p^{0}_{gg}(x_{1})-\frac{11}{6}{\cal D}_{0}(x_{1})+2{\cal D}_{1}(x_{1})
-\frac{11}{6x_{1}}\nonumber\\
&-&\frac{2(1-2x_{1}+x_{1}^2-x_{1}^3)}{x_{1}}\h(1,x_{1})
-\frac{2(1-x_{1}+x_{1}^2)^2}{x_{1}(1-x_{1})}\h(0,x_{1})+{\cal O}(\eps)\Bigg]\Bigg).\nonumber\\
\end{eqnarray}
The splitting kernels, $p_{gg}^{0}(x)$, and distributions ${\cal D}_{n}(x_{1})$ that appear above
are defined in Eqs.~(\ref{eq:pgg0}), and (\ref{eq:Dn})
respectively. The function $H(m_1,\hdots,m_{w};y)$ denotes the harmonic polylogarithms and their notation
is also described in Section~\ref{sec:appendixD}.

\subsubsection{Initial-Initial emitters}
The full set of integrated initial-initial three-parton tree-level antennae were computed in~\cite{Daleo:2006xa}.  For this kinematic configuration
the ${\cal F}_{3}^{0}$ antenna reads,
\begin{eqnarray}
{\cal F}^{0}_{12,j}(s_{\bar{1}\bar{2}})&=&\delta(1-x_{1})\delta(1-x_{2})\l(-{\bf I}_{gg}^{(1)}(\e,s_{\bar{1}\bar{2}})+\frac{1}{4}\pi^2\r)\nonumber\\
&+&\delta(1-x_{1})\Bigg(\left(\frac{s_{\bar{1}\bar{2}}}{\mu^2}\right)^{-\e}\left[-\frac{1}{2\e}p_{gg}(x_{2})\right]+{\cal D}_{1}(x_{2})
+ \frac{ \left(\log(2)-\h(-1;x_2)\right)}{1-x_2}\nonumber\\
&-&\frac{
   \left(x_2^3-x_2^2+2 x_2-1\right) \left(\log(2)-\h(1,x_2)-\h(-1;x_2)\right)}{x_2}\Bigg)\nonumber\\
&+&\left(-x_{2}^2+x_{2}-2+\frac{1}{x_{2}}\right){\cal D}_{0}(x_{1})
+ \frac{1}{2}{\cal D}_{0}(x_1){\cal D}_{0}(x_2)-\frac{3}{2(1+x_1) (1+x_2)}\nonumber\\
&+&\frac{2 \left(x_1^4-x_1^2+2\right) (1+x_1 x_2)^2}{(x_1+x_2)^4}
+\frac{ x_1^2- x_1+2+\frac{2}{(x_2+1) x_1}-\frac{1}{x_1}}{1-x_2}\nonumber\\
&-&\frac{2 \left(x_1^4+x_1^2+\left(x_1^2+3\right) x_1+2\right) (1+x_1 x_2)^2}{(1-x_2^2)
   (x_1+x_2)^4}\nonumber\\
&+&\frac{3 \left(x_1^3+x_1\right) \left(x_2^3+x_2\right) (1+x_1 x_2)}{(x_1+x_2)^4}\nonumber\\
&+&\frac{2 \left(x_2^2 x_1^4+\left(x_2^2+1\right)^2 x_1^2+x_2^2\right) (1+x_1  x_2)}{(x_1+x_2)^4}
+\frac{2 x_1 \left(x_1^2+3\right) (1+x_1 x_2)^2}{(x_2+1) (x_1+x_2)^4}\nonumber\\
&+&\frac{(1+x_1 x_2)}{x_1 (x_1+1) x_2 (x_2+1)}
+(x_1\leftrightarrow x_2) + {\cal O}(\e).
\end{eqnarray}

\subsection{One-loop three-parton antennae}
\label{subsec:oneloopantenna}
The one-loop antenna functions are obtained from the colour-ordered renormalised one-loop three-parton matrix elements according to Eq.~\eqref{eq:X1def} \cite{GehrmannDeRidder:2005cm}.
These contain explicit poles from the loop integration. The integrated forms for the final-final, initial-final and initial-initial cases are available in \cite{GehrmannDeRidder:2005cm}, \cite{Daleo:2009yj} and \cite{Gehrmann:2011wi} respectively.

\subsubsection{Final-Final emitters}
The gluon-gluon antenna functions are derived from the physical matrix elements for $H\to \textrm{(partons)}$~\cite{GehrmannDeRidder:2005aw}. At one-loop, the leading colour term reads~\cite{GehrmannDeRidder:2005cm}
\begin{eqnarray}
\label{eq:F31ff}
F_3^1(1_g,2_g,3_g)&=& 
2 \bigg( {\bf I}^{(1)}_{gg} (\e,s_{12}) + {\bf I}^{(1)}_{gg} (\e,s_{13})
+ {\bf I}^{(1)}_{gg} (\e,s_{23})- 2\,{\bf I}^{(1)}_{gg} (\e,s_{1 2 3}) \bigg) F_3^0(1,2,3) \nonumber \\
&-&
\bigg(  R(y_{12},y_{13}) 
+  R(y_{13},y_{23}) +  R(y_{12},y_{23})
 + \frac{11}{6} \h(0; |y_{12}|) + \frac{11}{6} \h(0; |y_{13}|)
\nonumber \\ &+& 
 \frac{11}{6} \h(0; |y_{23}|) \bigg)
F_3^0(1,2,3) 
  +  \frac{1}{3s_{12}}+  \frac{1}{3s_{13}}
+  \frac{1}{3s_{23}}+  \frac{1}{3s_{123}}
\end{eqnarray}
with $y_{ij}  = s_{ij}/s_{123}$ and where the colour-ordered infrared singularity operator ${\bom I}_{gg}^{(1)}$ was defined in (\ref{eq:Ione}). 
The function $R(y,z)$ is defined by
\begin{eqnarray}
\label{eq:RRyz}
R(y,z)  
&=& \h(0;|y|)\h(0;|z|) + \frac{\pi^2}{6} - \Theta(y<0)\Theta(z<0) \pi^2 
\nonumber \\
&& + \Theta(0\leq y \leq 1) \h(1,0;y) \nonumber \\
&& + \Theta(y<0) \left[ -\h\left(1,0;-\frac{y}{1-y}\right)
- \h\left(1,1;-\frac{y}{1-y}\right) \right] \nonumber \\
&& + \Theta(y>1) \left[ -\h\left(1,0;\frac{1}{y}\right)
- \h\left(0,0;\frac{1}{y}\right)-\frac{\pi^2}{3} \right] \nonumber \\
&& + \Theta(0\leq z \leq 1) \h(1,0;z) \nonumber \\
&& + \Theta(z<0) \left[ -\h\left(1,0;-\frac{z}{1-z}\right)
- \h\left(1,1;-\frac{z}{1-z}\right) \right] \nonumber \\
&& + \Theta(z>1) \left[ -\h\left(1,0;\frac{1}{z}\right)
- \h\left(0,0;\frac{1}{z}\right)-\frac{\pi^2}{3} \right] \;.
\end{eqnarray}
For the final-final case $s_{ij}=(p_{i}+p_{j})^2$ and the antenna is 
evaluated in the region where all $s_{ij}$ are positive.

The numerical implementation requires the partonic emissions to be ordered meaning that the two hard radiator partons defining the antenna are uniquely identified.
For $F_{3}^{1}(1,2,3)$ the decomposition into subantennae follows the same pattern of the tree-level antenna since it can be written as a function
proportional to its tree-level counterpart plus a function which is not singular in any unresolved limit
\begin{equation}
F_{3}^{1}(1,2,3)=f_{3}^{1}(1,3,2)+f_{3}^{1}(3,2,1)+f_{3}^{1}(2,1,3).
\end{equation}

The subantenna reads
\begin{eqnarray}
\label{eq:litlef31ff}
f_3^1(1_g,2_g,3_g)&=& 
2 \bigg( {\bf I}^{(1)}_{gg} (\e,s_{12}) + {\bf I}^{(1)}_{gg} (\e,s_{13})
+ {\bf I}^{(1)}_{gg} (\e,s_{23})- 2\,{\bf I}^{(1)}_{gg} (\e,s_{1 2 3}) \bigg) f_3^0(1,2,3) \nonumber \\
&-&
\bigg(  R(y_{12},y_{13}) 
+  R(y_{13},y_{23}) +  R(y_{12},y_{23})
 + \frac{11}{6} \h(0; |y_{12}|) + \frac{11}{6} \h(0; |y_{13}|)
\nonumber \\ &+& 
 \frac{11}{6} \h(0; |y_{23}|) \bigg)
f_3^0(1,2,3) 
  +  \frac{1}{6s_{12}}
+  \frac{1}{6s_{23}}+  \frac{1}{9s_{123}}.
\end{eqnarray}

\subsubsection{Initial-Final emitters}
The initial-final one-loop three-gluon antenna function can be obtained from its final-final counterpart \eqref{eq:F31ff} by the appropriate crossing of one of the particles from the final to the initial state, 
i.e.\ by making the replacements,
$s_{23}\to(p_2+p_3)^2>0$,
$s_{12}\to(p_1-p_2)^2<0$,
$s_{13}\to(p_1-p_3)^2<0$ and $s_{123}\to q^2=s_{12}+s_{13}+s_{23}<0$.

Some care needs to be 
taken in the continuation, since the final-final antenna function is 
renormalised at $\mu^2=q^2=s_{123}$, while the initial-final antenna function 
is renormalised at $\mu^2 = -q^2 = -s_{123}$~\cite{Daleo:2009yj}.  
With this in mind, and defining again the $y_{ij}=s_{ij}/s_{123}$, Eq.~(\ref{eq:F31ff}) can 
be taken over to the initial-final case as:
\begin{eqnarray}
F_3^1(\hat{1}_g,2_g,3_g)&=& 
2 \bigg( {\bf I}^{(1)}_{gg} (\e,s_{12}) + {\bf I}^{(1)}_{gg} (\e,s_{13})
+ {\bf I}^{(1)}_{gg} (\e,s_{23})- 2\,{\bf I}^{(1)}_{gg} (\e,s_{1 2 3}) \bigg) F_3^0(\hat{1},2,3) \nonumber \\
&-&
\bigg(  R(y_{12},y_{13}) 
+  R(y_{13},y_{23}) +  R(y_{12},y_{23})
 + \frac{11}{6} \h(0; |y_{12}|) + \frac{11}{6} \h(0; |y_{13}|)
\nonumber \\ &+& 
 \frac{11}{6} \h(0; |y_{23}|) \bigg)
F_3^0(\hat{1},2,3) 
  +  \frac{1}{3s_{12}}+  \frac{1}{3s_{13}}
+  \frac{1}{3s_{23}}+  \frac{1}{3s_{123}}\;.
\end{eqnarray}

Just as in the tree-level case, the one-loop initial-final antenna is decomposed symmetrically into subantennae containing only unresolved limits interpolated by the appropriate initial-final
phase space mapping,
\begin{equation}
F_{3}^{1}(\hat{1},2,3)=f_{3}^{1}(\hat{1},2,3)+f_{3}^{1}(\hat{1},3,2),
\end{equation}
\begin{eqnarray}
\label{eq:litlef31if}
f_3^1(\hat{1}_g,2_g,3_g)&=& 
2 \bigg( {\bf I}^{(1)}_{gg} (\e,s_{12}) + {\bf I}^{(1)}_{gg} (\e,s_{13})
+ {\bf I}^{(1)}_{gg} (\e,s_{23})- 2\,{\bf I}^{(1)}_{gg} (\e,s_{1 2 3}) \bigg) f_3^0(\hat{1},2,3) \nonumber \\
&-&
\bigg(  R(y_{12},y_{13}) 
+  R(y_{13},y_{23}) +  R(y_{12},y_{23})
 + \frac{11}{6} \h(0; |y_{12}|) + \frac{11}{6} \h(0; |y_{13}|)
\nonumber \\ &+& 
 \frac{11}{6} \h(0; |y_{23}|) \bigg)
f_3^0(\hat{1},2,3) 
  +  \frac{1}{3s_{12}}
+  \frac{1}{6s_{23}}+  \frac{1}{6s_{123}}\;.
\end{eqnarray}

\subsubsection{Initial-Initial emitters}
The initial-initial one-loop three-gluon antenna function is also obtained from \eqref{eq:F31ff} by crossing two of the gluons into the initial state with the replacements $s_{23}\to(p_{2}-p_{3})^2<0$, 
$s_{12}\to(p_{1}+p_{2})^2>0$, $s_{13}\to(p_{1}-p_{3})^2<0$ and $s_{123}\to
Q^2=s_{12}+s_{13}+s_{23}>0$. The expression yields:
\begin{eqnarray}
F_3^1(\hat{1}_g,2_g,\hat{3}_g)&=& 
2 \bigg( {\bf I}^{(1)}_{gg} (\e,s_{12}) + {\bf I}^{(1)}_{gg} (\e,s_{13})
+ {\bf I}^{(1)}_{gg} (\e,s_{23})- 2\,{\bf I}^{(1)}_{gg} (\e,s_{1 2 3}) \bigg) F_3^0(\hat{1},2,\hat{3}) \nonumber \\
&-&
\bigg(  R(y_{12},y_{13}) 
+  R(y_{13},y_{23}) +  R(y_{12},y_{23})
 + \frac{11}{6} \h(0; |y_{12}|) + \frac{11}{6} \h(0; |y_{13}|)
\nonumber \\ &+& 
 \frac{11}{6} \h(0; |y_{23}|) \bigg)
F_3^0(\hat{1},2,\hat{3}) 
  +  \frac{1}{3s_{12}}+  \frac{1}{3s_{13}}
+  \frac{1}{3s_{23}}+  \frac{1}{3s_{123}}\;.
\end{eqnarray}

Since the hard radiators are uniquely identified with the initial-state 
partons, no further decomposition is necessary. 

\label{sec:appendix}
\section{Large angle soft terms for the double real subtraction term $\dsigma^S_{NNLO}$}
As discussed in Sect.~\ref{sec:RVsub}, it is convenient to rewrite the large angle soft terms in a form where the first mapping is of final-final type \eqref{3to2FFmap}.  In this case, one should replace the wide angle soft contributions given in eqs.~(5.9),
(5.13) and (5.17) of Ref.~\cite{Glover:2010im} by eqs.~\eqref{eq:LAST1},
\eqref{eq:LAST2} and \eqref{eq:LAST3} respectively.

\subsection{IIFFFF topology}
\begin{eqnarray}
\lefteqn{{\rm d}\hat\sigma_{NNLO}^{e,X_6}= {\cal N}_{LO} \left(\frac{\alpha_s N}{2\pi}\right)^2\frac{\bar{C}(\epsilon)^2}{C(\epsilon)^2} 
{\rm d}\Phi_{4}(p_{3},\ldots,p_{6};p_{1},p_{2})
 \, \frac{2}{4!}\,\sum_{(ijkl)}\,\Bigg\{ }\nonumber \\
\ph{(ss2)}&&\phantom{+}\frac{1}{2}\Big(-S_{2l((il)j)}+S_{2l(il)}-S_{1l((kl)j)}+S_{1l(kl)}
+S_{((il)j)l((kl)j)}-S_{(il)l(kl)}\Big)\nonumber\\
&&\hspace{0.5cm}\times f_3^0(\widetilde{(il)}_g,j_g,\widetilde{(kl)}_g)
A_4^0(\hat{1}_g,\hat{2}_g,\widetilde{((il)j)}_g,\widetilde{((kl)j)}_g)
J_2^{(2)}(\widetilde{p_{(il)j}},\widetilde{p_{(kl)j}})\nonumber\\
\ph{ss1}&&+\frac{1}{2}\Big(-S_{2i((il)k)}+S_{2i(il)}-S_{1i((ij)k)}+S_{1i(ij)}
+S_{((il)k)i((ij)k)}-S_{(il)i(ij)}\Big)\nonumber\\
&&\hspace{0.5cm}\times f_3^0(\widetilde{(il)}_g,k_g,\widetilde{(ij)}_g)
A_4^0(\hat{1}_g,\hat{2}_g,\widetilde{((il)k)}_g,\widetilde{((ij)k)}_g)
J_2^{(2)}(\widetilde{p_{(il)k}},\widetilde{p_{(ij)k}})\nonumber\\
\ph{ss3}&&+\frac{1}{2}\Big(-S_{(il)l((kl)j)}+S_{(il)l(kl)}+S_{\bar{1}l((kl)j)}-S_{1l(kl)}\Big)\nonumber\\
&&\hspace{0.5cm}\times f_3^0(\hat{1}_g,j_g,\widetilde{(kl)}_g)
A_4^0(\hat{\bar{1}}_g,\hat{2}_g,\widetilde{(il)}_g,\widetilde{((kl)j)}_g)
J_2^{(2)}(\widetilde{p_{(il)}},\widetilde{p_{(kl)j}})\nonumber\\
\ph{ss4}&&+\frac{1}{2}\Big(-S_{(ij)j((kj)l)}+S_{(ij)j(kj)}+S_{\bar{1}j((kj)l)}-S_{1j(kj)}\Big)\nonumber\\
&&\hspace{0.5cm}\times f_3^0(\hat{1}_g,l_g,\widetilde{(kj)}_g)
A_4^0(\hat{\bar{1}}_g,\hat{2}_g,\widetilde{(ij)}_g,\widetilde{((kj)l)}_g)
J_2^{(2)}(\widetilde{p_{(ij)}},\widetilde{p_{(kj)l}})\nonumber\\
\ph{ss5}&&+\frac{1}{2}\Big(-S_{(lk)k((jk)i)}+S_{(lk)k(jk)}+S_{\bar{2}k((jk)i)}-S_{2k(jk)}\Big)\nonumber\\
&&\hspace{0.5cm}\times f_3^0(\hat{2}_g,i_g,\widetilde{(jk)}_g)
A_4^0(\hat{1}_g,\hat{\bar{2}}_g,\widetilde{((jk)i)}_g,\widetilde{(lk)}_g)
J_2^{(2)}(\widetilde{p_{(jk)i}},\widetilde{p_{(lk)}})\nonumber\\
\ph{ss6}&&+\frac{1}{2}\Big(-S_{(li)i((ji)k)}+S_{(li)i(ji)}+S_{\bar{2}i((ji)k)}-S_{2i(ji)}\Big)\nonumber\\
&&\hspace{0.5cm}\times f_3^0(\hat{2}_g,k_g,\widetilde{(ji)}_g)
A_4^0(\hat{1}_g,\hat{\bar{2}}_g,\widetilde{((ji)k)}_g,\widetilde{(li)}_g)
J_2^{(2)}(\widetilde{p_{(ji)k}},\widetilde{p_{(li)}})\nonumber\\
\ph{ss7}&&+\frac{1}{2}\Big(-S_{\bar{1}\tilde{i}\bar{2}}+S_{1i2}-S_{2i(ji)}+S_{\bar{2}\tilde{i}\widetilde{(ji)}}-S_{1i(il)}+S_{\bar{1}\tilde{i}\widetilde{(il)}}\Big)\nonumber\\
&&\hspace{0.5cm}\times F_3^0(\hat{1}_g,k_g,\hat{2}_g)
A_4^0(\hat{\bar{1}}_g,\hat{\bar{2}}_g,\widetilde{(ji)}_g,\widetilde{(il)}_g)
J_2^{(2)}(p_{\widetilde{(ji)}},p_{\widetilde{(il)}})\nonumber\\
\ph{ss8}&&+\frac{1}{2}\Big(-S_{\bar{1}\tilde{k}\bar{2}}+S_{1k2}-S_{2k(jk)}+S_{\bar{2}\tilde{k}\widetilde{(jk)}}-S_{1k(kl)}+S_{\bar{1}\tilde{k}\widetilde{(kl)}}\Big)\nonumber\\
&&\hspace{0.5cm}\times F_3^0(\hat{1}_g,i_g,\hat{2}_g)
A_4^0(\hat{\bar{1}}_g,\hat{\bar{2}}_g,\widetilde{(jk)}_g,\widetilde{(kl)}_g)
J_2^{(2)}(p_{\widetilde{(jk)}},p_{\widetilde{(kl)}})\Bigg\}.
\label{eq:LAST1}
\end{eqnarray}

\subsection{IFIFFF topology}
\begin{eqnarray}
\lefteqn{{\rm d}\hat\sigma_{NNLO}^{e,Y_6}={\cal N}_{LO} \left(\frac{\alpha_s N}{2\pi}\right)^2 \frac{\bar{C}(\epsilon)^2}{C(\epsilon)^2}
{\rm d}\Phi_{4}(p_{3},\ldots,p_{6};p_{1},p_{2})
 \, \frac{2}{4!}\,\sum_{P_C(j,k,l)}\Bigg\{ }\nonumber \\
\ph{(4s1)}&&\phantom{+}\left(-S_{\bar{2}\tilde{j}\widetilde{(kj)}}+S_{2j(kj)}
-S_{\bar{1}\tilde{j}\widetilde{(jl)}}+S_{1j(jl)}
-S_{2j1}+S_{\bar{2}\tilde{j}\bar{1}}\right)
F_3^0(\hat{1}_g,i_g,\hat{2}_g)\nonumber\\
&&\hspace{0.3cm}\times A^0_{4}(\hat{\bar{1}}_g,\hat{\bar{2}}_g,\widetilde{(kj)}_g,\widetilde{(jl)}_g)
{J}_{2}^{(2)}(p_{\widetilde{(kj)}},p_{\widetilde{(jl)}})\nonumber\\
\ph{(4s2)}&&\phantom{+}\left(-S_{\bar{2}\tilde{l}\widetilde{(kl)}}+S_{2l(kl)}
-S_{\bar{1}\tilde{l}\widetilde{(jl)}}+S_{1l(lj)}
-S_{2l1}+S_{\bar{2}\tilde{l}\bar{1}}\right)
F_3^0(\hat{1}_g,i_g,\hat{2}_g)\nonumber\\
&&\hspace{0.3cm}\times A^0_{4}(\hat{\bar{1}}_g,\hat{\bar{2}}_g,\widetilde{(kl)}_g,\widetilde{(lj)}_g)
{J}_{2}^{(2)}(p_{\widetilde{(kl)}},p_{\widetilde{(lj)}})\nonumber\\
\ph{(4s3)}&&+\frac{1}{2}\left(S_{\bar{2}\tilde{j}\widetilde{(jk)}}-S_{2j(jk)}
+S_{\bar{1}\tilde{j}\widetilde{(jk)}}-S_{1j(jk)}
+S_{2j1}-S_{\bar{2}\tilde{j}\bar{1}}\right)
F_3^0(\hat{1}_g,l_g,\hat{2}_g)\nonumber\\
&&\hspace{0.3cm}\times A^0_{4}(\hat{\bar{1}}_g,\widetilde{(ij)}_g,\hat{\bar{2}}_g,\widetilde{(jk)}_g)
{J}_{2}^{(2)}(p_{\widetilde{(ij)}},p_{\widetilde{(jk)}})\nonumber\\
\ph{(4s4)}&&+\frac{1}{2}\left(S_{\bar{2}\tilde{l}\widetilde{(lk)}}-S_{2l(lk)}
+S_{\bar{1}\tilde{l}\widetilde{(lk)}}-S_{1l(lk)}
+S_{2l1}-S_{\bar{2}\tilde{l}\bar{1}}\right)
F_3^0(\hat{1}_g,j_g,\hat{2}_g)\nonumber\\
&&\hspace{0.3cm}\times A^0_{4}(\hat{\bar{1}}_g,\widetilde{(il)}_g,\hat{\bar{2}}_g,\widetilde{(lk)}_g)
{J}_{2}^{(2)}(p_{\widetilde{(il)}},p_{\widetilde{(lk)}})\nonumber\\
\ph{4s5}&&+\frac{1}{2}\left(-S_{2j((kj)l)}+S_{2j(kj)}
-S_{1j(kj)}+S_{\bar{1}j(l(kj))}\right)
f_3^0(\hat{1}_g,l_g,\widetilde{(kj)}_g)\nonumber\\
&&\hspace{0.3cm}\times A^0_{4}(\hat{\bar{1}}_g,\widetilde{(ij)}_g,\hat{2}_g,\widetilde{((kj)l)}_g)
{J}_{2}^{(2)}(p_{\widetilde{(ij)}},p_{\widetilde{(kj)l)}})\nonumber\\
\ph{4s6}&&+\frac{1}{2}\left(-S_{2l((kl)j)}+S_{2l(kl)}
-S_{1l(kl)}+S_{\bar{1}l(j(kl))}\right)
f_3^0(\hat{1}_g,j_g,\widetilde{(kl)}_g)\nonumber\\
&&\hspace{0.3cm}\times A^0_{4}(\hat{\bar{1}}_g,\widetilde{(il)}_g,\hat{2}_g,\widetilde{((kl)j)}_g)
{J}_{2}^{(2)}(p_{\widetilde{(il)}},p_{\widetilde{(kl)j)}})\nonumber\\
\ph{4s7}&&+\frac{1}{2}\left(-S_{1j((kj)l)}+S_{1j(kj)}
-S_{2j(kj)}+S_{\bar{2}j(l(kj))}\right)
f_3^0(\hat{2}_g,l_g,\widetilde{(kj)}_g)\nonumber\\
&&\hspace{0.3cm}\times A^0_{4}(\hat{1}_g,\widetilde{(ij)}_g,\hat{\bar{2}}_g,\widetilde{((kj)l)}_g)
{J}_{2}^{(2)}(p_{\widetilde{(ij)}},p_{\widetilde{(kj)l)}})\nonumber\\
\ph{4s8}&&+\frac{1}{2}\left(-S_{1l((kl)j)}+S_{1l(kl)}
-S_{2l(kl)}+S_{\bar{2}l(j(kl))}\right)
f_3^0(\hat{2}_g,j_g,\widetilde{(kl)}_g)\nonumber\\
&&\hspace{0.3cm}\times A^0_{4}(\hat{1}_g,\widetilde{(il)}_g,\hat{\bar{2}}_g,\widetilde{((kl)j)}_g)
{J}_{2}^{(2)}(p_{\widetilde{(il)}},p_{\widetilde{(kl)j)}})\Bigg\}. 
\label{eq:LAST2}
\end{eqnarray}

\subsection{IFFIFF topology}
\begin{eqnarray}
\lefteqn{{\rm d}\hat\sigma_{NNLO}^{e,Z_6}= {\cal N}_{LO} \left(\frac{\alpha_s N}{2\pi}\right)^2\frac{\bar{C}(\epsilon)^2}{C(\epsilon)^2}
{\rm d}\Phi_{4}(p_{3},\ldots,p_{6};p_{1},p_{2})
 \, \frac{2}{4!}\,\Bigg\{ }\nonumber \\
\ph{(3s)}&&\phantom{+}\frac{1}{2}\left(-S_{1i2}+S_{\bar{1}\tilde{i}\bar{2}}
+S_{2i(ji)}-S_{\bar{2}\tilde{i}\widetilde{(ji)}}
+S_{1i(ji)}-S_{\bar{1}\tilde{i}\widetilde{(ji)}}\right)
F_3^0(\hat{1}_g,\hat{2}_g,l_g)\nonumber\\
&&\hspace{0.3cm}\times
A^0_{4}(\hat{\bar{1}}_g,\widetilde{(ji)}_g,\hat{\bar{2}}_g,\widetilde{(ik)}_g)
{J}_{2}^{(2)}(p_{\widetilde{(ji)}},p_{\widetilde{(ik)}})\nonumber\\
\ph{(4s)}&&\phantom{+}\frac{1}{2}\left(-S_{1j2}+S_{\bar{1}\tilde{j}\bar{2}}
+S_{2j(ij)}-S_{\bar{2}\tilde{j}\widetilde{(ij)}}
+S_{1j(ij)}-S_{\bar{1}\tilde{j}\widetilde{(ij)}}\right)
F_3^0(\hat{1}_g,\hat{2}_g,k_g)\nonumber\\
&&\hspace{0.3cm}\times
A^0_{4}(\hat{\bar{1}}_g,\widetilde{(ij)}_g,\hat{\bar{2}}_g,\widetilde{(jl)}_g)
{J}_{2}^{(2)}(p_{\widetilde{(ij)}},p_{\widetilde{(jl)}})\nonumber\\
\ph{(5s)}&&\phantom{+}\frac{1}{2}\left(-S_{1k2}+S_{\bar{1}\tilde{k}\bar{2}}
+S_{2k(kl)}-S_{\bar{2}\tilde{k}\widetilde{(kl)}}
+S_{1k(kl)}-S_{\bar{1}\tilde{k}\widetilde{(kl)}}\right)
F_3^0(\hat{1}_g,\hat{2}_g,j_g)\nonumber\\
&&\hspace{0.3cm}\times
A^0_{4}(\hat{\bar{1}}_g,\widetilde{(ik)}_g,\hat{\bar{2}}_g,\widetilde{(kl)}_g)
{J}_{2}^{(2)}(p_{\widetilde{(ik)}},p_{\widetilde{(kl)}})\nonumber\\
\ph{(6s)}&&\phantom{+}\frac{1}{2}\left(-S_{1l2}+S_{\bar{1}\tilde{l}\bar{2}}
+S_{2l(lk)}-S_{\bar{2}\tilde{l}\widetilde{(lk)}}
+S_{1l(lk)}-S_{\bar{1}\tilde{l}\widetilde{(lk)}}\right)
F_3^0(\hat{1}_g,\hat{2}_g,i_g)\nonumber\\
&&\hspace{0.3cm}\times
A^0_{4}(\hat{\bar{1}}_g,\widetilde{(jl)}_g,\hat{\bar{2}}_g,\widetilde{(lk)}_g)
{J}_{2}^{(2)}(p_{\widetilde{(jl)}},p_{\widetilde{(lk)}})\Bigg\}. 
\label{eq:LAST3}
\end{eqnarray}

\section{Splitting kernels}
\label{sec:appendixD}

In this Appendix, we give explicit forms for the four-dimensional space-like splitting kernels used in the paper.  Note that we systematically extract a factor of $N$ from the splitting kernels, and furthermore, retain only the leading colour contribution. The relevant gluonic
splitting kernels for the purpose of this paper read~\cite{Vogt:2004mw},
\begin{eqnarray}
\label{eq:pgg0}
p_{gg}^{0}(y)&=&  \frac{11}{6}\delta(1-y)+2{\cal D}_{0}(y)+\frac{2}{y}-4+2y-2y^2 ,\\
\label{eq:pgg0otimespgg0}
\l(p_{gg}^{0}\otimes p_{gg}^{0}\r)(y)&=&\l(\frac{121}{36}-\frac{2\pi^2}{3}\r)\delta(1-y)
	+\frac{11}{3}\l(2{\cal D}_{0}(y)+\frac{2}{y}-4+2y-2y^2\r)\nonumber\\&&
	+8{\cal D}_{1}(y)-\frac{4\h(0;y)}{1-y}+12-12y+\frac{44}{3}y^2-12y\,\h(0;y)+4y^2\,\h(0;y)\nonumber\\&&
	-\frac{44}{3y}-\frac{4\,\h(0;y)}{y},\\
\label{eq:pgg1}
p_{gg}^{1}(y) &=&  \l( \frac{8}{3} + 3 \* \z3 \r)\delta(1-y)
           + 27
          + (1+y) \* \bigg(
         \frac{11}{3} \* \h(0;y)
          + 8 \* \h(0,0;y)
       - \frac{27}{2}
          \bigg)\nonumber \\ &&
          + 2 \*\left(\frac{1}{1+y} -\frac{1}{y}-2-y-y^2\right)
\* \bigg(
            \h(0,0;y)
          - 2 \* \h(-1,0;y)
          - \z2
          \bigg)\nonumber \\ &&
          - \frac{67}{9} \* \bigg(\frac{1}{y}-y^2\bigg)
          - 12 \* \h(0;y)
       - \frac{44}{3} \*y^2\*\h(0;y)\nonumber \\ &&
          + \l(2{\cal D}_{0}(y)+\frac{2}{y}-4+2y-2y^2\r) \nonumber \\
	  && \times \bigg(
            \frac{67}{18}
          - \z2
          + \h(0,0;y)
          + 2 \* \h(1,0;y)
          + 2 \* \h(0,1;y)
          \bigg)           
\end{eqnarray}
where we have introduced the distributions,
\beq
{\cal D}_{n}(y)=\left(\frac{\ln^{n}(1-y)}{1-y}\right)_{+}
\label{eq:Dn}
\eeq
and adopted the following notation for the harmonic polylogarithms $\h(m_1,...,m_w;y)$, 
$m_j = 0,\pm 1$; the lowest-weight ($w = 1$) functions $\h(m;y)$ are 
given by
\beq
\label{eq:hpol1}
  \h(0;y)       \: = \: \ln y \:\: , \quad\quad
  \h(\pm 1;y) \: = \: \mp \, \ln (1 \mp y) \:\: .
\eeq
The higher-weight ($w \geq 2$) functions are recursively defined as
\beq
\label{eq:hpol2}
  \h(m_1,...,m_w;y) \: = \:
    \left\{ \begin{array}{cl}
    \displaystyle{ \frac{1}{w!}\,\ln^w y \:\: ,}
       & \quad {\rm if} \:\:\: m^{}_1,...,m^{}_w = 0,\ldots ,0 \\[2ex]
    \displaystyle{ \int_0^y \! dz\: f_{m_1}(z) \, \h(m_2,...,m_w;z)
       \:\: , } & \quad {\rm otherwise}
    \end{array} \right.
\eeq
with
\beq
\label{eq:hpolf}
  f_0(y)       \: = \: \frac{1}{y} \:\: , \quad\quad
  f_{\pm 1}(y) \: = \: \frac{1}{1 \mp y} \:\: .
\eeq
We use the {\tt CHAPLIN} Fortran library~\cite{Buehler:2011ev} to evaluate the  harmonic polylogarithms up to weight four numerically. {\tt CHAPLIN} is based on a reduction of harmonic polylogarithms to a minimal set of basis functions that are computed numerically using series expansions and provide fast and reliable numerical results.

\bibliographystyle{JHEP-2}
\bibliography{ref}
\end{document}